\documentclass[aps,prd,10pt,twocolumn,superscriptaddress,showpacs,longbibliography]{revtex4-1}

\pdfoutput=1
\usepackage{hyperref}
\usepackage{graphicx}
\usepackage{amsfonts,amsmath,amssymb,bm,bbm}
\usepackage{color}

\hypersetup{
    pdfnewwindow=true,    % links in new window
    colorlinks=true,      % false: boxed links; true: colored links
    linkcolor=blue,       % color of internal links
    citecolor=blue,       % color of links to bibliography
    filecolor=blue,       % color of file links
    urlcolor=blue         % color of external links
}

\usepackage{dcolumn}
\newcolumntype{d}[1]{D{.}{.}{-1}}
\newcolumntype{C}[1]{>{\centering\let\newline\\\arraybackslash\hspace{0pt}}m{#1}}
\usepackage{multirow}
\usepackage{makecell}
\usepackage{tabularx}
\usepackage{capt-of}
\usepackage{siunitx}
\usepackage{enumitem}  

\newcolumntype{.}{D{.}{.}{-1}}
\makeatletter
\newcolumntype{B}[3]{>{\boldmath\DC@{#1}{#2}{#3}}c<{\DC@end}}

\begin{document}

\title{Developing the MeV potential of DUNE:\\ Detailed considerations of muon-induced spallation and other backgrounds}

\author{Guanying Zhu}
\email{zhu.1475@osu.edu}
\thanks{\scriptsize \!\! \href{https://orcid.org/0000-0003-0031-634X}{0000-0003-0031-634X}}
\affiliation{Center for Cosmology and AstroParticle Physics (CCAPP), Ohio State University, Columbus, Ohio 43210, USA}
\affiliation{Department of Physics, Ohio State University, Columbus, Ohio 43210, USA}

\author{Shirley Weishi Li}
\email{shirleyl@slac.stanford.edu}
\thanks{\scriptsize \!\! \href{https://orcid.org/0000-0002-2157-8982}{0000-0002-2157-8982}}
\affiliation{Center for Cosmology and AstroParticle Physics (CCAPP), Ohio State University, Columbus, Ohio 43210, USA}
\affiliation{Department of Physics, Ohio State University, Columbus, Ohio 43210, USA}
\affiliation{SLAC National Accelerator Laboratory, Menlo Park, California 94025, USA}

\author{John F. Beacom}
\email{beacom.7@osu.edu}
\thanks{\scriptsize \!\! \href{https://orcid.org/0000-0002-0005-2631}{0000-0002-0005-2631}}
\affiliation{Center for Cosmology and AstroParticle Physics (CCAPP), Ohio State University, Columbus, Ohio 43210, USA}
\affiliation{Department of Physics, Ohio State University, Columbus, Ohio 43210, USA}
\affiliation{Department of Astronomy, Ohio State University, Columbus, Ohio 43210, USA}

\date{\today}

\begin{abstract}
The Deep Underground Neutrino Experiment (DUNE) could be revolutionary for MeV neutrino astrophysics, because of its huge detector volume, unique event reconstruction capabilities, and excellent sensitivity to the $\nu_e$ flavor.  However, its backgrounds are not yet known.  A major background is expected due to muon spallation of argon, which produces unstable isotopes that later beta decay.  We present the first comprehensive study of MeV spallation backgrounds in argon, detailing isotope production mechanisms and decay properties, analyzing beta energy and time distributions, and proposing experimental cuts.  We show that above a nominal detection threshold of 5-MeV electron energy, the most important backgrounds are --- surprisingly --- due to low-A isotopes, such as Li, Be, and B, even though high-A isotopes near argon are abundantly produced.  We show that spallation backgrounds can be powerfully rejected by simple cuts, with clear paths for improvements.  We compare these background rates to rates of possible MeV astrophysical neutrino signals in DUNE, including solar neutrinos (detailed in a companion paper [Capozzi {\it et al.} \href{https://arxiv.org/abs/1808.08232}{arXiv:1808.08232} [hep-ph]]), supernova burst neutrinos, and the diffuse supernova neutrino background.  Further, to aid trigger strategies, in the Appendixes we quantify the rates of single and multiple MeV events due to spallation, radiogenic neutron capture, and other backgrounds, including through pileup.  Our overall conclusion is that DUNE has high potential for MeV neutrino astrophysics, but reaching this potential requires new experimental initiatives. 
\end{abstract}

\maketitle

%%%%%%%%%%%%%%%%%%%%%%%%%%%%%%%%%%%%%%%%%%%%%%%%%%%%%%
%%%%%%%%%%%%%%%%%%%%%%%%%%%%%%%%%%%%%%%%%%%%%%%%%%%%%%

\section{Introduction}
\label{sec: introduction}

Astrophysical neutrinos are uniquely penetrating probes of their sources, whose extreme physical conditions in turn allow for new tests of neutrino properties.  In the MeV energy range, there are three important targets: solar neutrinos~\cite{Bahcall:1987jc, Robertson:2012ib, Vissani:2017dto, Capozzi:2018dat}, supernova burst neutrinos~\cite{Burrows:1990ts, Cei:2002mq, Mezzacappa:2005ju, Janka:2012wk, Scholberg:2017czd}, and the diffuse supernova neutrino background (DSNB)~\cite{Ando:2004hc, Beacom:2010kk, Lunardini:2010ab}.  Despite great achievements in solar neutrino studies, opportunities remain for detailed tests of astrophysics (e.g., the first detection of the $hep$ flux) and particle physics (e.g., resolving the discrepancy between reactor and solar mixing parameters).  The next Galactic core-collapse supernova will enable multi-flavor neutrino measurements, revealing details of explosion physics and testing neutrino mixing at high densities.  Meanwhile, the DSNB could be detected as a steady source, which would probe the core-collapse history and test black-hole versus neutron-star formation.  With new experiments, exciting progress on these and other topics could be made.

DUNE, the leading next-generation neutrino experiment in the United States~\cite{Acciarri:2016crz, Acciarri:2015uup, Strait:2016mof, Acciarri:2016ooe, Abi:2018dnh, Abi:2018alz, Abi:2018rgm}, offers such opportunities.  Its principal science goals are to measure CP violation and the mass ordering, to search for nucleon decay, and to detect the next Galactic supernova burst.  For MeV neutrino astrophysics, DUNE has three key advantages.  First, its far detector will be huge, 20~kton in total for two modules (eventually twice that) of liquid argon (LAr), comparable to the current largest MeV neutrino detector, Super-Kamiokande.  Second, with the Liquid Argon Time-Projection Chamber (LArTPC) technology, DUNE will have excellent capabilities in event reconstruction, enabling the separation of different types of events (e.g., electrons vs.~gammas).  Third, the charged-current detection channel in DUNE ($\nu_e + \, ^{40}{\rm Ar} \rightarrow e^- + \, ^{40}{\rm K}^*$) isolates the $\nu_e$ flavor.  Compared to elastic scattering ($\nu_{e,\mu,\tau} + \, e^- \rightarrow \nu_{e,\mu,\tau} + \, e^-$), the main $\nu_e$ detection channel in current experiments, the charged-current channel has a much larger cross section and a much sharper correlation between neutrino energy and electron energy.  Therefore, new results from DUNE should powerfully complement results from previous and ongoing experiments.

Understanding the detector backgrounds is an essential step for successful MeV neutrino programs in DUNE.  Above 5~MeV electron energy, the nominal detection threshold, a significant background rate is expected from muon-induced spallation.  (Another, from radiative neutron captures in the detector due to neutrons produced by radioactivities in the rock, given some consideration below, is detailed in Ref.~\cite{Capozzi:2018dat}.)  When cosmic-ray muons pass through the detector, they produce secondary particles, which then occasionally break argon nuclei and make other isotopes.  Unstable daughter isotopes decay later, emitting betas, which can mimic neutrino signals.  Prior spallation studies on LAr were incomplete.  Barker et al.~\cite{Barker:2012nb} focused on only high-A isotopes (near argon) and Gehman et al.~\cite{Schoberg} considered only isotopes produced by high-energy neutrons.  More recently, Franco et al.~\cite{Franco:2015pha} produced a thorough study of spallation backgrounds for low-energy ($\leq 1.3$~MeV) solar neutrinos, but did not provide the details needed for the higher energies (5--20~MeV) of DUNE.  

Building on the work of Li and Beacom~\cite{Li:2014sea, Li:2015kpa, Li:2015lxa}, our goals for this paper are to calculate the spallation backgrounds for DUNE in detail, understand their physical mechanisms, and use this understanding to develop cuts to reject backgrounds.  We compare these background rates to the signal rates for solar neutrinos, supernova burst neutrinos, and the DSNB, finding that spallation backgrounds can be well controlled.  We aim for a factor of $\approx2$ precision on isotope yields, which is appropriate given the hadronic uncertainties.  This is adequate to guide development of DUNE as a detector for MeV neutrino astrophysics.  Once there are measurements, which could begin soon with surface-based detectors, the uncertainties can be reduced via empirical calibrations.

This paper is organized as follows.  In Sec.~\ref{sec: SK}, we overview the physics of spallation, using Super-Kamiokande as a concrete example.  Our main results are in Sec.~\ref{sec: DUNE}, where we calculate the isotope yields, describe the production mechanisms, show the component background energy spectra and time distributions, and detail our proposed background-rejection methods.  Using the projected post-cut background levels, we discuss the possible MeV neutrino programs in DUNE in Sec.~\ref{sec: signal}, along with new results to aid trigger development
(with the details provided in the Appendixes).  Finally, we present our conclusions in Sec.~\ref{sec: conclusions}.

%%%%%%%%%%%%%%%%%%%%%%%%%%%%%%%%%%%%%%%%%%%%%%%%%%%%%%
%%%%%%%%%%%%%%%%%%%%%%%%%%%%%%%%%%%%%%%%%%%%%%%%%%%%%%

\section{Overview of spallation}
\label{sec: SK}

In this section, we review the physics of spallation isotope production by cosmic-ray muons, which is now understood, due to Refs.~\cite{Li:2014sea, Li:2015kpa, Li:2015lxa}.  Though those papers focus on the water-based detector Super-Kamiokande, the results can be widely applied with modifications (e.g., muon flux, detector materials, etc.), including to the DUNE.  The most important concepts are as follows:

\begin{enumerate}[label=(\arabic*)]

\item Almost all isotopes are made by muon secondaries, not directly by muons.  (This point was known earlier~\cite{Galbiati:2004wx, Galbiati:2005ft}.)

\item Almost all of these secondaries are made in showers, which are relatively rare along muon tracks.

\item Almost all the isotope-producing secondaries are made in hadronic showers, which are even rarer.  ($^{11}$C, a dominant background isotope in oil, is made in electromagnetic showers.)

\item The positions of decaying isotopes produced by spallation can be constrained by localizing the preceding showers.  (This point was found empirically in Ref.~\cite{Bays:2011si}, though the physical reason was unknown.)  One could go further by identifying the showers that are hadronic.  

\end{enumerate}
We now detail spallation processes, beginning with cosmic-ray muon energy loss~\cite{Patrignani:2016xqp, Li:2014sea, Li:2015kpa, Li:2015lxa}.

Muons lose energy in two ways: ionization when interacting with atomic electrons and radiative processes when interacting with atomic nuclei.  The ionization losses have a typical value of $\approx2$~MeV~g$^{-1}$~cm$^{2}$, moderately depending on muon energy and the medium material.  These losses can be separated into a restricted energy loss from soft collisions and delta-ray production from hard collisions, where delta rays can be almost as energetic as the parent muon.  Radiative losses produce most secondary particles, and the rate rises with muon energy.  For muons up to several hundred GeV, the dominant radiative processes are pair production and bremsstrahlung.  Photonuclear interactions, a low-Q$^{2}$ analog to deep inelastic scattering, are less frequent. 

The first-generation particle production is closely associated with muon energy loss.  The most abundantly produced particles are electrons from delta-ray production, followed by electrons and positrons from pair production.  There are also some gamma rays, made mostly through bremsstrahlung.  Even though muons mainly lose energy electromagnetically, a small number of hadrons are made through photonuclear interactions.  The dominant hadrons are pions, which are almost equally distributed among the three charges.  In Super-Kamiokande, there are 3.6, 0.4, 0.04, and 0.003 daughter particles above 0.1, 1, 10, and 100~GeV per vertical muon (of track length 32.2~m)~\cite{Li:2015kpa}, showing that shower frequencies are small.

Isotopes are born primarily in showers induced by secondaries.  There are two types of showers.  Electrons, positrons, and gammas make electromagnetic showers, which typically have no hadronic components, aside from some low-energy neutrons made through ($\gamma,n$) interactions.  Pions induce hadronic showers, producing roughly equal numbers of $\pi^{+}$, $\pi^{-}$, $\pi^{0}$ in each generation, in analogy to $e^{+}$, $e^{-}$, $\gamma$ for electromagnetic showers.  Hadronic showers always have large electromagnetic components, because neutral pions decay promptly to gammas.  For Super-Kamiokande solar neutrino analyses, the most dangerous background isotopes are produced in the less frequent hadronic showers, where pions and neutrons are very efficient at making isotopes, due to the strong interactions.  There are also many unstable isotopes made from the more frequent electromagnetic showers, but those tend to have harmless decays, e.g., $^{15}$O, produced by $(\gamma,n)$, decays by electron capture.  Overall, isotope production in Super-Kamiokande is a rare process, with the most abundantly produced background isotope, $^{16}$N, having a yield of $\simeq 0.006$ per muon~\cite{Li:2014sea}.

A small fraction (e.g., $\approx7\%$ in Super-Kamiokande~\cite{Li:2014sea}) of isotopes are made by stopping muons.  Once $\mu^+$ are brought to rest, they simply decay.  However, once $\mu^-$ are brought to rest, nearly all undergo atomic capture, an electromagnetic process in which electrons are ejected, because muons are bound more tightly by a factor $\approx m_\mu/m_e$~\cite{Konopinski1950, Ponomarev:1973ya, Measday:2001yr}.  Of muons in atomic orbits, an appreciable fraction undergo nuclear capture, a weak process that converts $\mu^- + p \rightarrow \nu_\mu + n$, often removing several low-energy nucleons from the nucleus~\cite{Konopinski1950, Ponomarev:1973ya, Measday:2001yr}.  Because stopping muons enter the detector with only a few GeV, they have vanishing radiative losses and hence do not produce isotopes along their tracks.  Therefore, powerful cuts on isotopes produced by stopping muons can be made by concentrating cuts at the ends of their tracks.

The precision of predicting isotope yields, which is mostly limited by the uncertainties in hadronic processes, is typically a factor of $\approx2$.  For example, in Super-Kamiokande~\cite{Super-Kamiokande:2015xra}, the {\tt FLUKA}-predicted yields of some isotopes agree with measurements within a few tens of percent; some are off by a factor $\approx2$--3.  In Borexino~\cite{Bellini:2013pxa}, {\tt FLUKA} predictions also agree well with experimental measurements.  A few tens of percent agreement is found for some isotopes, but a factor of 2--4 for some others.  As for the predicted yields from {\tt GEANT4}, a factor of $\approx2$ agreement with data is observed for some isotopes, while a few differ by a factor of $\approx10$.  Overall, a factor of $\approx2$ precision is adequate as isotope yields usually differ by orders of magnitude.  Because the decay time profiles and energy spectra are known from laboratory data, all that is needed is the yield constants.  Theory is needed to get the predicted yields close enough to identify the key physical processes and to develop cuts, and then these predictions can be refined with experimental measurements.

There are multiple ways to cut the background betas from unstable-isotope decays.  The basic concepts of spallation cuts can be explained in a simplified picture.  While neutrino signals are uniform in the detector, spallation backgrounds are highly correlated with muons.  One strategy is a cylinder cut, where one discards all events inside a cylinder of radius $R$ surrounding each muon track in a time window of duration $\Delta t$.  The values of $R$ and $\Delta t$ are chosen such that both cut efficiency and the resulting deadtime are acceptable.  For example, in Super-Kamiokande solar neutrino analyses, $\simeq 90\%$ of the isotopes can be rejected with $\simeq 20\%$ deadtime via a likelihood-based version of this approach.  Separately, one can also cut on shower energy, because isotope yields rise with increasing muon energy losses.  In Super-Kamiokande, about 60\% of the spallation yields could be cut by rejecting the 2\% of muons with the highest energy losses~\cite{Li:2015kpa}.  Finally, more advanced cuts use reconstructions of the shower profiles~\cite{Bays:2011si, Li:2015lxa}.  Instead of cutting a whole cylinder for each muon, one could only cut where isotopes are made --- the rare shower regions.  This technique is under development for solar neutrino detection in Super-Kamiokande.  

%%%%%%%%%%%%%%%%%%%%%%%%%%%%%%%%%%%%%%%%%%%%%%%%%%%%%%
%%%%%%%%%%%%%%%%%%%%%%%%%%%%%%%%%%%%%%%%%%%%%%%%%%%%%%

\section{Spallation backgrounds in DUNE}
\label{sec: DUNE}

We calculate muon-induced spallation backgrounds for MeV astrophysics studies in DUNE.  Under some reasonable assumptions about the detector properties, we present our simulation inputs, the calculated isotope production rates, and the component background spectra.  We show that the spallation backgrounds are low after a simple two-step cut we propose, and could be improved.

%%%%%%%%%%%%%%%%%%%%%%%%%%%%%%%%%%%%%%%%%%%%%%%%%%%%%%

\subsection{Basic facts of DUNE}
\label{sec: DUNE_detector}

Located 4850~ft (4300~m.w.e.)~underground in the Homestake mine, the DUNE far detector will have two 10-kton (fiducial) modules of LArTPC deployed by 2024~\cite{Acciarri:2016crz}, and two more modules later.

The LArTPC technique that DUNE will use is superb in tracking and calorimetry performance~\cite{Marchionni:2013tfa, Acciarri:2016crz, Acciarri:2015uup, Strait:2016mof, Acciarri:2016ooe, Abi:2018dnh, Abi:2018alz, Abi:2018rgm}.  Charged particles cause ionization and excitation of argon atoms.  The ionization electrons then drift to wire planes at a speed of $\simeq 1.6$~mm~$\rm\mu s^{-1}$ under an applied electric field of $\simeq 500$~V~cm$^{-1}$ and form a two-dimensional (2-D) particle track~\cite{Acciarri:2016ooe}.  Combined with the timing information from prompt scintillation light emitted by argon excimer states, one can reconstruct the three-dimensional (3-D) image. 

For all three astrophysical sources we consider, the neutrinos either elastically scatter off electrons or have charged-current interactions with argon nuclei.  In the elastic-scattering channel, the final state is an electron.  In the charged-current channel, there would be one electron plus multiple gamma rays from $^{40}\rm K^{*}$ de-excitations.  These gamma rays do not typically overlap with the outgoing electron in space, because the 14-cm radiation length~\cite{Patrignani:2016xqp} is much larger than the position resolution in DUNE ($\simeq0.5$~cm~\cite{Acciarri:2016ooe}).  The ability to detect these gamma rays has recently been demonstrated for ArgoNeuT~\cite{Acciarri:2018myr}.  A precise gamma-ray energy reconstruction would aid neutrino energy reconstruction, and help signal and background separation (details below).  However, we conservatively assume no ability to separate charged-current events and elastic-scattering events, and take the signal for both channels to be one outgoing electron, following Ref.~\cite{Capozzi:2018dat}. 

Given such neutrino signals, we take the backgrounds to be just the betas from spallation isotope decays, including the radioactive decay types of $\beta$, $\beta+\gamma$, and $\beta+n$, of which the first is vastly dominant.  We especially focus on the energy range above 5~MeV electron energy, which could be a reasonable choice (see Ref.~\cite{Acciarri:2016ooe}), although the energy threshold could vary in different analysis programs.  We also remark on the background rates below 5~MeV to help with trigger design.  In addition, we expect good energy reconstructions for the spallation betas, because their energies ($\lesssim 20$~MeV) are well below the electron critical energy of 45~MeV~\cite{NIST}, and hence radiative losses are minimal~\cite{Acciarri:2017sjy}.  When smearing the background spectra in Sec.~\ref{sec:DUNE_aftercut}, we use a 7\% energy resolution~\cite{Acciarri:2016ooe}, of which the specific value has little effect on the continuum spectra, except for the tails.  In the Appendixes, we show results for 20\% energy resolution.

Our assumption of no tagging on the charged-current de-excitation gamma rays is conservative for two reasons.  First, if when detecting an electron, one could reliably detect an associated gamma ray, then this would allow better reconstruction of the neutrino energy in charged-current events, moving the signal spectrum significantly to higher energies ($\simeq 4$~MeV) while changing background spectrum much less ($\simeq 1$~MeV), greatly improving the physics case, because the backgrounds would be much less relevant.  Second, this would also allow one to easily separate charged-current events (mostly $\beta+\gamma$) from spallation events (mostly $\beta$), based on topology, again greatly improving the physics potential.

%%%%%%%%%%%%%%%%%%%%%%%%%%%%%%%%%%%%%%%%%%%%%%%%%%%%%%

\subsection{Setup of the calculation}
\label{sec: DUNE_fluka}

We use the Monte Carlo code {\tt FLUKA}~\cite{Bohlen:2014buj, Ferrari:2005zk} (version 2011.2x-1) to simulate muon propagation in liquid argon.  {\tt FLUKA} is a well-known package for simulating particle transport and interactions in matter on an event-by-event basis.  For hadron-nucleus interactions, {\tt FLUKA} uses its own {\tt PEANUT} model~\cite{fasso1994fluka, Ferrari:1993xr}.  It describes target nuclei with a local Fermi gas model, and hadronic inelastic interactions in a Generalized IntraNuclear Cascade (GINC) approach, where the cross sections used are a mixture of tabulated data and parametrized fits.  The GINC step continues until all nucleons are below 30--100~MeV and all non-nucleons (typically pions) are decayed or absorbed.  Then the preequilibrium stage takes over, which is mostly based on Geometry Dependent Hybrid Model.  At the end of the preequilibrium stage, a compound nucleus (Z, N) with known momentum and excitation energy is left, starting from which the evaporation/fission/fragmentation stage is modeled.  When the excitation energy of the residual nucleus is below the particle emission threshold, the remaining energy is released through gamma emission.
  
In our simulations, the {\tt PRECISIOn} card is used.  All relevant electromagnetic processes and hadronic processes are included, such as ionization, bremsstrahlung, pair production, photonuclear, Compton scattering, pion production and transport, photo-disintegration, low-energy neutron interactions, etc.  We switch on {\tt EVAPORAT} and {\tt COALESCE} through the {\tt PHYSICS} cards to enable accurate residual nuclei scoring.  When we calculate isotope production yields, the {\tt RADDECAY} card is switched off to make sure that only the isotopes produced by muon spallation are counted, i.e., not including daughter nuclei from spallation isotope decays (such as $^{39}$Ar from $^{39}$Cl decay).  It is later switched on when we simulate radioactive decays.

%+++++++++++++++++++++++++++++FIGURE++++++++++++++++++++++++++++++++++++++++%
\begin{figure}[!t]
\includegraphics[width=\columnwidth]{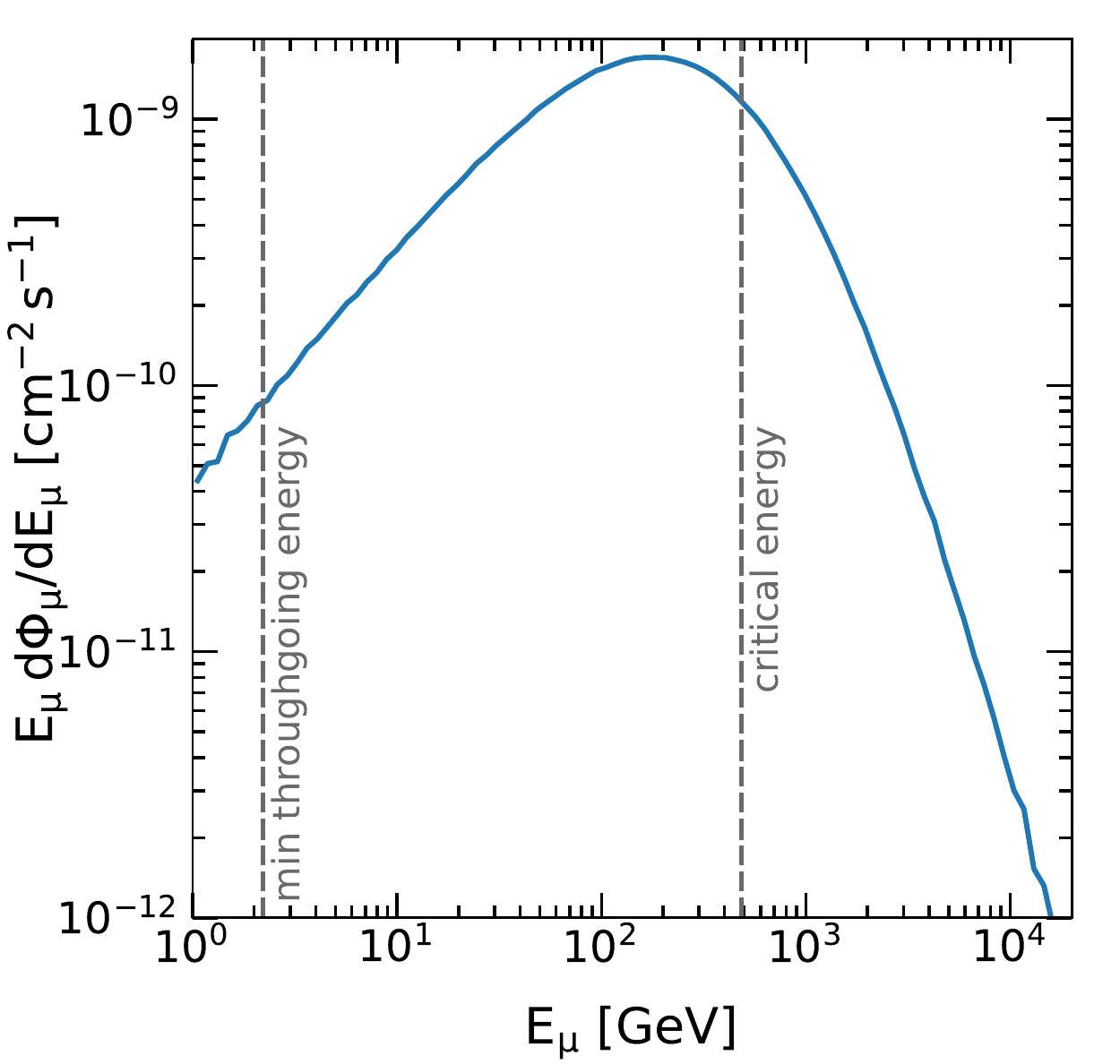}
\caption{Simulated cosmic-ray muon flux at DUNE (4850~ft underground) as a function of muon total energy~\cite{Vitaly, Kudryavtsev:2008qh, Klinger:2015kva}.  The dashed line near 2~GeV corresponds to the minimum-ionization energy loss for a vertical muon passing through DUNE's fiducial volume.  The dashed line at 484~GeV shows the muon critical energy in liquid argon.}
\label{fig: prm_mu}
\end{figure}
%++++++++++++++++++++++++++++++++++++++++++++++++++++++++++++++++++++++++++++%

The first main input is the detector setup.  Each module is a box of liquid argon (active volume), with dimensions 58~m (l) $\times$ 14.5~m (w) $\times$ 12~m (h).  To match the 10-kton fiducial mass, for which our results are calculated, we assume that the fiducial volume has dimensions 56~m (l) $\times$ 12.5~m (w) $\times$ 10~m (h).  The detector material is natural argon, consisting of 99.6\% $^{40}$Ar, 0.3\% $^{36}$Ar, and 0.1\% $^{38}$Ar.  Chemical impurities (water, air, etc.) and radioactive impurities ($^{39}$Ar, $^{42}$Ar) have tiny abundances, so we ignore them as possible targets for muons.  Outside of the active volume, we include 2~m of rock to induce full showers but not significantly affect the muon spectrum.  The rock chemical composition follows that given in Ref.~\cite{rogers:1990geology}.  In reality, there is $\approx1$~m of LAr cryostat layer outside of the active volume.  We have verified that our calculated isotope yields in the fiducial volume are not affected by more than a few tens of percent if we included that extra LAr layer in the simulation.  This is expected, because the 2~m of rock has enabled full shower development, and the production of muon secondaries is nearly material-independent. 

Figure~\ref{fig: prm_mu} shows the other main input, the simulated cosmic-ray muon spectrum, averaged over zenith angles, at the DUNE underground site, based on the simulations of Kudryavtsev et al.~\cite{Vitaly, Kudryavtsev:2008qh, Klinger:2015kva}.  In their calculations, the sea-level muon flux follows Gaisser's formula~\cite{Gaisser:1990vg}, modified for large zenith angles and prompt muon flux with the best fit to the LVD data~\cite{Aglietta:1998nx}; muon propagation throughout the rock is then carefully modeled in {\tt MUSIC/MUSUN}~\cite{Antonioli:1997qw, Kudryavtsev:2008qh}.  The good agreement with the measured muon flux by the Davis experiment~\cite{Cherry:1983dp} and the Majorana Demonstrator~\cite{Abgrall:2016cfi} shows that this simulated muon spectrum should be reliable.  We plot the spectrum as $E\,d\Phi/dE = 2.3^{-1}\,d\Phi/d\log_{10}E$, so that the relative heights at different energy decades correctly represent their relative contributions to the total flux of $\Phi_{\mu} = 5.66 \times 10^{-9}$~cm$^{-2}$~s$^{-1}$.  The muon rate in four modules of DUNE will be $\simeq 0.2$~Hz, roughly 10 times lower than that for Super-Kamiokande.  The muons have an average energy of 283~GeV.  The minimum kinetic energy required for a muon to vertically pass through DUNE's fiducial volume is $\simeq 2$~GeV, so $\simeq 2\%$ of the muons would stop in the detector.  The muon critical energy is 484~GeV in argon~\cite{Patrignani:2016xqp}; muons at higher energy dominantly lose energy through radiative processes, and hence are more likely to make showers and produce isotopes.  

There are two main simplifications we adopt for the primary muons.  One is that we only simulate $\mu^{-}$, because the energy losses and hence isotope production rates of $\mu^{+}$ are almost identical.  The only difference comes for stopping muons (details in Sec.~\ref{sec: SK}), for which we could correct the relevant isotope yields with the expected $\mu^{+}/\mu^{-}$ ratio of 1.38~\cite{Vitaly, Kudryavtsev:2008qh, Klinger:2015kva}, but we choose not to because of negligible differences (details in Sec.~\ref{sec: DUNE_yield}).

The other simplification is about muon injection.  We inject single muons above the rock, vertically downward at the center of the detector.  (Muons that miss the detector are discussed separately below.)  In reality, muons arrive with a variety of angles and positions, which can be easily measured.  Once the muon track is localized, there is no difference in the analysis procedure compared to a vertically throughgoing muon.  All that matters is the muon track length, because isotope production is a Poisson process.  According to {\tt MUSIC/MUSUN}~\cite{Vitaly, Kudryavtsev:2008qh, Klinger:2015kva}, most muons are downward going.  They have a mean zenith angle of $\langle\cos\theta_z\rangle\simeq 0.9$, resulting an average path length of $h/\langle\cos\theta_z\rangle\simeq 11$~m, very close to the fiducial-volume height ($h=10$~m).  Besides the single-muon case, there can be two muons appearing in a readout time window ($\simeq 5.4$~ms~\cite{Acciarri:2016ooe}), either due to muon bundles from cosmic-ray showers or an accidental coincidence ($\simeq 1$ per day per module).  If these muons are far from each other, then they can be treated separately.  If they are close, such that their possible showers could overlap, then one could apply a wider cylinder cut.  The results are not appreciably affected by our simplifications.

With the specified inputs, we expect three main outputs from {\tt FLUKA}: the isotope yields, the energy spectra of the isotope decay products, and the time and spatial correlations between the decay secondaries and the muons.  We record the first from the {\tt RESNUCLEi} card, which directly returns per-muon isotope yields, and the latter two from a modified {\tt mgdraw.f} subroutine.  

In addition to the {\tt RESNUCLEi} card, isotope yields can be recorded from subroutine {\tt usrrnc.f} and {\tt mgdraw.f} as well.  In {\tt usrrnc.f}, isotopes are identified with the arguments {\tt IZ} (atomic numbers) and {\tt IA} (mass numbers).  In {\tt mgdraw.f}, we note that isotopes are divided into two categories, both of which should be accounted for, to get the correct yields.  Some isotopes are characterized with the arguments {\tt ICRES} (residual nucleus atomic number) and {\tt IBRES} (mass number).  Some isotopes, mostly with small mass numbers, are treated as heavy ions, whose charge and mass number are stored in the arguments {\tt ICHEAV} and {\tt IBHEAV}.  As expected, these three counting methods return the same values of isotope yields. 

%%%%%%%%%%%%%%%%%%%%%%%%%%%%%%%%%%%%%%%%%%%%%%%%%%%%%%

\subsection{Predicted isotope yields}
\label{sec: DUNE_yield}

%+++++++++++++++++++++++++++++TABLE++++++++++++++++++++++++++++++++++++++++%
\begin{table*}[!p]
\begin{center}
\caption{Simulated isotope production yields in DUNE.  The conversion factor from the 5th to 6th column is 3423.  Yields above 0.01 per muon are quoted with three digits after the decimal; smaller yields are expressed in scientific notation.  {\bf Top block (upper part)}: Isotopes that survive the 250-ms cut (discussed below) and make $\gtrsim$ 1 event~yr$^{-1}$~(10~kton)$^{-1}$ above 5~MeV, sorted by Q values.  Isotopes making more than 1\% of the total are in bold.  {\bf Top block (lower part)}: Isotopes eliminated by the 250-ms cut (making $\lesssim$ 1 event~(10~yr)$^{-1}$~(10~kton)$^{-1}$ above 5~MeV), sorted by Q values.  {\bf Bottom block:} Isotopes that do not cause backgrounds above 5 MeV, sorted by yields, showing only those with $\geq$ 0.01 per muon.}  
\setlength{\tabcolsep}{4.5pt}
\setlength{\extrarowheight}{1.364pt}
\begin{tabular}{p{0.17\textwidth} d{.2} d{2.3} c d{.3} d{.3} }
\Xhline{2\arrayrulewidth}\\[-1.5em]\Xhline{2\arrayrulewidth}
\multicolumn{1}{l}{Isotope} &
\multicolumn{1}{C{1.515cm}}{Q value} &
\multicolumn{1}{C{1.64cm}}{Half-life} &
\multicolumn{1}{C{3.2cm}}{Decay mode} &
\multicolumn{1}{C{3.cm}}{Yield} &
\multicolumn{1}{C{3.cm}}{Yield}
\\[-0.1em]
\multicolumn{1}{l}{} &
\multicolumn{1}{C{1.515cm}}{[MeV]} &
\multicolumn{1}{C{1.64cm}}{[s]} & 
\multicolumn{1}{C{3.2cm}}{} &
\multicolumn{1}{C{3.cm}}{[per vertical muon]} &
\multicolumn{1}{C{3.cm}}{[day$^{-1}$~(10~kton)$^{-1}$]} 
\\[1pt] 
\Xhline{1.1\arrayrulewidth}\\[-1.25em]\Xhline{1.1\arrayrulewidth}
$\mathbf{^{8}B}$ &   16.96  &   0.77  &   $\beta^{+}$    &    \multicolumn{1}{B{.}{.}{-1}}{9.3\mathrm{\times10^{-5}}}    &   \multicolumn{1}{B{.}{.}{-1}}{0.32}   \\
$^{9}$C &   15.47  &   0.13  &   $\beta^{+}$    &   9.4\mathrm{\times10^{-6}}    &   0.032   \\
$^{18}$N &   13.90 &   0.62  &   $\beta^{-}$    &   1.0\mathrm{\times10^{-5}}    &   0.034   \\
$\mathbf{^{9}Li}$ &   13.61  &   0.18  &   $\beta^{-}$(49\%), $\beta^{-}n$(51\%)    &    \multicolumn{1}{B{.}{.}{-1}}{2.9\mathrm{\times10^{-4}}}    &    \multicolumn{1}{B{.}{.}{-1}}{0.99}   \\
$\mathbf{^{8}Li}$ &   12.98   &   0.84  &   $\beta^{-}$    &   \multicolumn{1}{B{.}{.}{-1}}{1.4\mathrm{\times10^{-3}}}    &    \multicolumn{1}{B{.}{.}{-1}}{4.8}  \\
$\mathbf{^{11}Be}$ &   11.51   &   13.76  &    $\beta^{-}$(55\%), $\beta^{-}\gamma$(45\%)   &    \multicolumn{1}{B{.}{.}{-1}}{1.0\mathrm{\times10^{-4}}}    &    \multicolumn{1}{B{.}{.}{-1}}{0.34}   \\
$^{8}$He &   10.66   &   0.12 &    $\beta^{-}\gamma$(84\%), $\beta^{-}n$(16\%)   &   7.2\mathrm{\times10^{-5}}   &   0.25   \\
$\mathbf{^{16}N}$ &   10.42   &   7.13  &   $\beta^{-}$(28\%), $\beta^{-}\gamma$(72\%)     &    \multicolumn{1}{B{.}{.}{-1}}{4.6\mathrm{\times10^{-4}}}     &    \multicolumn{1}{B{.}{.}{-1}}{1.6}  \\
$\mathbf{^{15}C}$ &   9.77   &   2.45  &   $\beta^{-}$(37\%), $\beta^{-}\gamma$(63\%)    &    \multicolumn{1}{B{.}{.}{-1}}{1.2\mathrm{\times10^{-4}}}     &    \multicolumn{1}{B{.}{.}{-1}}{0.41}   \\
$\mathbf{^{26}Na}$ &   9.35   &   1.07  &   $\beta^{-}\gamma$   &    \multicolumn{1}{B{.}{.}{-1}}{1.1\mathrm{\times10^{-4}}}   &    \multicolumn{1}{B{.}{.}{-1}}{0.38}   \\
$^{27}$Na &   9.01   &   0.30  &   $\beta^{-}\gamma$   &   2.1\mathrm{\times10^{-5}}     &   0.072   \\
$^{17}$N &   8.68   &   4.17  &   $\beta^{-}$(5\%), $\beta^{-}n$(95\%)   &   1.3\mathrm{\times10^{-4}}    &   0.45   \\
$^{30}$Al &   8.57   &   3.62  &   $\beta^{-}$    &   4.1\mathrm{\times10^{-4}}    &   1.4   \\
$^{23}$F &   8.48   &   2.23  &   $\beta^{-}$(30\%), $\beta^{-}\gamma$(70\%)     &   1.2\mathrm{\times10^{-5}}    &   0.041   \\
$^{16}$C &   8.01   &   0.75  &   $\beta^{-}n$   &   2.1\mathrm{\times10^{-5}}   &   0.072   \\
$\mathbf{^{31}Al}$ &   7.99   &   0.64  &   $\beta^{-}$(65\%), $\beta^{-}\gamma$(35\%)   &   \multicolumn{1}{B{.}{.}{-1}}{9.7\mathrm{\times10^{-5}}}     &    \multicolumn{1}{B{.}{.}{-1}}{0.33}   \\
$^{29}$Mg &   7.61   &   1.30  &   $\beta^{-}$(27\%), $\beta^{-}\gamma$(73\%)   &   1.8\mathrm{\times10^{-5}}     &   0.062   \\
$\mathbf{^{40}Cl}$ &   7.48   &   81  &   $\beta^{-}$(85\%), $\beta^{-}\gamma$(15\%)    &    \multicolumn{1}{B{.}{.}{-1}}{7.9\mathrm{\times10^{-3}}}     &    \multicolumn{1}{B{.}{.}{-1}}{27}   \\
$^{20}$F &   7.02   &   11.16  &   $\beta^{-}\gamma$   &   5.8\mathrm{\times10^{-4}}     &   2.0   \\
$^{34}$P &   5.38   &   12.43  &   $\beta^{-}\gamma$    &  3.4\mathrm{\times10^{-3}}   &   12   \\ 
$\mathbf{^{38}Cl}$ &   4.92   & 2234 &   $\beta^{-}$  &    \multicolumn{1}{B{.}{.}{-1}}{0.031}     &    \multicolumn{1}{B{.}{.}{-1}}{110}   \\
\hline
$^{14}$B &   20.64  &   0.012  &   $\beta^{-}\gamma$    &   1.1\mathrm{\times10^{-5}}     &   0.038   \\
$^{11}$Li &   20.62  &   0.009  &   $\beta^{-}n$    &   1.4\mathrm{\times10^{-5}}    &   0.048   \\
$^{12}$N &   17.34  &   0.011  &   $\beta^{+}$    &   6.5\mathrm{\times10^{-6}}     &   0.022   \\
$^{13}$O &   16.75  &   0.009 &   $\beta^{+}$    &   6.6\mathrm{\times10^{-7}}    &   0.002  \\
$^{13}$B &   13.44   &   0.017  &   $\beta^{-}$(92\%), $\beta^{-}\gamma$(8\%)   &   2.2\mathrm{\times10^{-4}}     &   0.75   \\
$^{12}$B &   13.37   &   0.020  &   $\beta^{-}$    &   5.5\mathrm{\times10^{-4}}    &   1.9   \\
$^{32}$Al &   13.02   &   0.032  &   $\beta^{-}(85\%)$, $\beta^{-}\gamma$(15\%)    &   7.9\mathrm{\times10^{-6}}     &   0.027   \\
$^{12}$Be &   11.71   &   0.021 &    $\beta^{-}$  &   3.5\mathrm{\times10^{-5}}     &   0.120   \\
\Xhline{1.1\arrayrulewidth}\\[-1.25em]\Xhline{1.1\arrayrulewidth}
$^{41}$Ar &  2.49  &  6.6\mathrm{\times10^{3}}  & $\beta^{-}$    &   0.474   &   1600 \\
$^{39}$Ar &  0.57  &  8.5\mathrm{\times10^{9}}   & $\beta^{-}$    &   0.354   &   1200 \\
$^{38}$Ar &    &    &  stable     &   0.274   &   940 \\
$^{37}$Cl &    &    &   stable  &   0.048   &   160 \\
$^{39}$Cl &  3.44  & 3.3\mathrm{\times10^{3}}  &  $\beta^{-}$    &   0.044   &   150 \\
$^{34}$S &    &    &  stable   &   0.034   &   120 \\
$^{36}$Cl & 0.71   & 9.5\mathrm{\times10^{12}}   &  $\beta^{-}$    &   0.032   &   110 \\
$^{35}$S &  0.17   & 7.5\mathrm{\times10^{6}}    &  $\beta^{-}$   &   0.024   &   82 \\
$^{36}$S &    &    & stable    &   0.024   &   82 \\
$^{37}$Ar &  0.81  &   3.0\mathrm{\times10^{6}}   &   EC  &   0.030   &   100 \\
$^{35}$Cl &    &    &  stable    &   0.014   &   48 \\
$^{33}$S &    &    &  stable    &   0.012   &   41\\
$^{32}$P &  1.71  & 1.2\mathrm{\times10^{6}}    &  $\beta^{-}$   &   0.010   &   34 \\
$^{33}$P &  0.25  &  2.2\mathrm{\times10^{6}}   &  $\beta^{-}$   &   0.010   &   34 \\
$^{30}$Si &    &    &  stable      &   0.010   &   34\\
\Xhline{1.1\arrayrulewidth}\\[-1.25em]\Xhline{1.1\arrayrulewidth}
$\mathbf{sum\ (top\ block,\ upper\ part)}$  &    &    &     &   \multicolumn{1}{B{.}{.}{-1}}{0.046}   &   \multicolumn{1}{B{.}{.}{-1}}{160}\\
$\mathrm{sum\ (top\ block,\ lower\ part)}$  &    &    &     &   0.001  &   3\\
sum (bottom block) &    &    &     &   1.394   &   4700\\
sum (total) &    &    &     &   1.441   &   4900\\
\Xhline{2\arrayrulewidth}\\[-1.25em]\Xhline{2\arrayrulewidth}
\end{tabular}
\label{tab: isotopes}
\end{center}    
\end{table*}
%+++++++++++++++++++++++++++++++++++++++++++++++++++++++++++++++++++++++++++%

The isotope yields reveal key features of spallation production processes in argon.  We find that there are about 100 different isotopes produced in DUNE, but only about 10 isotopes contribute significantly to the background rate above 5~MeV.  The underlying physics can be explained by a two-group structure.  Below, we summarize our results, highlighting the differences in three characteristics between high-A (Ar, Cl, S, etc.)~isotopes and low-A (Li, Be, B, etc.)~isotopes, with corresponding details given in Table~\ref{tab: isotopes}. 

The first difference concerns decay properties.  In general, beta-decay Q values and half-lives follow Sargent's rule, $t_{1/2}\propto Q^{-5}$~\cite{Konopinski1950}, although nuclear-structure differences cause deviations.  In DUNE, high-A isotopes mostly have small Q values ($\lesssim 2$--3~MeV) and long half-lives (minutes to years).  Only a few isotopes from the high-A group can decay to betas above 5~MeV, among which $^{40}$Cl is the one with the largest Q value (7.48~MeV).  In contrast, low-A isotopes often have large Q values ($\approx10$--20~MeV) and short half-lives (milliseconds to seconds).  The largest Q value in DUNE is 21~MeV, for $^{14}$B and $^{11}$Li, which determines the end of the background energy range.  We show in Sec.~\ref{sec:DUNE_aftercut} that these two isotopes and a few others with large Q values are completely rejected after a 250-ms cut, due to their short half-lives.  

The second difference is the production mechanism.  High-A isotopes are mainly produced by neutrons and gammas, while pions are only responsible for $\lesssim 10$\% of them.  Some dominant production processes for those isotopes are for example, $^{40}$Ar$(n,\,\gamma)$$^{41}$Ar, $^{40}$Ar$(\gamma,\,n)$$^{39}$Ar, $^{40}$Ar($n$,\,$p$+$2n$)$^{38}$Cl, and $^{40}$Ar$(n,\,\alpha+3n)$$^{34}$S.  In contrast, low-A isotopes are mostly made from pions and neutrons breaking $^{40}$Ar into pieces.  For example, the ratio of $^{8}$Li,  $^{12}$B, and $^{13}$B produced by $(\pi^+ : \pi^- : n) \simeq (1 : 2 : 1),\, (3 : 4 : 4)$, and $(2 : 3 : 3)$, respectively.  Typically, a dominant production channel does not exist.

The third difference is production yields.  High-A isotopes typically have large yields.  The most abundantly produced one is $^{41}$Ar, which has a yield of 0.47 per muon, corresponding to 1600 per day in one module.  Next is $^{39}$Ar, with a yield of 0.35 per muon.  These two together comprise 82\% of the unstable isotope yields.  In contrast, low-A isotopes are made much more rarely.  The highest yield from the low-A group is $\simeq 1.4\times10^{-3}$ per muon ($\simeq 5$ per day) from $^{8}$Li, while a typical yield is even smaller, of order $10^{-4}$ or $10^{-5}$ per muon.           

Table~\ref{tab: isotopes} shows the decay properties (from Refs.~\cite{NNDC, TOI}) and predicted production yields of selected isotopes in DUNE.  The average production yield of all isotopes is $\simeq 1.5$ per vertical muon, of which $\simeq 1$ per muon are beta-unstable, although the production rate of troublesome isotopes is much lower, as described below.  The statistical variation of this  average isotope yield is negligible, given the huge number ($\simeq 4\times10^9$) of primary muons we use in the simulation.  In the top block, we list isotopes relevant to the backgrounds above 5~MeV.  Its upper part contains the isotopes that make at least 1 background event per year, among which each of the 10 isotopes in bold individually makes more than 1\% of the total background.  Its lower part contains the isotopes that will be decimated by the 250-ms cut (details below).  In the bottom block, we list selected isotopes that either are stable or have small Q values.  Overall, we see that most backgrounds above 5~MeV are from low-A isotopes (near oxygen), while abundantly produced high-A isotopes (near argon) are more important for the energy range below 5~MeV.
 
The production yields shown in Table~\ref{tab: isotopes} are dominantly due to throughgoing muons, with only a small fraction from stopping $\mu^{-}$.  Once $\mu^{-}$ are stopped in DUNE, 76\% of the time they make isotopes though nuclear capture.  For most isotopes, the yields from $\mu^{-}$ capture are orders of magnitude lower than those due to throughgoing muons.  However, 36\% of $^{40}$Cl and 8\% of $^{38}$Cl are made by $\mu^{-}$ capture.  With the suggested $\mu^{+}/\mu^{-}$ ratio of 1.38, the yields of $^{40}$Cl and $^{38}$Cl would differ by $\approx20\%$ and $\approx5\%$, respectively, which is below the precision of our calculation, so we neglect the muon charge ratio correction in our subsequent discussions.   

We compare our predicted isotope yields with those of previous papers.  Franco et al.~\cite{Franco:2015pha} used {\tt FLUKA} to simulate the spallation backgrounds in a 100-ton fiducial LArTPC, which is inside a scintillator sphere within a water tank.  Their dark-matter-detection-style detector and DUNE have very different strengths.  The detector in Ref.~\cite{Franco:2015pha} has a very low energy threshold, and excellent intrinsic radio purity.  Although this kind of detector drifts charge, the main detection strategy is collecting light.  In contrast, DUNE is a ``tracking'' detector, primarily collecting charge, and the much larger volume ensures much larger statistics, although the energy threshold in DUNE is higher.  Nevertheless, the detection material in both detectors is LAr, so we could compare our calculated spallation yields with the 75 isotope decay rates listed in their Table A.  For the relative yields, there is good agreement for most isotopes, although several less important isotopes show discrepancies that can be explained.  Some of those ($^{9}$C, $^{32}$Al, $^{30}$S, etc.)~have large yield uncertainties in Franco et al.'s simulation, due to their low statistics.  Others are extremely long-lived isotopes ($^{14}$C, $^{32}$Si, $^{39}$Ar, etc.), so the comparison between our production rates and their decay rates is not valid, because these isotopes will not be in equilibrium.  For the absolute yields, if we take their muon track length as the detector height, then our yields are a factor of 2--3 higher.  Our results should agree when normalized by volume or muon track length, as the muon spectrum is similar, the target is similar (the LAr is the same; while the presence of surrounding material is important, its composition is not), and we both use {\tt FLUKA} (which can have some changes over time).  

To investigate this discrepancy further, we have done our own simulations for the detector setup of Ref.~\cite{Franco:2015pha}, and find yields per volume or muon track length comparable to ours for DUNE, i.e., higher than those of Ref.~\cite{Franco:2015pha} by the factor of 2--3 noted above.  However, we can recover their results if we ignore the surrounding material, i.e., if we start the muons just outside the fiducial volume instead of outside the full detector.  The physical reason for the difference would then be that showers are not fully developed.  We have discussed this with the lead author of Ref.~\cite{Franco:2015pha}, who agrees this may be a possible explanation for their results being different, and this is being investigated further.

We also compare our isotope yields with previous results that are from a different simulation package.  Barker et al.~\cite{Barker:2012nb} used {\tt GEANT4} and listed about 20 high-A isotope production rates in their Table VI.  For half of them, including some important ones such as $^{40}$Cl, $^{38}$Cl, $^{34}$P, etc., we find a good agreement, within a factor of $\approx2$.  For some rarely produced isotopes ($^{37}$P,  $^{33}$Cl,  $^{35}$Ar, etc.), our yields are a factor $\approx0.1$ of theirs.  However, even with Barker et al.'s yields, those isotopes would still be unimportant, due to other much more frequently produced isotopes that have similar decay Q values.  For $^{39}$Ar and $^{41}$Ar, when we remove the rock layer and count in the active volume, as in Ref.~\cite{Barker:2012nb}, our yields are a factor $\approx10$ of theirs.  We note the correlations of their production channels, i.e., $^{40}$Ar$(n,\gamma)$$^{41}$Ar, $^{40}$Ar$(\gamma,n)$$^{39}$Ar.  Nevertheless, these two isotopes are not particularly problematic due to their low Q values.  We think it is possible that the nuclear breakup models in {\tt FLUKA} and {\tt GEANT4} differ enough to cause the factor of $\approx10$ differences for certain isotope yields, as has been seen in Ref.~\cite{Bellini:2013pxa}.  Gehman et al.~\cite{Schoberg} took muon-induced neutrons coming from the rock as the primary particles, so their yields are subdominant to ours (details in Sec.~\ref{sec:DUNE_aftercut}).   

In summary, we believe our spallation predictions for LAr to be the most complete and accurate available.  We are pursuing plans to measure the most important spallation yields in near-surface argon detectors, which will help normalize {\tt FLUKA} predictions.  In addition, new reaction measurements (e.g., Ref.~\cite{Bacon}) will help.

%%%%%%%%%%%%%%%%%%%%%%%%%%%%%%%%%%%%%%%%%%%%%%%%%%%%%%

\subsection{Overview of spallation spectrum in DUNE}
\label{sec: DUNE_beta}

%+++++++++++++++++++++++++++++FIGURE++++++++++++++++++++++++++++++++++++++++%
\begin{figure*}[!t]
\begin{center}                  
\includegraphics[width=\columnwidth]{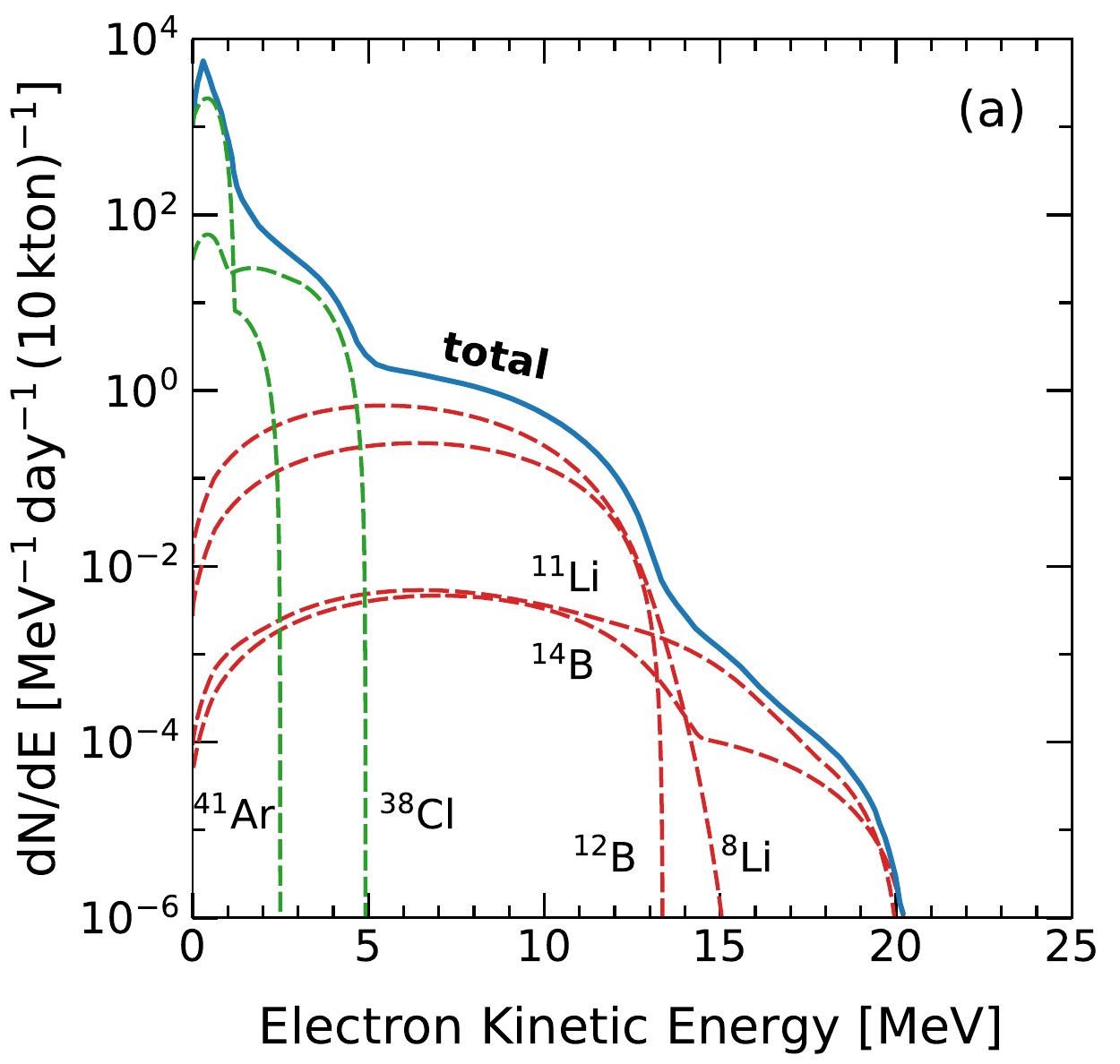}
\hspace{0.25cm}
\includegraphics[width=\columnwidth]{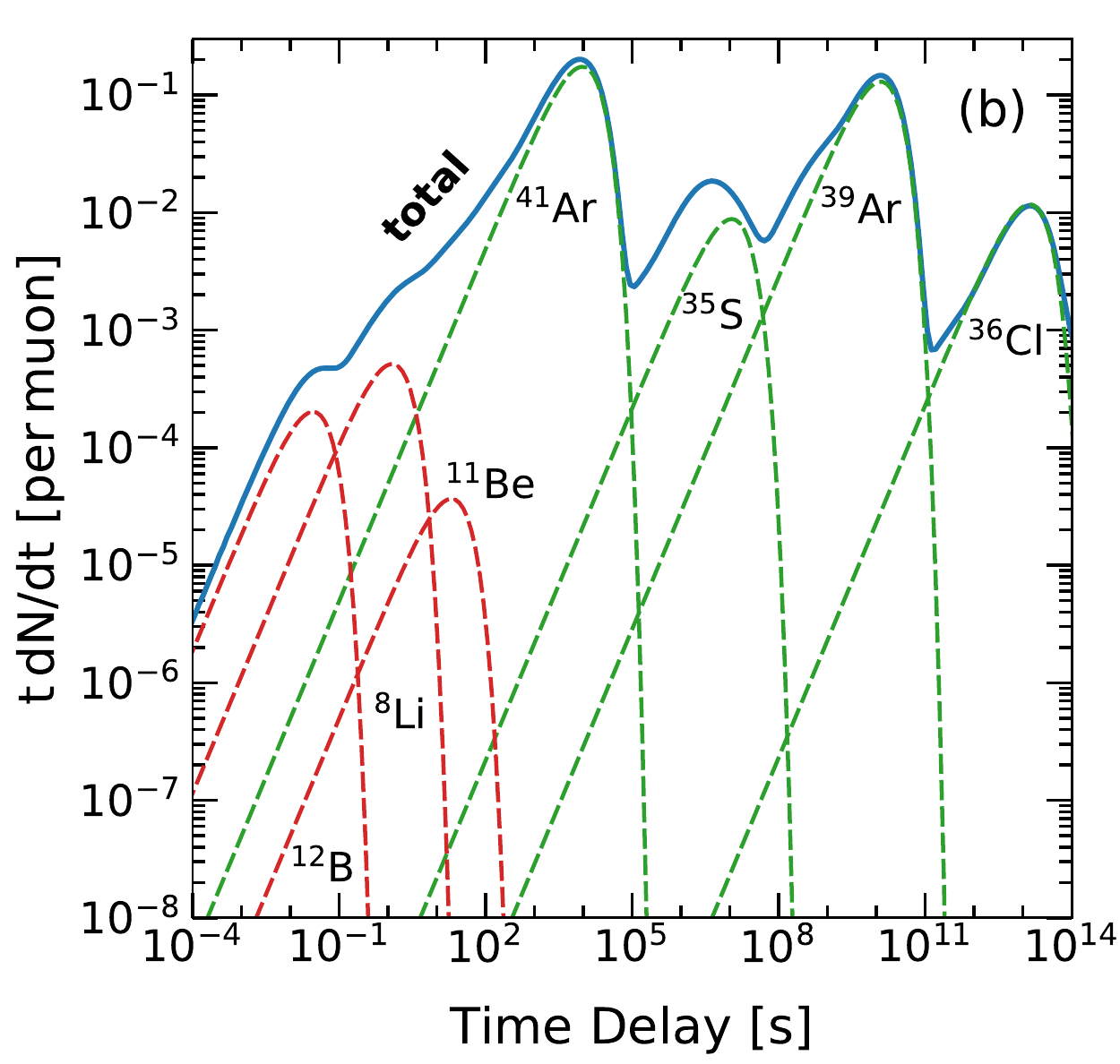}
\caption{Spallation spectra in DUNE before cuts (assuming equilibrium between isotope production and decay).  {\bf Left panel (a):} Energy spectra of spallation betas (no energy smearing yet).  Blue solid line: total spallation rate.  Green dashed lines: betas from selected high-A isotopes.  Red dashed lines: betas from selected low-A isotopes.  {\bf Right panel (b):} Time profile of spallation betas, normalized per vertical muon, with the same labeling conventions.}
\label{fig: spallation_all}
\end{center}
\end{figure*}
%++++++++++++++++++++++++++++++++++++++++++++++++++++++++++++++++++++++++++++%

Given the predicted isotope yields and the beta-decay properties of each isotope, we calculate the all-isotope background energy spectra and time distributions from {\tt FLUKA}.  We hold off on applying energy resolution until the next subsection, to provide a useful starting point in case the DUNE energy resolution turns out to be better than assumed below.

Getting the individual beta spectra right is important.  For most decays, the spectrum is specified completely by the Q value.  For some isotopes, extra care is needed.  For example, $^{36}$P has a Q value of 10~MeV, but the probability to decay to the branch with the highest endpoint energy (7~MeV) is only 1\%~\cite{NNDC}.  Both $^{8}$Li and $^{8}$B have special spectral shapes, because they decay to the $^{8}$Be 2$^{+}$ continuum state, which has an excitation energy of 3~MeV and a width of 1.5~MeV~\cite{Bahcall:1996qv, Winter:2004kf, Bhattacharya:2006ah}.  {\tt FLUKA} simulates their decays correctly.  We find two problems when cross checking the decay spectra from {\tt FLUKA} with the analytic spectra given the nuclear data from Refs.~\cite{NNDC, TOI}.  One is $^{11}$Li, for which the endpoint energy in {\tt FLUKA} is incorrect.  The other is $^{11}$Be, which {\tt FLUKA} does not simulate.  For these, we use analytic spectra. 

In addition, some decays are more complex than just beta decays, producing also gamma rays or neutrons.  Gammas produce electrons through Compton scattering or pair production, with radiation length 14~cm~\cite{Patrignani:2016xqp}.  Neutrons typically travel many meters and eventually get captured on $^{40}$Ar, making gammas, which then Compton-scatter or pair-produce electrons.  Unlike in water or scintillator, here the original beta and the $\gamma$ (or n) induced electrons from the same decay can be well separated, due to the good position resolution in DUNE ($\simeq 0.5$~cm).  Thus, we only focus on the betas from spallation isotope decays, and ignore the energy deposited by the accompanying gammas or neutrons, if any.

Figure~\ref{fig: spallation_all} (left panel) shows the total spallation beta spectrum in DUNE, with several components shown individually to explain the breakpoints near 2, 5, 15, and 20~MeV.  The rate scales corresponding to these breakpoints are roughly $10^{4}$, $10^{1}$, $10^{-1}$, and $10^{-4}$, spanning almost 10 orders of magnitude, showing the large variations among the isotope yields.  From the individual components, we see that high-A isotopes (near argon) have large yields and dominate the spectrum at low energy, while low-A isotopes (near oxygen) have small yields and dominate the spectrum at high energy.  The clear separation of high-A and low-A isotopes shown here makes visual the points made above about Table~\ref{tab: isotopes}.  Only 0.2\% of the spallation betas appear above 5 MeV, corresponding to 0.002 per muon or 7 events per day in each module.  

%+++++++++++++++++++++++++++++FIGURE++++++++++++++++++++++++++++++++++++++++%
\begin{figure*}[!t]
\begin{center}                  
\includegraphics[width=\columnwidth]{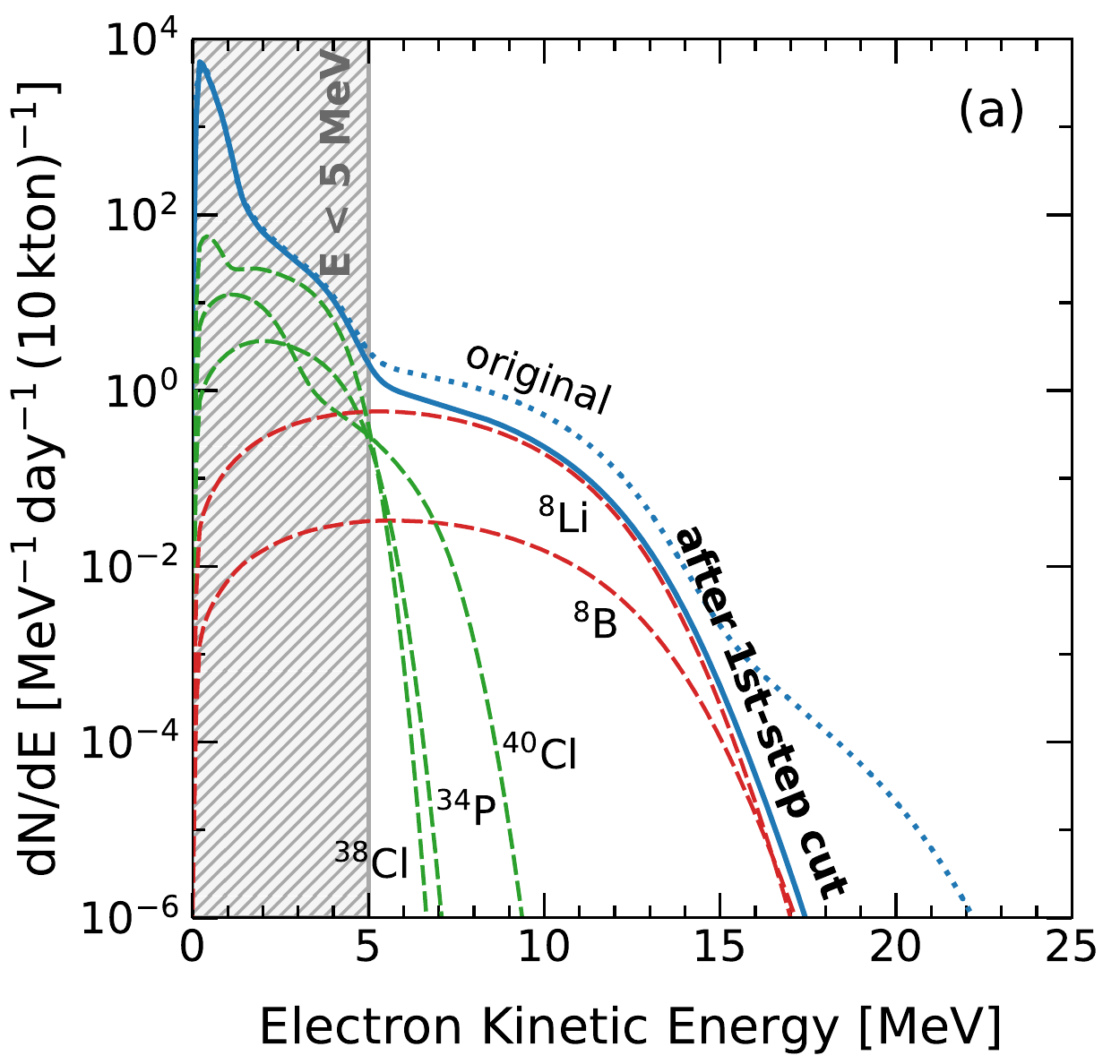}
\hspace{0.25cm}
\includegraphics[width=\columnwidth]{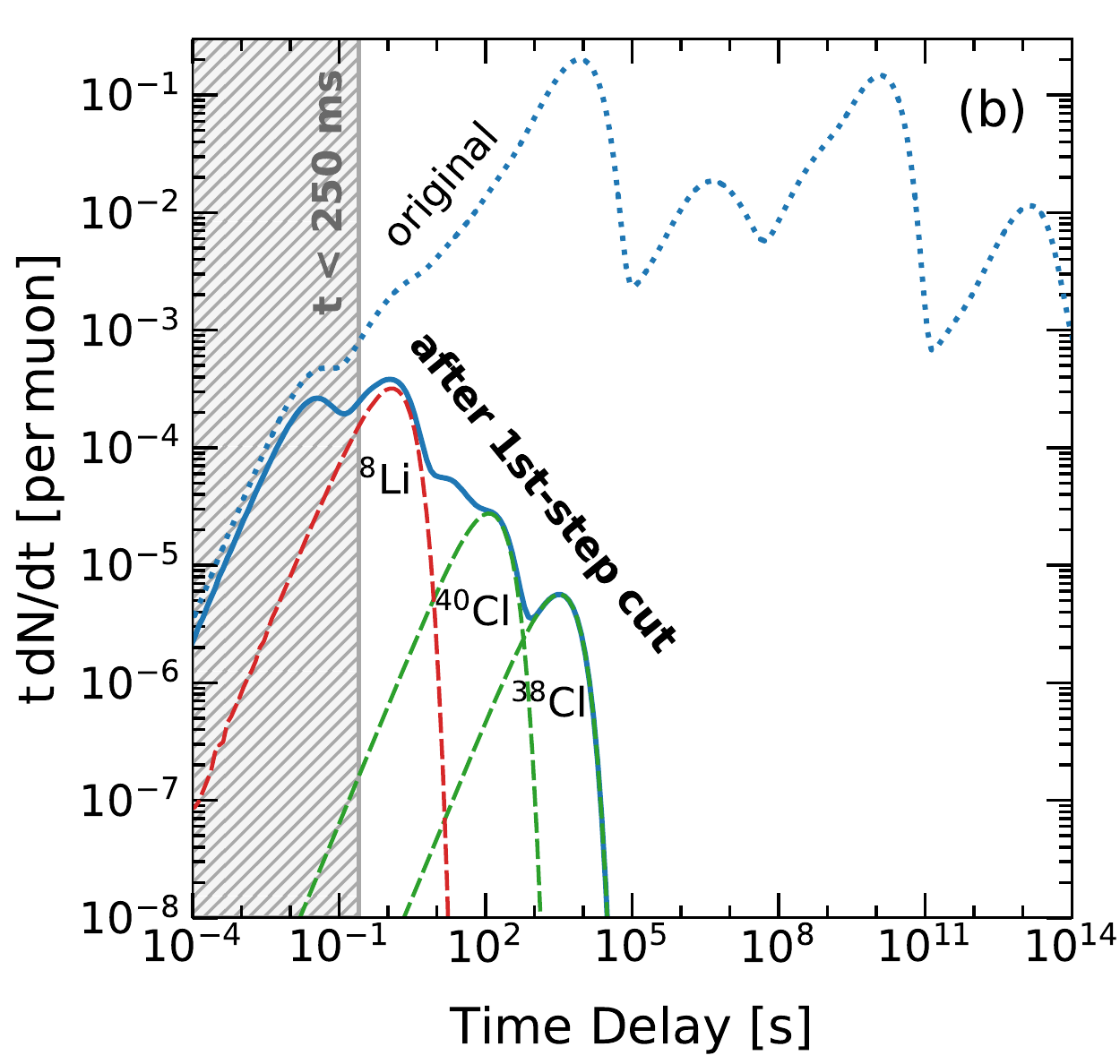}
\caption{Spallation spectra in DUNE after the 1st-step cut (discard events with $t < 250$~ms), assuming a 5-MeV energy threshold.  Events in the gray shaded regions are discarded, which introduces 1\% deadtime.  {\bf Left panel (a):} Energy spectra of spallation betas (with nominal 7\% energy resolution).  Blue dotted line: pre-cut spallation spectrum.  Blue solid line: post-cut spallation spectrum.  Green dashed lines: betas from selected high-A isotopes.  Red dashed lines: betas from selected low-A isotopes.  {\bf Right panel (b):} Time profile of spallation betas, normalized per vertical muon, with the same labeling conventions.} 
\label{fig: spallation_simple_cut}
\end{center}
\end{figure*}
%++++++++++++++++++++++++++++++++++++++++++++++++++++++++++++++++++++++++++++%

Figure~\ref{fig: spallation_all} (right panel) shows the spallation isotope decay time distributions relative to the cosmic-ray muons.  Because we use a log time axis, we plot the decay profile for each isotope with half-life $t_{1/2}$ as $t\,dN/dt = 2.3^{-1}\,dN/d\log_{10}t\propto \left(t/t_{1/2}\right)\times2^{-t/t_{1/2}}$, which shows the number of decays per log time.  We see this generic shape for different isotopes, while the decay time scales vary greatly.  High-A isotopes are typically long-lived.  Even though they are less important for higher energies in DUNE, they could form a steady-state background once production and decay reach equilibrium, which may affect the trigger rate.  For example, $^{41}$Ar (Q = 2.5~MeV) saturates after half a day of exposure, resulting a steady decay rate of 1600 events per day in each module of DUNE.  Another high-A isotope, $^{39}$Ar (Q = 0.6~MeV), has a similar production rate, but its half-life is extremely long (268~yrs).  We expect that $^{39}$Ar made from spallation has a decay rate of only 3 per day after one year of operation, while it would increase to 30 per day once DUNE has run for 10 years, which is still insignificant compared to the rate ($\approx10$~MHz) due to the pre-existing $^{39}$Ar in the atmospheric argon that DUNE will use~\cite{Marchionni:2013tfa}.  Low-A isotopes, the dominant source for the backgrounds at higher energies, are typically short-lived.  This is a crucial point for successful background cuts.  

%%%%%%%%%%%%%%%%%%%%%%%%%%%%%%%%%%%%%%%%%%%%%%%%%%%%%%

\subsection{Spallation backgrounds after cuts}
\label{sec:DUNE_aftercut}

Even though the detector performance and analysis procedures of DUNE in the MeV energy range are not available yet, it would be valuable to estimate how much backgrounds can be reduced.  In this subsection, we propose a two-step spallation cut as an example.  We smear the energy spectra with 7\% resolution, and choose 5~MeV as the energy threshold (i.e., the after-smearing background events below 5~MeV are automatically rejected).  

\textbf{The 1st-step cut} is a time cut, discarding events with $t < 250$~ms for each whole module with a muon, assuming an energy threshold of 5~MeV.  The time cut mainly rejects those short-lived low-A isotopes that dominate at higher energies.  We choose 250~ms as long enough to reject those backgrounds and short enough to not introduce significant deadtime.  In addition, this means that time resolution becomes irrelevant as the 250 ms is much longer than the few-ms readout time.  Because the muon rate is low ($\simeq 0.05$~Hz per module), the deadtime from this cut is only $\simeq 1\%$.  With only the time cut, long-lived high-A isotopes that dominate the isotope production yields still exist, as shown in Fig.~\ref{fig: spallation_all} (right panel).  However, having a 5-MeV threshold rejects almost all of them, because those high-A isotope decays are dominantly at lower energies, as shown in Fig.~\ref{fig: spallation_all} (left panel).  Thus, after the 1st-step cut, one should have a softer energy spectrum and a much smaller background rate.  The concepts behind this simple cut proposed here could work well for detectors besides DUNE.

Figure~\ref{fig: spallation_simple_cut} (left panel) shows the spallation background energy spectrum after the 1st-step cut.  Note that these spectra now have energy resolution included.  Above 5~MeV, 48\% of the backgrounds are rejected, which is entirely due to the 250-ms cut.  Its more important effect is to lower the endpoint of the background spectrum.  Pre-cut, the spectrum at high energies ($\gtrsim 18$~MeV) is dominated by $^{11}$Li and $^{14}$B, both of which have a Q value of 21~MeV and a half-life of 10~ms.  Post-cut, only a fraction $\approx10^{-7}$ of them remain, corresponding to a negligible decay rate of $\approx10^{-6}$ per year.  Now, the residual backgrounds at high energies are mostly due to $^{8}$Li, with $^{8}$B becoming comparable near the endpoint; at low energies, a few high-A isotopes are important.  Even though some of them (such as $^{38}$Cl and $^{34}$P) barely decay to betas above 5~MeV, they could be visible due to large production yields and energy smearing effect. 

Figure~\ref{fig: spallation_simple_cut} (right panel) shows the spallation background time profile after the 1st-step cut.  Above 250~ms, 99.9\% of the backgrounds are rejected, which is entirely due to the 5-MeV threshold.  Once we focus on those higher-energy events, there are no decays at large time delays any more, which motivates our follow-up cuts. 

\textbf{The 2nd-step cut} is a cylinder cut for throughgoing muons, discarding events with $R < 2.5$~m and $t < 40$~s, and a sphere cut for stopping muons, discarding events with $R < 3$~m and $t < 10$~min.  As noted, in DUNE it will be easy to determine the muon positions.  Through the correlations between the background events and the parent muons, we can cut the backgrounds further.  A key variable is the perpendicular distance of the backgrounds to the muon track.  

\begin{figure}[!t]
\includegraphics[width=\columnwidth]{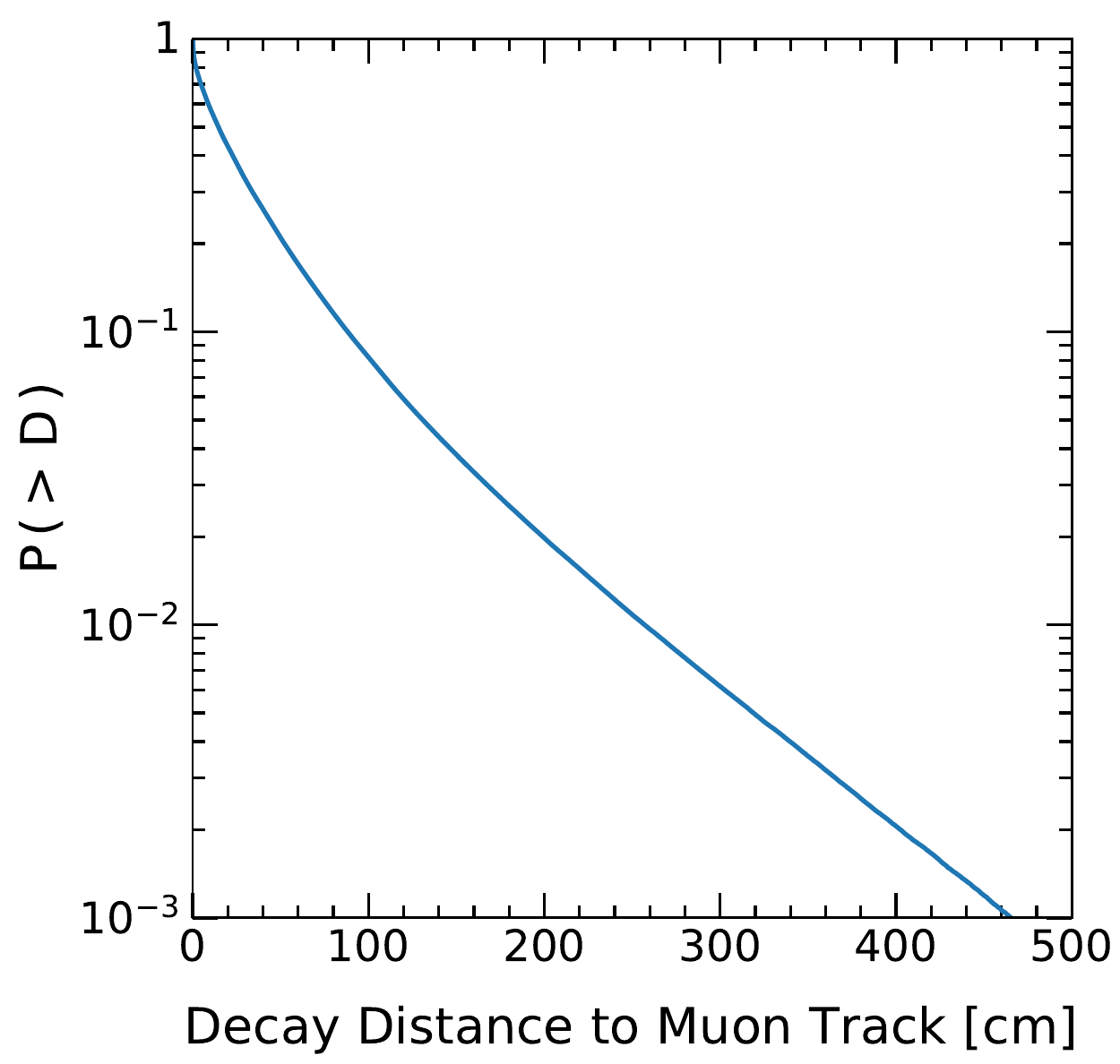}
\caption{Reverse cumulative distribution of spallation betas that survive from the 1st-step cut in perpendicular distance to the muon track.}
\label{fig: decay_distance}
\end{figure}

Figure~\ref{fig: decay_distance} shows the cumulative distance distribution of spallation betas relative to the muon track.  We find that, on average, 99\% of the decays happen within 2.5~m.  Isotopes decay nearly at the same places as where they are born, so the distribution shown here also reveals the isotope parent particle ($\rm\gamma,n,\pi$) absorption distances.  Neutron behavior in LAr is highly energy dependent.  High-energy neutrons efficiently make isotopes.  They typically die within a few meters, as shown in Fig.~\ref{fig: decay_distance}, because of large inelastic hadronic cross sections.  Low-energy neutrons ($\lesssim 10$~MeV) do not make isotopes except for $^{41}$Ar.  They have to travel much longer distances to lose enough energy through elastic scattering, and eventually get captured via $^{40}$Ar$(n,\gamma)$$^{41}$Ar or escape.  Once they are captured, the emitted gammas could cause backgrounds (details in Appendix.~\ref{sec: radio_neutron}).  

The measured isotope decay distances relative to the muon track could differ from that shown in Fig.~\ref{fig: decay_distance}, due to the potential movements of the isotopes prior to their decays.  One cause could be fluid motion in LAr, which is expected to have a speed $\lesssim 3$~cm~s$^{-1}$~\cite{Abi:2018alz}.  In addition, spallation isotope ions could drift under the electric field.  Those isotopes could be created in a fully ionized state, but would quickly catch electrons from argon until they are singly or doubly ionized, and drift with a speed typically $\lesssim1$~cm~s$^{-1}$, about 5 orders of magnitude lower than that of electrons~\cite{Santorelli:2017aut}.  With such speeds, most background isotopes in Table~\ref{tab: isotopes} would only drift for negligible distances, $\lesssim 10$~cm, before decay.  Thus, our proposed cuts below would not be affected at all.  Two exceptions are for $^{40}$Cl ($\rm t_{1/2} = 81$~s) and $^{38}$Cl ($\rm t_{1/2} = 2234$~s).  Because of their long half-lives, the nominal drift distances would be of order 1 and 10~m, respectively.  In principle, one could develop likelihood-based techniques that take into account the isotope drift and the maximum drift distances allowed in the detector to reject these two isotopes.  Our simple cuts do little to reject these two long-lived isotopes, which means even though they drift out the cylinder, our post-cut backgrounds would not be appreciably affected.  Therefore, the simulated distances shown in Fig.~\ref{fig: decay_distance} should fulfill our purpose.

The cylinder cut is especially useful for rejecting short-lived isotopes.  The cut efficiency and resulting detection deadtime can be estimated with the predicted spatial and time distributions of the backgrounds to the parent muons.  The remaining background flux after an $(R, \Delta t)$ cylinder cut is
%%%%%%%%%%%%%%%
\begin{multline}
\frac{dN'}{dE}(E) = \sum_{i}\, \frac{dN_{i}}{dE}(E) \\
\times \big(P_{i}\,(D>R) + P_{i}\,(D<R)\times\,P_{i}\,(t>\Delta t) \big),
\label{cut_efficiency}
\end{multline}
%%%%%%%%%%%%%%
where $\sum_{i}$ denotes a sum over all background isotopes.  For each isotope, $P_{i}\,(D>R)$ represents the fraction of decays outside the cylinder of radius $R$.  The averaged value over all background isotopes in DUNE can be extracted from Fig.~\ref{fig: decay_distance}.  Similarly, for each isotope with half-life $ t_{1/2}$, $P_{i}\,(t>\Delta t)= 2^{-\Delta t/t_{1/2}}$ represents the fraction of the decays outside the time window of duration $\Delta t$.  The resulting deadtime given the muon flux and the detector geometry described in Sec.~\ref{sec: DUNE_fluka} is accordingly
%%%%%%%%%%%%%%%
\begin{equation}
\begin{aligned}
f(R, \Delta\,t)  & \simeq \Phi_{\mu}\, \pi\, R^{2}\, \Delta t \\
& \simeq 1.8\times10^{-4} \left(\frac{R}{1\, {\rm m}}\right)^2 \left(\frac{\Delta t}{1\, {\rm s}}\right). \\
\end{aligned}
\end{equation}
%%%%%%%%%%%%%%
In DUNE, the (2.5~m, 40~s) cylinder cut in the 2nd step would cause a 4\% deadtime.   

%+++++++++++++++++++++++++++++FIGURE++++++++++++++++++++++++++++++++++++++++%
%++++++++++++++++++++++++++++++++++++++++++++++++++++++++++++++++++++++++++++%

%+++++++++++++++++++++++++++++FIGURE++++++++++++++++++++++++++++++++++++++++%
\begin{figure*}[!t]
\begin{center}                  
\includegraphics[width=\columnwidth]{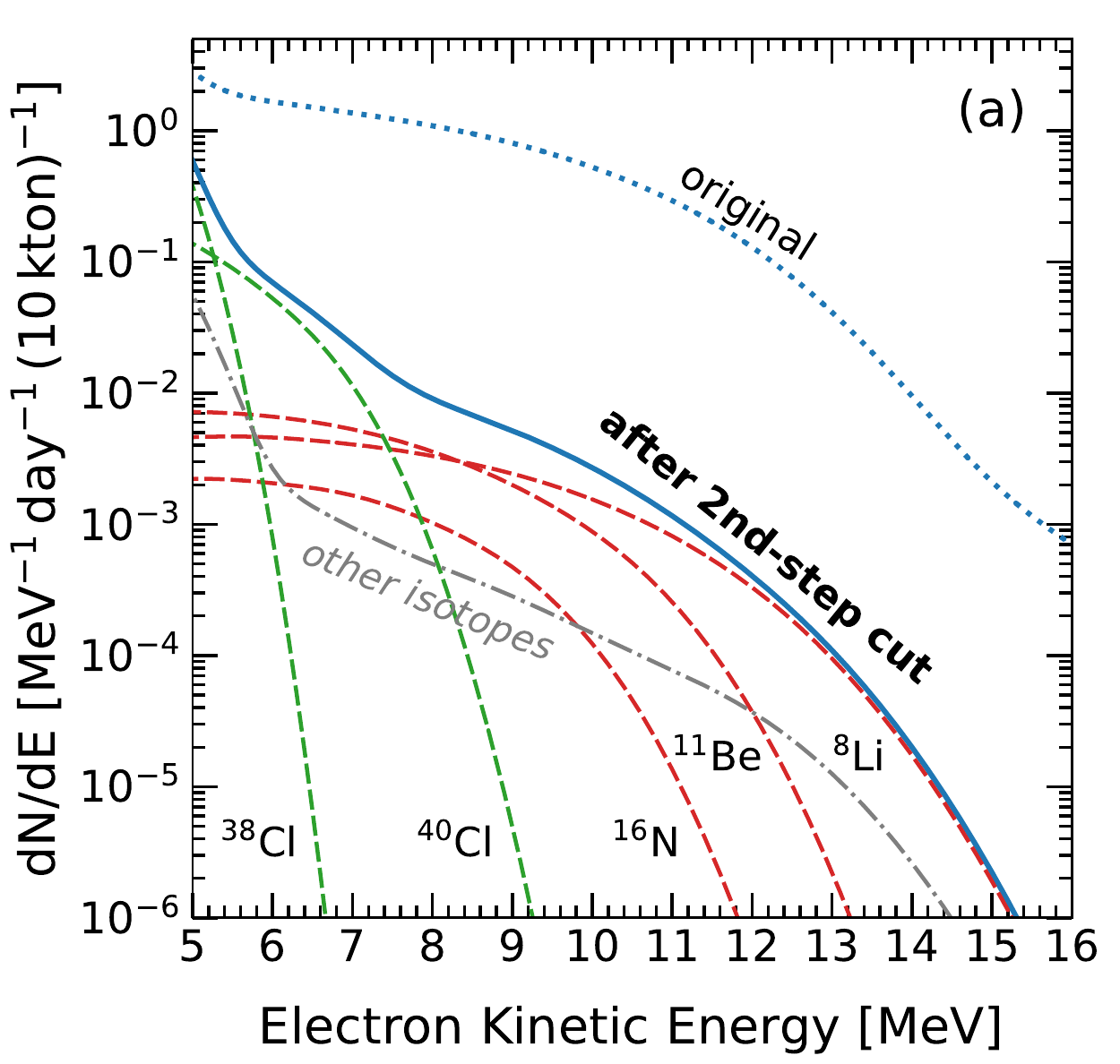}
\hspace{0.25cm}
\includegraphics[width=\columnwidth]{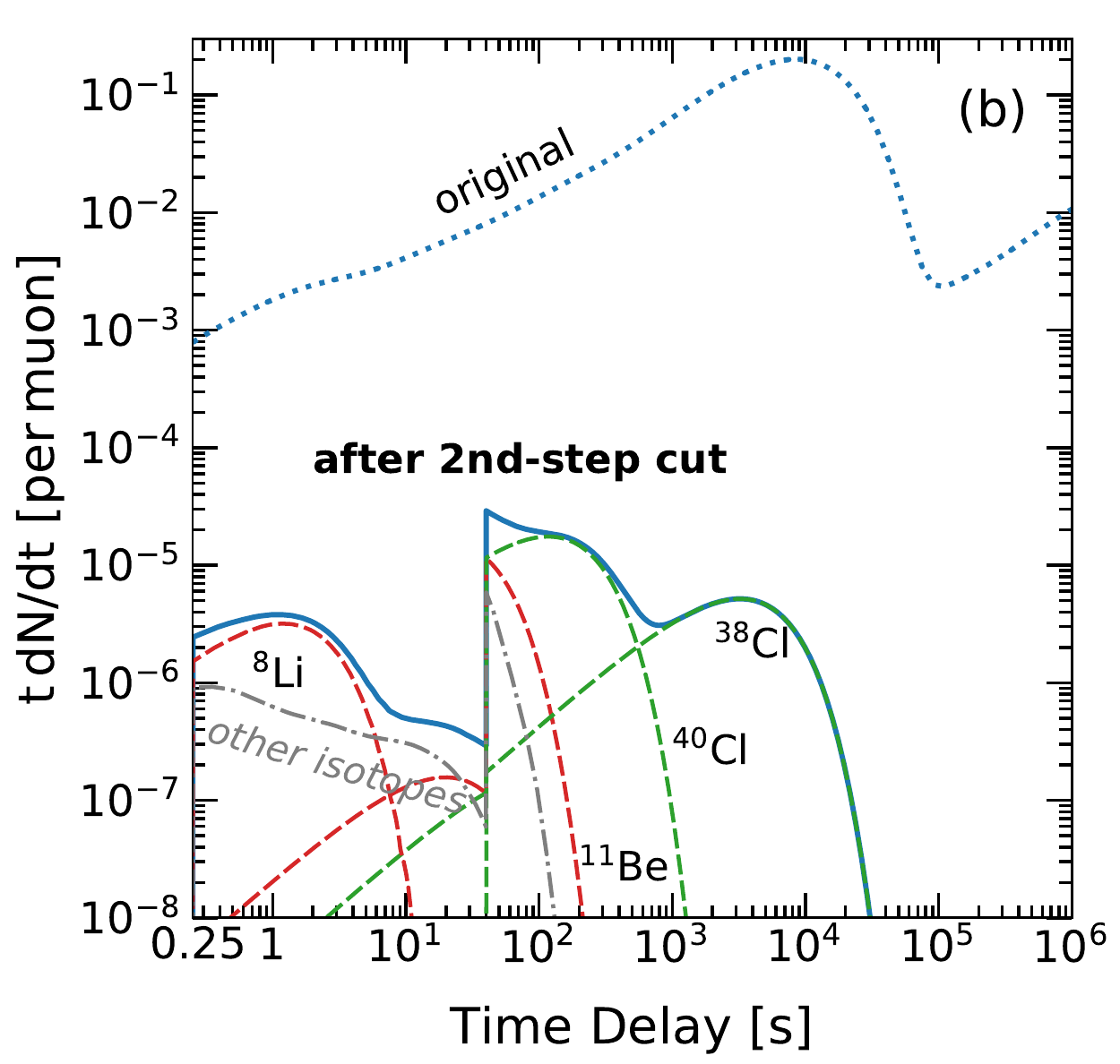}
\caption{Spallation spectra in DUNE after both the 1st-step and 2nd-step cuts.  The total deadtime is 5\%.  Note the change of energy and time ranges compared to Fig.~\ref{fig: spallation_simple_cut}.  {\bf Left panel (a):} Energy spectra of spallation betas (including 7\% energy resolution).  Blue dotted line: pre-cut spallation spectrum.  Blue solid line: post-cut spallation spectrum.  Green dashed lines: betas from selected high-A isotopes.  Red dashed lines: betas from selected low-A isotopes.  Gray dash-dotted line: remaining post-cut backgrounds from the isotopes not listed individually.  {\bf Right panel (b):} Time profile of spallation betas, normalized per vertical muon, with the same labeling style.  The sharp feature at 40~s is due to the time duration of the cylinder cut.} 
\label{fig: spallation_final_cut}
\end{center}
\end{figure*}
%++++++++++++++++++++++++++++++++++++++++++++++++++++++++++++++++++++++++++++

The stopping-muon cut can better reject those relatively long-lived isotopes, as one can afford a larger time window than for the cylinder cut.  Stopping muons have much lower rates than throughgoing muons.  In DUNE, the per-module stopping muon rate is only 0.001~Hz.  With the (3~m, 10~min) stopping-muon cut in the 2nd step, all the 36\% of $^{40}$Cl made through $^{40}$Ar$(\mu,\nu_\mu)$$^{40}$Cl could be removed, and the deadtime is $\lesssim 1\%$.

Figure~\ref{fig: spallation_final_cut} (left panel) shows the spallation energy spectrum after the 2nd-step cut.  Under this simple cut, only $\simeq 3\%$ of the backgrounds above 5~MeV remain.  The post-cut flux in the low energy end is dominantly from $^{38}$Cl and $^{40}$Cl.  In the more important energy range above 8~MeV, $^{8}$Li and $^{11}$Be make $\simeq 86\%$ of the remaining backgrounds.  This cut efficiency is basically limited by the cylinder time window $\Delta t$ for high-A isotopes (low-energy spectrum) and by the cylinder radius $R$ for low-A isotopes (high-energy spectrum).  This can be understood from Eq.~(\ref{cut_efficiency}).  For all isotopes, the decay fraction outside of the cylinder, $P\, (D>2.5\,\rm m)$, is typically $\simeq 1\%$, as shown in Fig.~\ref{fig: decay_distance}.  While the decay fraction outside of the time window, $P\, (t>40\,\rm s)$, could be as big as 76\% for high-A isotopes with half-lives of order 100~s, and as small as $10^{-12}$ for low-A isotopes with half-lives of order 1~s (and 6\% for 10~s).  This suggests that a set of cylinders might be useful if one wants to have a good cut efficiency in the entire energy range.

Figure~\ref{fig: spallation_final_cut} (right panel) shows the spallation time profile after the 2nd-step cut.  The decays at extreme time delays ($< 250$~ms and $\gtrsim 10^{4}$~s) are already gone after the 1st-step cut, as shown in Fig.~\ref{fig: spallation_simple_cut}.  In the remaining time window, the cylinder cut helps reject the backgrounds at small time delays ($< 40$~s), appearing as a sharp drop in Fig.~\ref{fig: spallation_final_cut} (right panel).  For events at large time delays ($> 40$~s), only the stopping-muon cut has an effect, rejecting 36\% of $^{40}$Cl.  Now, the residual spectrum seems to tell us that the high-A isotopes, $^{38}$Cl and $^{40}$Cl, are the most important ones.  However, note that the majority of the energy range (8--15~MeV) is still covered by the low-A isotopes, as shown in Fig.~\ref{fig: spallation_final_cut} (left panel).

%+++++++++++++++++++++++++++++FIGURE++++++++++++++++++++++++++++++++++++++++%
\begin{figure*}[t]
\centering 
\includegraphics[width=2\columnwidth]{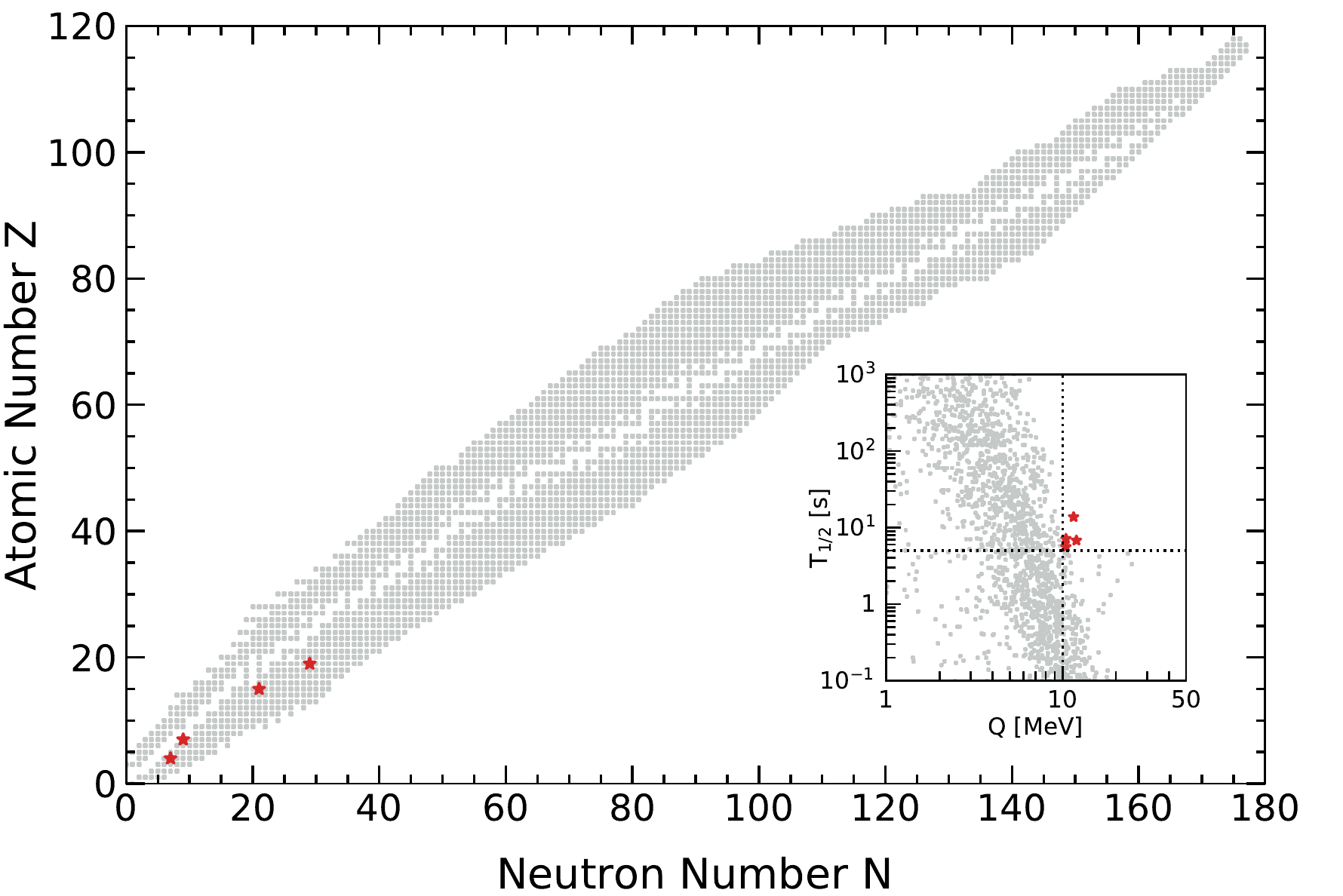}
\caption{Chart of beta-unstable nuclides with known Q values and half-lives T$_{1/2}$ (data from Refs.~\cite{Audi_2017, Wang_2017, Sonzogni}).  We seek to identify isotopes that could be hard to cut in detectors for high-energy ($> 5$ MeV) neutrinos; here we are agnostic about detector materials or isotope production rates.  Red stars identify nuclides with $Q > 10$~MeV and $T_{1/2} > 5$~s.  These are extreme outliers to Sargent's rule (see inset).  Gray squares identify remaining nuclides, which could be irrelevant or important before cuts, but which would be at most somewhat important after cuts.  Of the four isotopes with red stars, two are important in DUNE: $^{11}$Be (Q = 11.51~MeV, T$_{1/2}$ = 13.76~s) and $^{16}$N (Q = 10.42~MeV, T$_{1/2}$ = 7.13~s).  The other two, which may or may not be important in any realistic detector, are: $^{36}$P (Q = 10.41~MeV, T$_{1/2}$ = 5.6~s) and $^{48}$K (Q = 11.94~MeV, T$_{1/2}$ = 6.8~s).}
 \label{fig: nuclide_chart}
\end{figure*}
%++++++++++++++++++++++++++++++++++++++++++++++++++++++++++++++++++++++++++++%

This two-step spallation cut is very encouraging, and improvements are possible.  Under this simple cut, the backgrounds are already rejected by a factor of $\simeq 1.7\times10^{4}$, resulting an acceptable deadtime of 5\%, with 1\% from the 1st-step cut and 4\% from the 2nd-step cut.  One improvement would be using a shower cut.  Isotopes are dominantly born in hadronic showers, which are rare and have special characteristics (greater fluctuations, transverse size, and muon and neutron counts~\cite{Li:2016kra}), and hence should be easily identified in LAr.  Those showers usually extend $\approx5$~m (less than the detector height) and, more importantly, not every muon showers.  Compared to the cylinder cut used here, cutting the shower region for rare hadronic showers would enable a smaller cut volume and a smaller cut frequency, and hence allow stronger cuts.  

We assume that the efficiency for detecting muons that enter the active volume is 100\%.  In principle, a muon could clip just the corner of the active volume or pass through an inactive region between wire planes (in the single-phase design), failing to trigger the detector.  Indeed, this may happen, but in such cases, the production of isotopes in the active volume will be heavily suppressed.  As noted above, the vast majority of isotopes are made by secondary particles made in showers, and typically, these showers are quite energetic.  To make isotopes in the active volume, a muon must generally shower in the active volume, and this can identify the presence of a muon even in cases when the muon itself is not detected.
  
So far, our focus is on the muons coming into the detector, but there can be muons that miss the detector but send in secondary particles.  Neutrons are especially dangerous, as they could enter the detector invisibly.  Low-energy neutrons can get captured and emit gammas in the fiducial volume.  However, their rate is much lower than that from radioactivities in the rock (details in Appendix.~\ref{sec: radio_neutron}).  High-energy neutrons can make isotopes, but they must be from the muons that are close to the detector edge, because the isotope production probability drops significantly when its distance to the muon track gets larger, as shown in Fig.~\ref{fig: decay_distance}.  Taking the 1 m of active LAr shielding that is outside of the fiducial volume into account, the isotope yield in the fiducial volume per muon-in-rock would be $\lesssim 7\%$ of the yield per muon-in-detector.  From simulation, we find that the isotope production yields in the fiducial volume from the muon-induced neutrons in the rock are typically 3--4 orders of magnitude lower than those from throughgoing muons.  They could be cut further by recognizing the electrons and gammas accompanying the incoming neutrons.  If we cut incoming electromagnetic showers with deposited energies larger than 50~MeV, we find that only $\simeq 20\%$ of the isotopes remain.  Thus, we ignore isotope production due to muons that miss the detector.  

Interestingly, the isotopes that matter most in LAr are many of the same isotopes that matter most in water- or scintillator-based detectors: low-A isotopes of Li, Be, B, etc.\ (and for water, also $^{16}$N).  This may suggest a hidden universality --- regardless of the detector material, some of the same isotopes show up as problems.

Figure~\ref{fig: nuclide_chart} confirms our hypothesis.  We take problematic isotopes to be those rare ones with both large Q values and long half-lives.  Among thousands of unstable nuclides, only four meet this criteria, and, as expected, they are extreme outliers to Sargent's rule.  In making this list, we discard isotopes that decay primarily by nucleon emission or by electron capture, as these decay modes will not be important in typical MeV neutrino detectors.  We do include those isotopes that decay by $\beta + \gamma$ decay, as some detectors can register gamma rays as well.  Of the four isotopes we identify, $^{11}$Be is a known problem for both scintillator-based and water-based detectors, and $^{16}$N is a known problem for water-based detectors.  The commonality of problematic isotopes for argon-, water-, and scintillator-based detectors does not end there.  There are several others that, while not identified in Fig.~\ref{fig: nuclide_chart}, are somewhat important after cuts due to their high production rates.  These are typically in the range near carbon and oxygen.  Overall, Fig.~\ref{fig: nuclide_chart} gives helpful general guidance on spallation backgrounds.

%%%%%%%%%%%%%%%%%%%%%%%%%%%%%%%%%%%%%%%%%%%%%%%%%%%%%%
%%%%%%%%%%%%%%%%%%%%%%%%%%%%%%%%%%%%%%%%%%%%%%%%%%%%%%

\section{MeV Potential of DUNE}
\label{sec: signal}

The potential of MeV neutrino astrophysics in DUNE is not fully studied.  For high-energy ($\gtrsim 5$~MeV) solar neutrinos, Ref.~\cite{Capozzi:2018dat} is the first comprehensive study, although there have long been discussions of detecting solar neutrinos with LAr~\cite{Bahcall:1986ry, Arneodo:2000fa, Franco:2015pha, Ioannisian:2017dkx}.  On the supernova neutrino front, even though dedicated efforts have been made to explore physics opportunities and predict experimental signals in DUNE~\cite{Mirizzi:2015eza, Ankowski:2016lab, Nikrant:2017nya, Seadrow:2018ftp, Acciarri:2016crz, Acciarri:2015uup, Strait:2016mof, Acciarri:2016ooe, Abi:2018dnh, Abi:2018alz, Abi:2018rgm}, more work is needed to understand backgrounds, triggers, reconstruction capabilities, etc., and similarly for the diffuse supernova neutrino background (DSNB) in LAr~\cite{Cocco:2004ac, Lunardini:2010ab, Jeong:2018yts, Moller:2018kpn}. 

One essential step forward is to understand detector backgrounds.  Below, we focus on spallation backgrounds, both pre- and post-cut, as well as neutron-capture backgrounds and other backgrounds, including through pileup.  We evaluate their impact on MeV program of DUNE from two aspects: the absolute background rates compared to the signal rates and how backgrounds would affect trigger design.

%%%%%%%%%%%%%%%%%%%%%%%%%%%%%%%%%%%%%%%%%%%%%%%%%%%%%%

\subsection{Solar neutrinos}

We first summarize the predicted solar neutrino signals from Ref.~\cite{Capozzi:2018dat}, where 100~kton-year exposure, 5-MeV energy threshold, and 7\% energy resolution are assumed.  The two signal channels are elastic-scattering interaction $\nu_{e,\mu,\tau} + \, e^- \rightarrow \nu_{e,\mu,\tau} + \, e^-$, and charged-current interaction $\nu_e + \, ^{40}{\rm Ar} \rightarrow e^- + \, ^{40}{\rm K}^*$.  They can be well separated by a forward-cone angular cut, because elastic-scattering events are forward-peaked whereas charged-current events are nearly isotropic.  For $^{8}$B signals, inside the cone, the elastic-scattering channel dominates ($\approx10^4$ events); outside the cone, the charged-current channel dominates ($\approx10^5$ events).  For $hep$ signals, the sensitivity is largely from the charged-current events above 11~MeV that are outside the cone ($\approx150$ events).  Given the background rates we explain below, DUNE would measure $\sin^2\theta_{12}$ and $\Delta m^{2}_{21}$ with a factor of $\simeq 1.5$ and $\simeq 3$ better precision, respectively, than all combined solar experiments to date, a factor of $\simeq 1.6$ better precision on $^{8}$B flux than from SNO, and make the first detection of the $hep$ flux, with a precision of 11\%. 

\begin{figure}[t]
\includegraphics[width=\columnwidth]{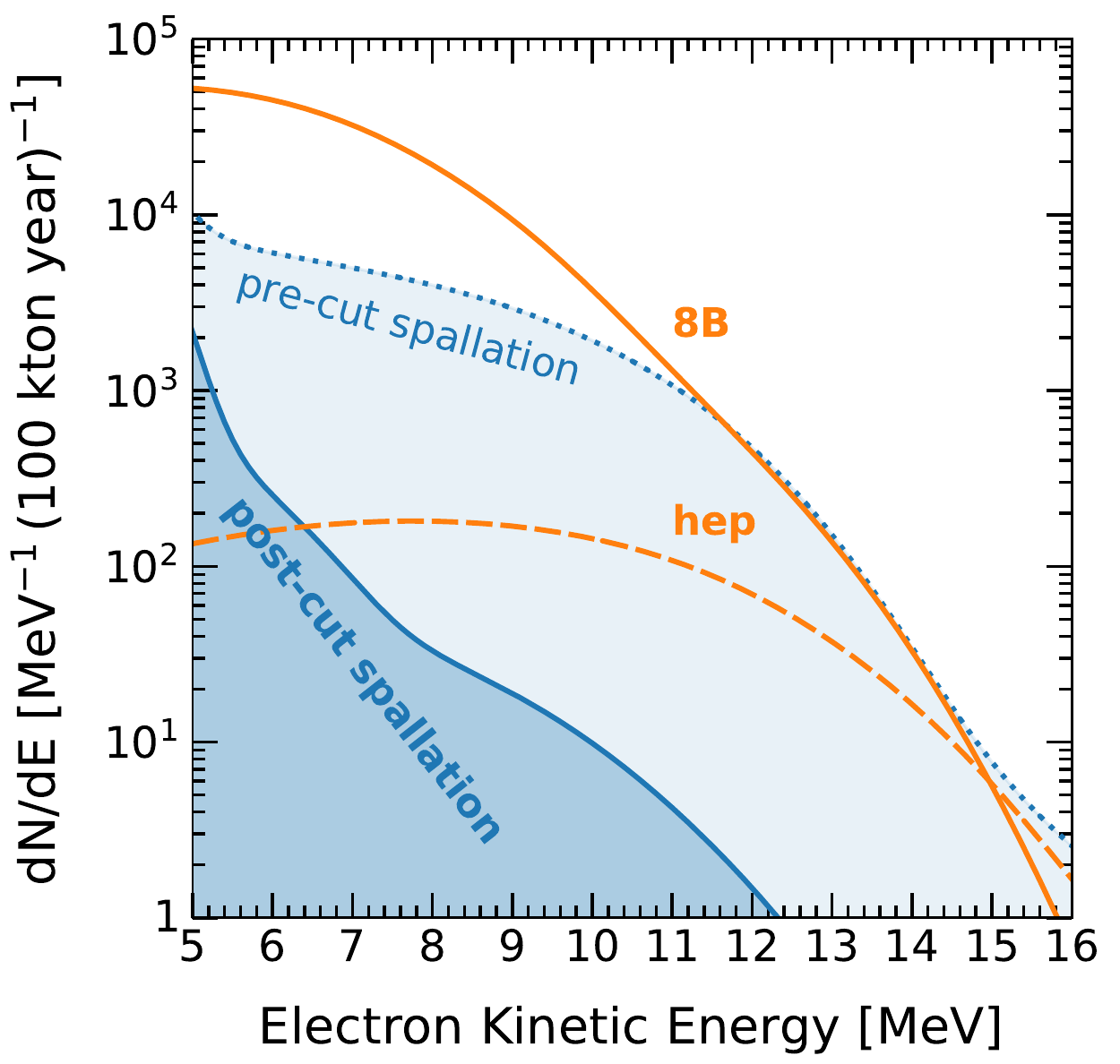}
\caption{Solar neutrino signal rates~\cite{Capozzi:2018dat} and spallation background rates in DUNE, with nominal 7\% energy resolution.  (Other backgrounds, due especially to neutron capture, which dominates at low energy, are not shown.)  The deadtime imposed by the spallation cut is not accounted for the signal rates.  Orange solid line: total $^{8}$B signal rates from charged-current channel and elastic-scattering channel.  Orange dashed line: total $hep$ signal rates from the two channels.  Light blue shaded: pre-cut spallation spectrum.  Dark blue shaded: post-cut spallation spectrum (same as Fig.~\ref{fig: spallation_final_cut}).}
\label{fig: solar}
\end{figure}

Figure~\ref{fig: solar} shows the solar neutrino signals and the spallation backgrounds.  For illustration purposes, we show only the combined signal rate from the two detection channels (details in Ref.~\cite{Capozzi:2018dat}).  Pre-cut, spallation backgrounds are subdominant but important.  There are $\simeq 2.4\times10^{4}$ background events above 5~MeV in 100~kton-year.  After the two-step cut we propose in Sec.~\ref{sec:DUNE_aftercut}, the backgrounds are reduced to $\simeq 700$ events above 5~MeV, as shown in Fig.~\ref{fig: solar}.  These post-cut backgrounds are negligible compared to both $^{8}$B and $hep$ event rates, and the imposed deadtime is only 5\%.  For this and the next figure, we show in the Appendixes versions where we adopt 20\% energy resolution.

%+++++++++++++++++++++++++++++FIGURE++++++++++++++++++++++++++++++++++++++++%
%++++++++++++++++++++++++++++++++++++++++++++++++++++++++++++++++++++++++++++%

%%%%%%%%%%%%%%%%%%%%%%%%%%%%%%%%%%%%%%%%%%%%%%%%%%%%%%

\subsection{Supernova neutrinos}
\label{sec: SN}

The expected counts (mostly $\nu_e$) from a supernova at 10~kpc are large, $\approx800$ in 10~s in each module~\cite{Acciarri:2015uup}.  This only accounts for neutrino interactions in the detector.  In principle, there could be another detection channel, due to radiative captures of neutrons produced by supernova neutrino interactions in the surrounding rock~\cite{Cline:1993rx, Duba:2008zz}.  Unfortunately, this process has a subdominant rate.

If we have independent information on when a supernova is happening, spallation backgrounds are negligible.  The total background rate is $\simeq 0.4$ in 10~s in each module, and it reduces to $\simeq 0.001$ if only the backgrounds above 5~MeV are counted.  On top of that, the backgrounds could be rejected further if cuts (e.g., as proposed in Sec.~\ref{sec:DUNE_aftercut}) are applied.  The harder case, where one waits to trigger on a supernova, is discussed below.   

%%%%%%%%%%%%%%%%%%%%%%%%%%%%%%%%%%%%%%%%%%%%%%%%%%%%%%

\subsection{Diffuse supernova neutrino background}

The DSNB is the flux of neutrinos emitted by all core-collapse supernovae throughout the universe.  While being a unique probe for both stellar astrophysics and neutrino physics, the DSNB has not been detected.  Currently, the strongest $\bar\nu_{e}$ flux limit is set by Super-Kamiokande~\cite{Malek:2002ns, Bays:2011si, Zhang:2013tua}.  Future progress could be made by the joint efforts from next-generation experiments.

Three ingredients are needed to calculate the DSNB event rates~\cite{Beacom:2010kk}.  The first is the supernova neutrino emission spectrum.  For each flavor, the neutrino spectrum can be approximated by a Fermi-Dirac distribution, where the two parameters, total energy $E_{\nu, tot}$ and temperature $T$, should be determined from experiments, although there are oft-quoted estimates.  The second is the cosmic supernova rate.  This is closely related to the star-formation rate, which has been measured.  The third is the neutrino interaction cross sections with argon.  For the charged-current interaction that dominates the DSNB signal rates, the uncertainty on the cross section is energy dependent and not well quantified, though certainly larger than 10\%~\cite{Capozzi:2018dat}.  Following Ref.~\cite{Horiuchi:2008jz}, which provides the star-formation rate from Ref.~\cite{Hopkins:2006bw}, we calculate the DSNB $\nu_e$ flux for neutrinos at $T$ = 4, 5, and 6~MeV, and then calculate the DSNB signal rates with the cross section from Refs.~\cite{GilBotella:2003sz, Scholberg:2012id}. 

\begin{figure}[!t]
\includegraphics[width=\columnwidth]{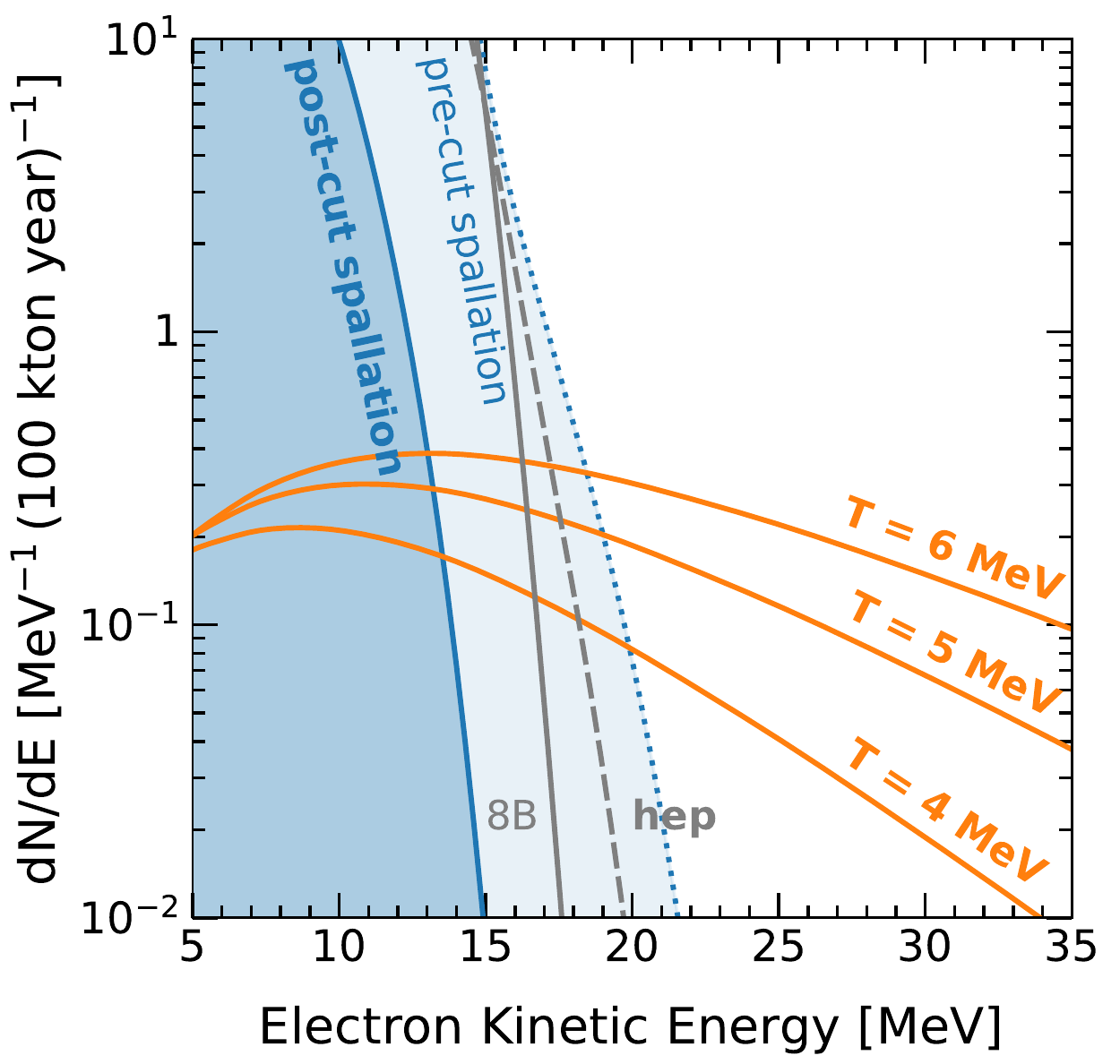}
\caption{DSNB $\nu_e$ signal rates and background rates in DUNE, with nominal 7\% energy resolution.  Orange solid lines: DSNB signal rates, assuming neutrino temperatures of 4, 5, and 6~MeV.  Light blue shaded: pre-cut spallation spectrum.  Dark blue shaded: post-cut spallation spectrum (same as Fig.~\ref{fig: spallation_final_cut}).  Gray solid line: predicted $^{8}$B event rates~\cite{Capozzi:2018dat}.  Gray dashed line: predicted $hep$ event rates~\cite{Capozzi:2018dat}.}
\label{fig: DSNB}
\end{figure}

Figure~\ref{fig: DSNB} shows our calculated DSNB $\nu_e$ signal rates, together with the spallation backgrounds and solar neutrinos, which are treated as backgrounds in this case.  Another background, due to atmospheric neutrinos, arising at $\approx40$~MeV~\cite{Cocco:2004ac}, is not shown here.  Pre-cut, the spallation backgrounds and the $hep$ events are comparable above $\simeq 15$~MeV, resulting a low-energy threshold for the DSNB signals at $\simeq 17$~MeV.  After the two-step cut, the spallation backgrounds above 15 MeV are completely rejected.  Unfortunately, it would not lower the energy threshold for the DSNB because of the $hep$ events.  In principle, the $hep$ elastic-scattering rates can be reduced by $\approx80$\% with a forward cone cut.  However, it would not significantly help the DSNB search, due to the remaining much larger $hep$ charged-current rates.  Taking 17 MeV as the threshold, the event rates of DSNB in an exposure of 100~kton-year would be about 1, 2, and 5 events, for neutrinos at $T$ = 4, 5, and 6 MeV.  The event rates might be higher, $\approx10$ in 100~kton-year, if one considers neutrino oscillations, such as in Ref.~\cite{Cocco:2004ac}, where the after-mixing neutrino spectrum is equivalent to assuming $T \simeq 8$~MeV, which is unrealistically high.  In summary, detecting DSNB in DUNE would be challenging.

\subsection{Trigger considerations}

Now that we have discussed potential signals and backgrounds, it is important to also consider the trigger requirements to collect the data.  In an effort to aid DUNE trigger design, we provide --- in the Appendixes --- extensive new results on rates, spectra, and multiplicities of low-energy events.  We consider both absolute rates and the possibility of pileup, where multiple below-threshold events combine to appear as one above-threshold event.  As above, we focus on detected energies above 5 MeV.

%+++++++++++++++++++++++++++++FIGURE++++++++++++++++++++++++++++++++++++++++%
%++++++++++++++++++++++++++++++++++++++++++++++++++++++++++++++++++++++++++++%

Though the trigger details are unknown, it is expected that there will be one kind of trigger for single low-energy events, e.g., solar neutrinos, and another for a burst of events, e.g., a supernova, which may allow a lower energy threshold per event.  For the burst-event trigger, a limiting factor will likely be data storage, as it is expected that all data will be recorded for say minutes in event of a possible burst, and that backgrounds will be negligible during a real burst.  However, backgrounds could lead to a considerable amount of fake triggers, due to the long waiting period for a burst.  Our focus is thus on the supernova trigger, as this is what sets the rates of low-energy data that DUNE can tolerate.  If these requirements are met, then a trigger for solar neutrinos would be possible.

In the Appendixes, we show that the only background that is significant is that due to the capture of radiogenic neutrons from the rock, which has a very large rate~\cite{Capozzi:2018dat}.  For trigger considerations the following are much less important: muon-generated activity in the detector (spallation decays and neutron captures), muon-generated activity in the rock (neutron captures), and radioactivities in the detector.  Further, here we show that pileup of low-energy backgrounds can also be neglected.

%%%%%%%%%%%%%%%%%%%%%%%%%%%%%%%%%%%%%%%%%%%%%%%%%%%%%%
%%%%%%%%%%%%%%%%%%%%%%%%%%%%%%%%%%%%%%%%%%%%%%%%%%%%%%

\section{Conclusions}
\label{sec: conclusions}
DUNE could provide a precious opportunity for MeV neutrino astrophysics.  A key step to probe DUNE's MeV potential is to understand the detection backgrounds.

We calculate the muon-induced spallation background in DUNE.  Using the Monte Carlo code {\tt FLUKA} and theoretical insights, we detail the physical mechanism of isotope production, calculate the isotope yields, evaluate the isotope decay energy and time profiles, and develop cuts to reduce the backgrounds.  Complementary to previous work on scintillation detectors and water-Cherenkov detectors, we provide a thorough understanding of the spallation backgrounds in argon.    

The uncertainty of our simulation is likely to be around a factor of 2, mostly due to the uncertainties of hadronic processes.  This is good enough in the sense that isotope yields vary by orders of magnitude.  The calibration could be done in situ, and the isotope yields in argon could be checked with detectors such as MicroBooNE.

We are the first to show explicitly that the essential difference between argon and oil or water is revealed by the two-group isotope production mechanism.  In DUNE, high-A isotopes (e.g., Ar, Cl, S) are abundantly produced.  However, these isotopes typically have small decay Q values ($\lesssim 2$--3~MeV), and hence dominantly produce betas well below the expected energy threshold.  Low-A isotopes (e.g., Li, Be, B), despite their small yields, are important background sources at higher energies, due to their large decay Q values ($\approx10$--20~MeV).  This two-group production reveals a hidden universality that the most important isotopes for all target nuclei will mostly be the same ones, such as Li, Be, B, and N.

The decay properties of those low-A isotopes are the key for us to design the cuts, so that the backgrounds become controllable.  While there are many unknowns for the future MeV programs in DUNE, we propose a two-step spallation cut, based on reasonable assumptions of the detector, showing that the spallation backgrounds can be greatly reduced.  Our 1st-step cut is discarding all events with $t < 250$~ms relative to each muon, assuming a 5-MeV energy threshold.  After this cut, the endpoint of the background is lowered down from $\simeq 20$~MeV to $\simeq 15$~MeV, and the per-module deadtime is only $\simeq 1$\%.  The 2nd-step cut is a cylinder cut for throughgoing muons, discarding events with $R < 2.5$~m and $t < 40$~s, and using a sphere cut for stopping muons.  After the 2nd step, the background rates above 5~MeV are reduced from $\simeq 7$ per day per module to $\simeq 0.2$ per day per module, and the total deadtime is only $\simeq 5$\%.      

The background cuts can be further improved.  We briefly note some possibilities.  In principle, one could cut muons with large energy loss to reduce the deadtime, because isotope yields are roughly proportional with muon energy losses.  In addition, one could use a shower profile cut.  Li and Beacom~\cite{Li:2014sea, Li:2015kpa, Li:2015lxa} show that most isotopes are made in rare hadronic showers, which should be easily recognized in DUNE.  In this approach, the deadtime would be reduced at least by a factor of a few.  

As a last step, we evaluate how the backgrounds would affect MeV programs in DUNE.  One aspect is understanding the absolute background rates compared to the signal rates.  For solar neutrinos, the pre-cut spallation backgrounds have comparable rates to the signals.  However, a simple two-step cut could make the spallation backgrounds totally subdominant~\cite{Capozzi:2018dat}.  For supernova neutrinos, the spallation background has a negligible rate compared to the intense burst.  For the DSNB, $hep$ solar neutrinos turn out to be the limiting background, the pre-cut spallation backgrounds are of comparable importance though.

Another aspect is to aid trigger design for supernova neutrino detection --- the primary MeV program in DUNE.  (Our results are detailed in the Appendixes.)  In this regard, the most significant background is not spallation isotope decays, but rather neutron captures on argon, where the neutrons are produced by U/Th decays in the rock.  This neutron capture background could also affect the solar-neutrino program proposed in Ref.~\cite{Capozzi:2018dat}, but we showed that it could be avoided by setting a high electron-energy threshold of $\simeq 7$~MeV (for 7\% energy resolution; higher if worse).  This would still allow a strong solar-neutrino program, but it would take somewhat longer to accumulate statistics.  For the supernova-neutrino program, to have an acceptable trigger rate, the threshold would need to be at least $\simeq 8$~MeV, which could mean that a faint supernova would be missed.  As shown in Ref.~\cite{Capozzi:2018dat}, these problems could be solved by even modest passive shielding.  If there is no shielding, the deleterious effects can be reduced if the energy resolution is good, which depends on a robust light-detection system~\cite{Acciarri:2016crz, Acciarri:2015uup, Strait:2016mof, Acciarri:2016ooe, Abi:2018dnh, Abi:2018alz, Abi:2018rgm}.  To fully realize its potential, DUNE must take new steps to ensure robust detector capabilities at MeV energies, as detailed above and in Ref.~\cite{Capozzi:2018dat}.

%%%%%%%%%%%%%%%%%%%%%%%%%%%%%%%%%%%%%%%%%%%%%%%%%%%%%%%%
%%%%%%%%%%%%%%%%%%%%%%%%%%%%%%%%%%%%%%%%%%%%%%%%%%%%%%%%

\begin{acknowledgments}

We are grateful for discussions with Alexander Friedland, Cristiano Galbiati, Ralph Massarczyk, Elizabeth Ricard-McCutchan, Fred Sarazin, Matt Strait, Mark Vagins, and especially Francesco Capozzi, David Caratelli, Davide Franco, Josh Klein, Dongming Mei, Kate Scholberg, Michael Smy, and Michel Sorel.  In particular, we thank Vitaly Kudryavtsev for providing the muon simulation code and the muon data at the DUNE site, and Alejandro Sonzogni for providing the radioactive decay data for Fig.~\ref{fig: nuclide_chart}.  The work of all authors was supported by NSF Grant PHY-1714479 awarded to JFB.  SWL was also supported by an Ohio State Presidential Fellowship, and later at SLAC by the Department of Energy under Contract No.\ DE-AC02-76SF00515.  

We speak for ourselves as theorists, not on behalf of the DUNE Collaboration. This work is based on our ideas, our calculations, and publicly available information.

\end{acknowledgments}

%%%%%%%%%%%%%%%%%%%%%%%%%%%%%%%%%%%%%%%%%%%%%%%%%%%%%%%%
%%%%%%%%%%%%%%%%%%%%%%%%%%%%%%%%%%%%%%%%%%%%%%%%%%%%%%%%

\appendix

\section{Isotope Production Yields}

%+++++++++++++++++++++++++++++FIGURE++++++++++++++++++++++++++++++++++++++++%
\begin{figure}[!t]
\includegraphics[width=\columnwidth]{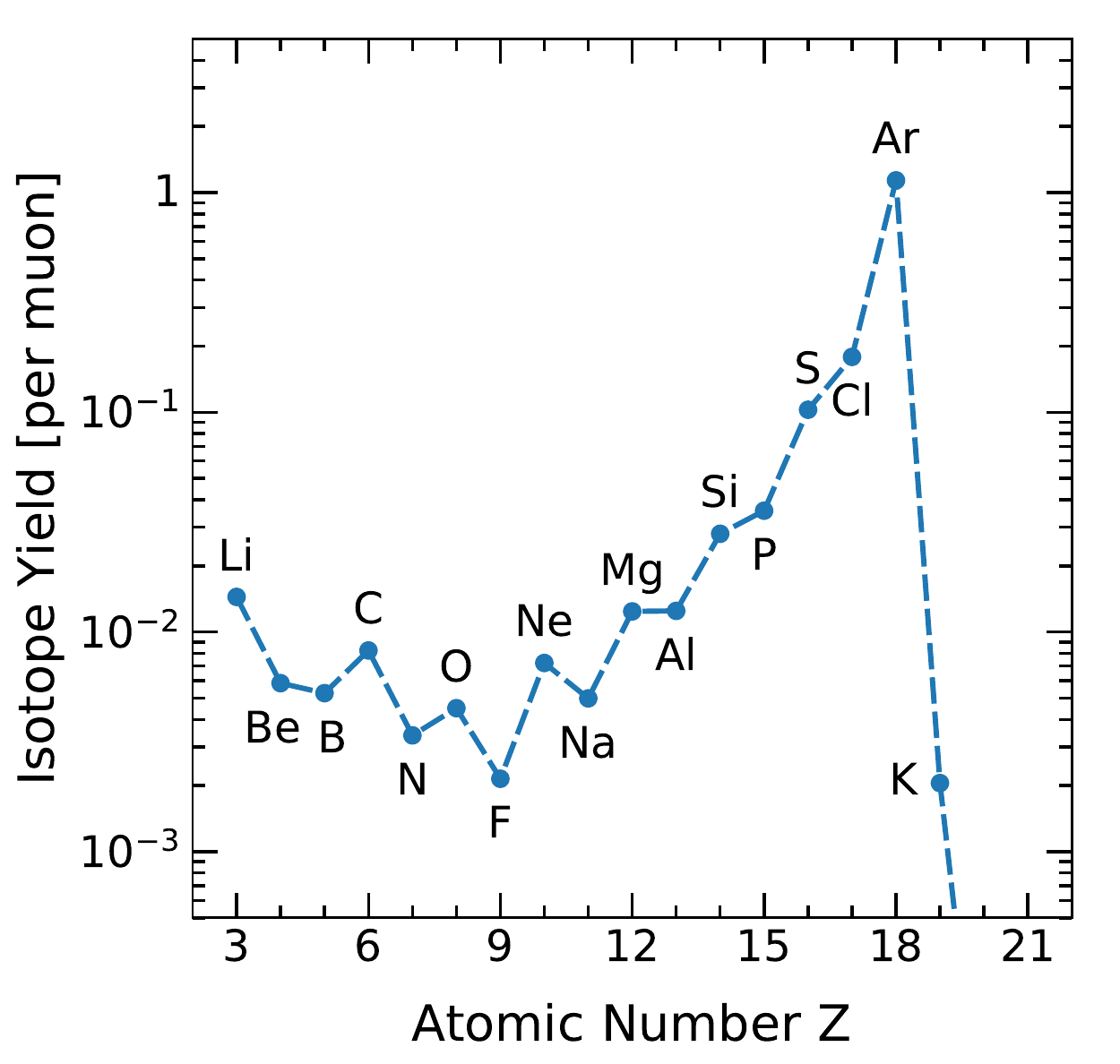}
\caption{Simulated isotope production yields in DUNE.  Both stable and unstable isotopes are counted.}
\label{fig: yield_vs_z}
\end{figure}
%++++++++++++++++++++++++++++++++++++++++++++++++++++++++++++++++++++++++++++%

Figure~\ref{fig: yield_vs_z} shows the dependence of predicted all-isotope production yields in DUNE on atomic number.  The high-A isotopes are much more abundant, though we have shown that they are not very important.  In contrast, the low-A isotopes are much less abundant, but are very important. 

\section{Effects of Different Assumed Energy Resolution}

\begin{figure}[!b]
\includegraphics[width=\columnwidth]{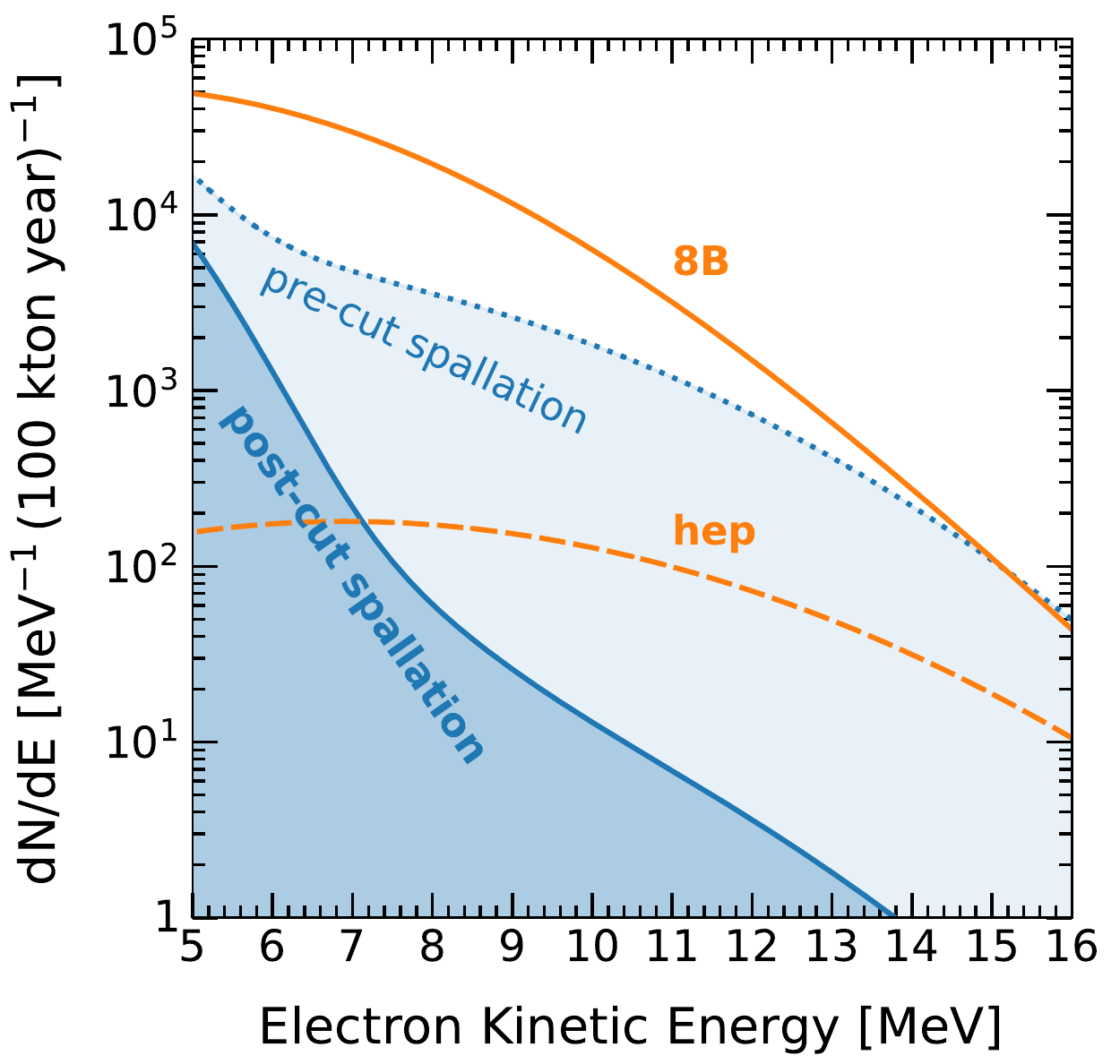}
\caption{Same as Fig.~\ref{fig: solar}, but with 20\% energy resolution.}
\label{fig: solar_20}
\end{figure}

\begin{figure}[!b]
\includegraphics[width=\columnwidth]{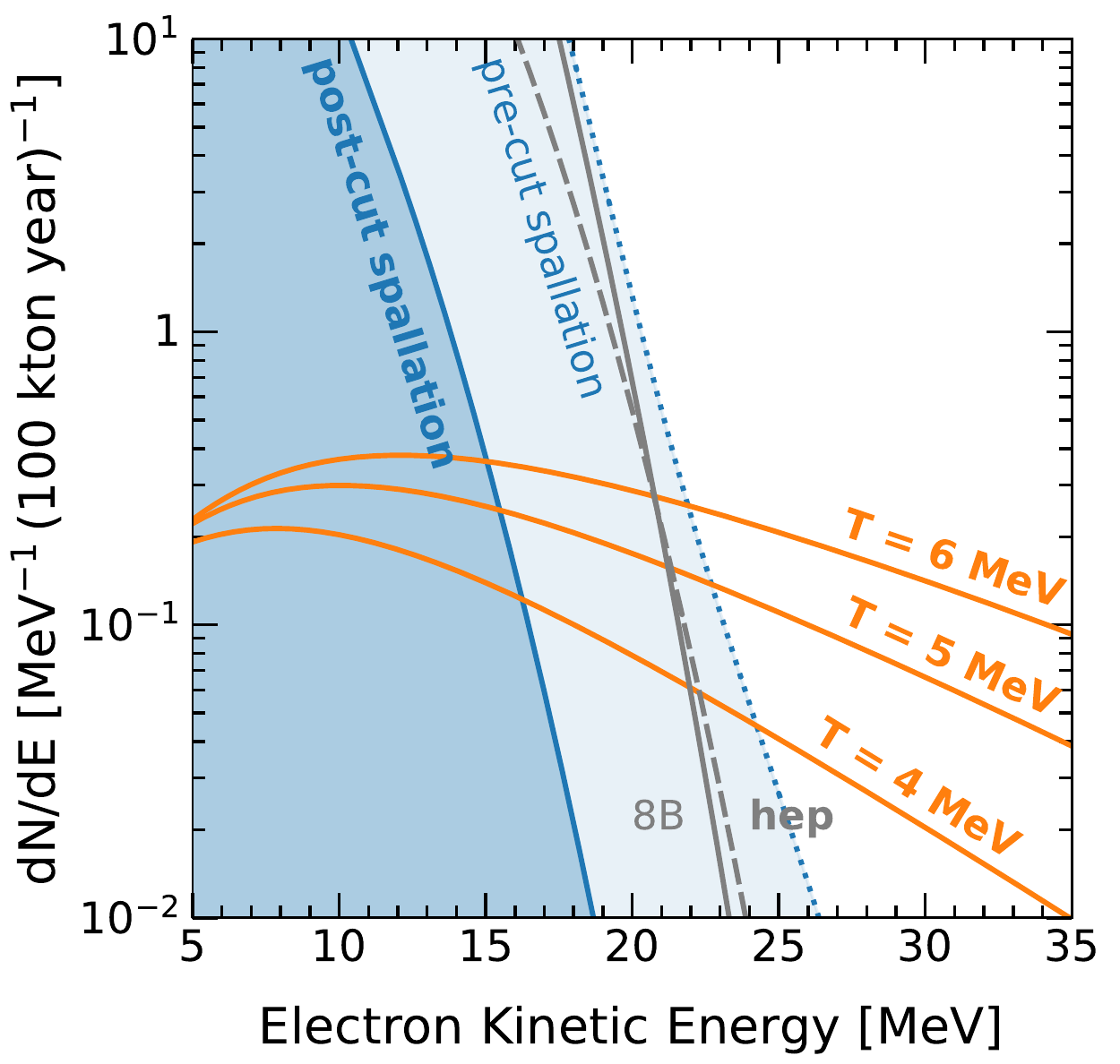}
\caption{Same as Fig.~\ref{fig: DSNB}, but with 20\% energy resolution.}
\label{fig: DSNB_20}
\end{figure}

We show the effects of worse energy resolution than the 7\% assumed in the main text.  Figures~\ref{fig: solar_20} and~\ref{fig: DSNB_20} are the same as Figs.~\ref{fig: solar} and~\ref{fig: DSNB}, but instead with 20\% energy resolution.  Though the spectra are somewhat wider, our basic conclusions are unchanged.

%+++++++++++++++++++++++++++++FIGURE++++++++++++++++++++++++++++++++++++++++%

%++++++++++++++++++++++++++++++++++++++++++++++++++++++++++++++++++++++++++++%
%+++++++++++++++++++++++++++++FIGURE++++++++++++++++++++++++++++++++++++++++%
%++++++++++++++++++++++++++++++++++++++++++++++++++++++++++++++++++++++++++++%

\section{Radiogenic Versus Cosmogenic Neutrons}

\begin{figure}[!t]
\includegraphics[width=\columnwidth]{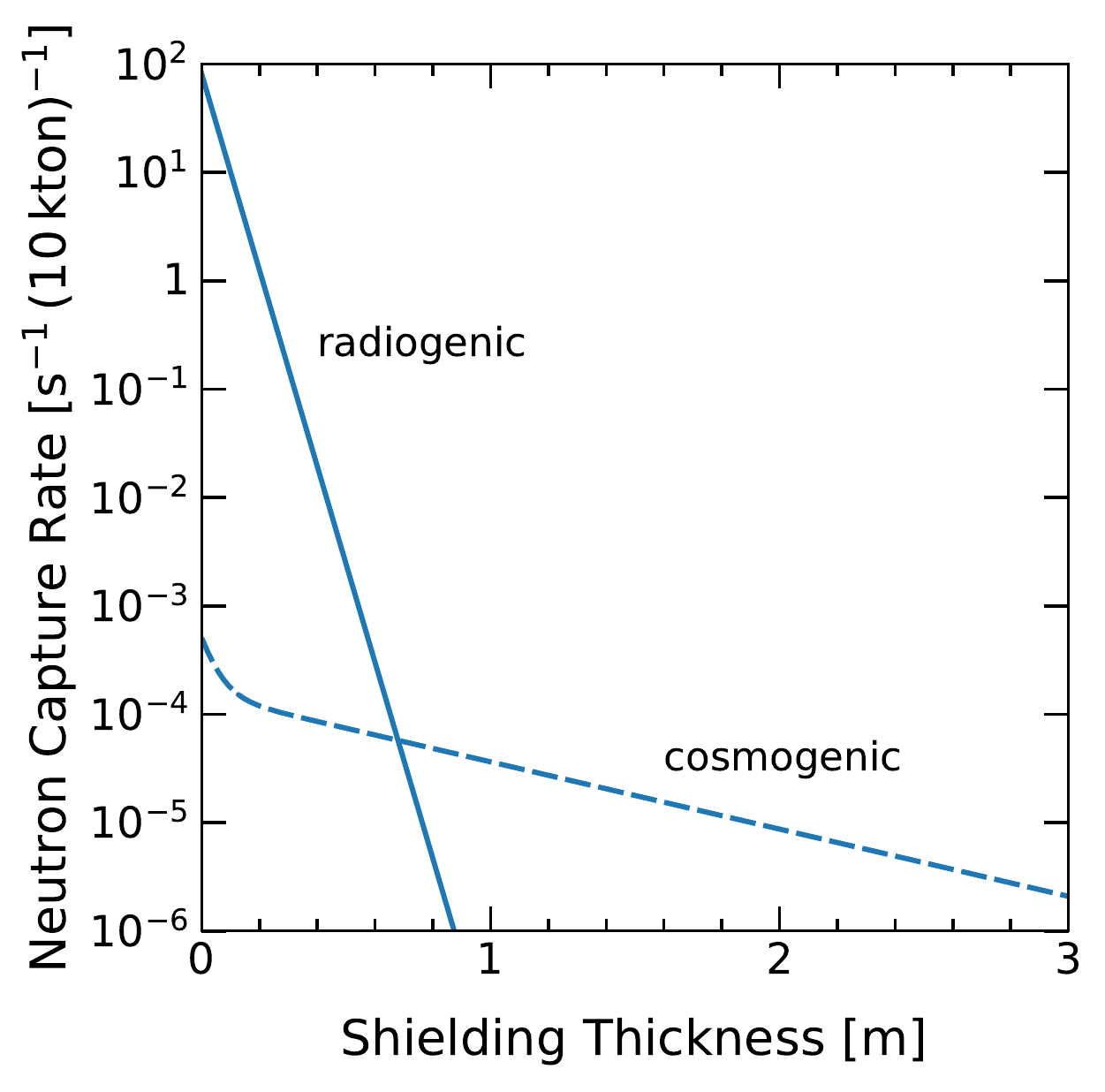}
\caption{Neutron capture rates in the DUNE fiducial volume.  Solid line:  Neutrons produced by radioactivities (U/Th) in the rock.  In our simulation, the radioactivity neutrons are injected uniformly and isotropically throughout the rock.  Dashed line:  Neutrons produced by muons passing through the rock.  In our simulation, the muons are injected vertically downward into the rock, counting only those that cannot be seen by the detector.}
\label{fig: radio_vs_cosmo}
\end{figure}

We compare the neutron-capture backgrounds due to radiogenic~\cite{Capozzi:2018dat} and cosmogenic sources.  Figure~\ref{fig: radio_vs_cosmo} shows the neutron-capture rates in the DUNE fiducial volume with different assumed thicknesses of passive water/oil/plastic shielding.  With zero or modest shielding, the dominant neutron source is radioactivities (assumed 3.43~ppm $^{238}$U and 7.11~ppm $^{232}$Th) in rock, which produce MeV-range neutrons at high rates.  These can be stopped efficiently by the shielding.  In argon, the neutron capture rate is $\propto e^{-x / \lambda}$, where $x$ is the shielding thickness, and the length scale is $\lambda \simeq 5$~cm.  Muons in the rock, unseen by the detector, produce GeV-range neutrons with low rates.  Because more shielding is needed to stop those energetic neutrons, the capture rate in argon is still $\propto e^{-x / \lambda}$, but with $\lambda \simeq 70$~cm.  As a high-energy neutron propagates, it can make low-energy neutrons that capture more easily.  We take this into account in our simulation.  However, this process is not very efficient, with a 100-MeV neutron in argon leading to only $\approx1$ neutron capture (neglecting escape, so the real number is lower).  The efficiency is low because, beyond the binding-energy cost of removing nucleons, often the ejecta are protons or light nuclei, and typically the ejected nucleons or light nuclei have substantial kinetic energies.  At 1~GeV, the number of captures increases, but only to $\approx4$, which is even less efficient in terms of neutron captures per energy injected.  We conclude that only in detectors with significant shielding are the cosmogenic neutrons more important than the radiogenic neutrons.

%+++++++++++++++++++++++++++++FIGURE++++++++++++++++++++++++++++++++++++++++%
%++++++++++++++++++++++++++++++++++++++++++++++++++++++++++++++++++++++++++++%

\section{Trigger Considerations}

The supernova trigger design is ongoing, so we make some reasonable assumptions.  First, we adopt a readout window of 5.4~ms --- one drift time before a trigger and one after --- following Ref.~\cite{Josh}.  Second, we assume a conservative energy resolution of 20\%, which may be more realistic at the trigger level.  Third, we assume an energy threshold of 5~MeV.  This means that any event below 5~MeV is invisible at the trigger level, unless there is pileup, which is discussed in detail below.

Under such assumptions, we consider two types of supernova triggers, based on the burst characteristics.  For a supernova at 10~kpc, the expected counts are $\approx1$--2 per 5.4-ms readout window in the first $\approx1$~s of the supernova, and falling thereafter.  Type I triggers on near-continuous tracks, i.e., individual electron tracks above 5~MeV, in several readout windows.  Type II triggers on many tracks in a single readout window, which is probably more suitable for the start of a supernova at a smaller distance.  For both types, determining the value of $n$, i.e., the number of 5.4-ms bins or the number of tracks per bin, is not trivial, because if $n$ is too small, there would be too many fake triggers due to backgrounds; if $n$ is too large, one may miss the first tens of milliseconds of supernova events.  Below, we detail how backgrounds would affect these two types of supernova neutrino trigger designs.  We report the calculated fake trigger rates in units of month$^{-1}$~(10~kton)$^{-1}$, as each module is independent in terms of trigger, and once per month is likely the scale to determine whether a fake rate is acceptable. 

In short, we find that muon-generated backgrounds are unimportant, but those due to radiogenic neutrons from the rock are important.

%%%%%%%%%%%%%%%%%%%%%%%%%%%%%%%%%%%%%%%%%%%%%%%%%%%%%%

\subsection{Muon activity in the detector}

%+++++++++++++++++++++++++++++FIGURE++++++++++++++++++++++++++++++++++++++++%
\begin{figure}[!t]
\includegraphics[width=\columnwidth]{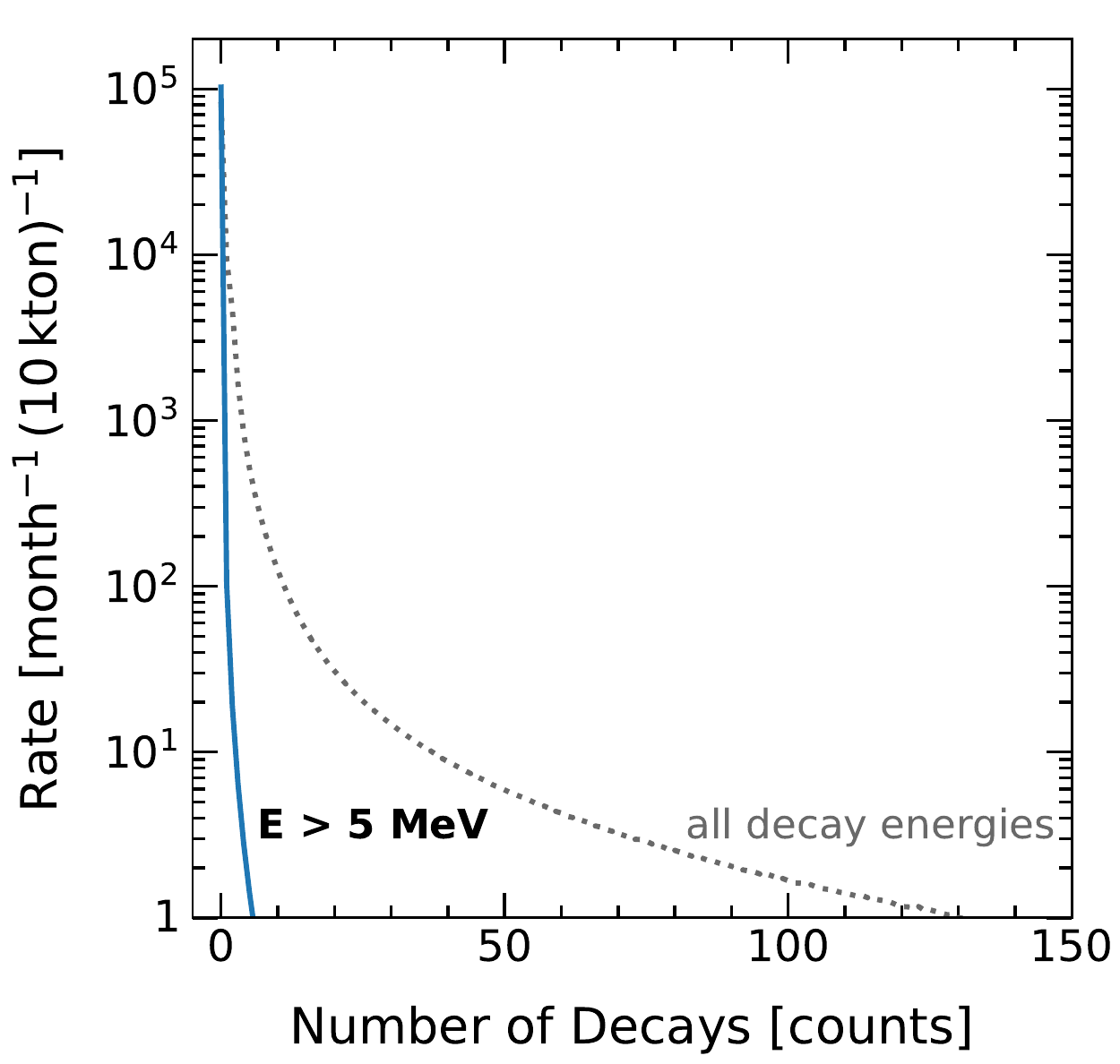}
\caption{Spallation isotope decay multiplicities per muon in DUNE.  Bin size is 1 count.  Gray dotted line: all spallation betas.  Blue solid line: spallation decays above the 5-MeV threshold (in electron energy).}
\label{fig: SN}
\end{figure}
%++++++++++++++++++++++++++++++++++++++++++++++++++++++++++++++++++++++++++++%

One potential fake trigger scenario is the aftermath of a cosmic-ray muon.  Throughgoing muons easily trigger the detector, due to the typical GeV-range energy deposition.  In the 2.7-ms drift window after the muon trigger, charges collected on the wire planes are dominantly from shower particles, and hence the detector is very active.  We ignore this time bin in our following discussions, as it is already triggered by a muon and it is likely to be discarded to clear the charges.  In the next 5.4-ms readout bin, charge depositions are mostly from spallation isotope decays, with a small fraction from neutron captures.  The neutron capture rate, on average, is $\simeq 0.5$ per muon.  Most captures happen with $\tau\simeq 0.35$~ms ($\simeq 30\%$ of captures are at energies above $\simeq 0.01$~MeV and hence sooner).  In the first 2.7~ms of this readout bin, the neutron capture rate is $\approx10$ per month, but the rate to have an electron above 5~MeV (i.e., a visible event), is much lower, $\approx0.5$ per month.  In the second 2.7~ms of this readout bin, the neutron capture rate is vanishing.  In all the following 5.4-ms readout bins, any associated charge depositions are purely from spallation isotope decays.  The detector module will become relatively quiet until another muon comes in, which is typically $\simeq 20$~s later.  Understanding the spallation decay multiplicities in the readout windows after a muon is important for supernova trigger design.

Figure~\ref{fig: SN} shows the spallation decay multiplicity distributions due to throughgoing muons.  The multiplicity does not follow a Poisson distribution, because isotopes are made in showers sampled from a broad energy range.  For the decays below 5~MeV, even though their multiplicities are high, we find that pileup is negligible, following our calculations below.  Thus, only decays above 5~MeV are relevant.  To approximately take into account the energy smearing effect, we use 4~MeV as the pre-smearing threshold in our analysis, which is conservative even for a 20\% energy resolution.     

Table~\ref{spallation_5.4ms}, in its first three rows, lists the rates of selected decay multiplicities for different energy thresholds.  To quantify their impact on the triggers, we further look into the per-readout multiplicities.  For a 5-MeV threshold, in the fourth row of Table~\ref{spallation_5.4ms}, we list the rates of decay multiplicities from 1 to 5 in the 2.7--8.1~ms bin, which has the maximum decay yield for $t > 2.7$~ms.  (Here we define $t = 0$ as when a muon triggers.)  The rate to have 1 decay from a muon in that bin is 9~month$^{-1}$, corresponding to a probability of $\simeq 10^{-4}$ for each muon.  We note that for any trigger, it must have more than 1 event.  For a Type I trigger, the rate to have a following decay in the next 5.4~ms is negligible, $\simeq 10^{-3}$~month$^{-1}$, so the spallation background is not a concern.  For a Type II trigger, it is not a problem either, because the rate to have more than 1 decay per readout is tiny, $\lesssim 0.3$~month$^{-1}$, as shown in Table~\ref{spallation_5.4ms}.  For a 10-MeV threshold, the trigger rates are even smaller.  Overall, even under conservative assumptions, spallation decays from throughgoing muons would not affect the supernova trigger.

%+++++++++++++++++++++++++++++TABLE++++++++++++++++++++++++++++++++++++++++%
\begin{table}[!t]
\centering
\setlength{\extrarowheight}{2.pt}
\caption{Approximate rates of spallation isotope decay multiplicities per muon in DUNE.  The rates are in units of month$^{-1}$~(10~kton)$^{-1}$; those lower than 1 event~(10 yr)$^{-1}$~(10~kton)$^{-1}$ are not shown.}
\begin{tabular}{lccccc}
\Xhline{2\arrayrulewidth}\\[-1.25em]\Xhline{2\arrayrulewidth}
\multicolumn{1}{l}{\multirow{2}{*}{}}&
\multicolumn{5}{c}{Number of decays} \\
\cline{2-6}
\multicolumn{1}{c}{} &
\multicolumn{1}{C{0.87cm}}{1} & 
\multicolumn{1}{C{0.87cm}}{2} & 
\multicolumn{1}{C{0.87cm}}{3} & 
\multicolumn{1}{C{0.87cm}}{4} & 
\multicolumn{1}{C{0.87cm}}{5}  \\ 
\Xhline{1.1\arrayrulewidth}\\[-1.25em]\Xhline{1.1\arrayrulewidth}
all decay energies                      &     $1\times10^4$      &     5$\times$10$^3$   &    2$\times$10$^3$          &      800              &            500     \\
$E > 5$~MeV                             &      100          &     20                           &    6                                    &       3                 &            1          \\
$E > 10$~MeV                           &      20            &     1                             &    0.1                                 &       0.03            &      0.01          \\
\hline
$E > 5$~MeV in 5.4~ms          &       9             &     0.3                          &    0.03                                &     --           &          --     \\
$E > 10$~MeV in 5.4~ms        &       1            &       --                           &    --                            &        --         &      --          \\
\Xhline{2\arrayrulewidth}\\[-1.25em]\Xhline{2\arrayrulewidth}
\label{spallation_5.4ms}
\end{tabular}
\end{table}

%++++++++++++++++++++++++++++++++++++++++++++++++++++++++++++++++++++++++++++%

%%%%%%%%%%%%%%%%%%%%%%%%%%%%%%%%%%%%%%%%%%%%%%%%%%%%%%

\subsection{Muon activity in the rock}

Similar to the throughgoing-muon case, there can be muons that pass through rock near the detector.  Their spallation betas (entering the detector) are also not a concern for the supernova triggers.  Note that our results for throughgoing muons are calculated for the pre-cut backgrounds, so recognizing the muon track or not does not matter.  In that sense, the spallation multiplicities in the detector due to the muons in the rock would have a similar distribution as shown in Fig.~\ref{fig: SN} and Table~\ref{spallation_5.4ms}, but have a much smaller normalization, due to the much lower yields of isotope in the detector (details in Sec.~\ref{sec:DUNE_aftercut}).  The neutron capture backgrounds due to these muons in the rock would not be a concern either, because their rates of $\simeq 5\times10^{-4}$~s$^{-1}$ are $\approx5$ orders of magnitude lower than those due to radioactivities in the rock, as shown in Fig.~\ref{fig: radio_vs_cosmo}, and one can further cut these neutrons from the near-detector muons by recognizing the accompanying electromagnetic showers that enter the detector.  

%%%%%%%%%%%%%%%%%%%%%%%%%%%%%%%%%%%%%%%%%%%%%%%%%%%%%%

%+++++++++++++++++++++++++++++TABLE++++++++++++++++++++++++++++++++++++++++%

\begin{table}[!h]
\centering
\caption{Approximate rates of fake Type I triggers due to neutron-capture backgrounds, where the background electrons appear in successive 5.4-ms bins.  A 20\% energy resolution is assumed.  Different shielding thicknesses and detector energy thresholds are used.  The rates are in units of month$^{-1}$~(10~kton)$^{-1}$; those lower than 1 event~(10 yr)$^{-1}$~(10~kton)$^{-1}$ are not shown.}
\label{neutron_typeI}
\centering
\setlength{\extrarowheight}{2.pt}
\begin{tabular}{cccccc}
\Xhline{2\arrayrulewidth}\\[-1.25em]\Xhline{2\arrayrulewidth}
\multicolumn{1}{C{1.35cm}}{Shielding} &
\multicolumn{1}{C{1.35cm}}{$\rm E_{thr}$} &
\multicolumn{4}{C{4.6cm}}{Number of 5.4-ms bins} \\
\cline{3-6}
\multicolumn{1}{C{1.35cm}}{[cm]} &
\multicolumn{1}{C{1.35cm}}{[MeV]} &
\multicolumn{1}{C{1.15cm}}{1} & 
\multicolumn{1}{C{1.15cm}}{2} & 
\multicolumn{1}{C{1.15cm}}{3} & 
\multicolumn{1}{C{1.15cm}}{4} \\
\Xhline{1.1\arrayrulewidth}\\[-1.25em]\Xhline{1.1\arrayrulewidth}
\multirow{3}{*}{0~cm} & 5           &   1$\times$10$^7$                   &   4$\times$10$^5$     &   1$\times$10$^4$  &   300  		  \\
                                   & 6           &   3$\times$10$^6$   &   1$\times$10$^4$                     &    90         &   0.5  		   \\
                                   & 7           &   4$\times$10$^5$   &    300 		     	       &     0.2  	  &   -- 	             \\                                   
                                   & 8           &   3$\times$10$^4$   &    2 		     	       &      --         &   -- 		      \\
                                   & 9           &   2$\times$10$^3$   &    -- 		                &     --          &  --   		       \\
                                   & 10         &   60			   &   -- 		    	        &     --          &  --  	                 \\
\hline
\multirow{3}{*}{20~cm} & 5           &   1$\times$10$^5$     		   	&     30                      &   --   	   &   --                     \\
                                     & 6           &   2$\times$10$^4$ 	&     1  		        &   --   	   &   --                      \\
                                     & 7           &   3$\times$10$^3$      &      0.02  		&   --            &   --	                 \\
                                     & 8           &   300			        &       -- 		         &   --  	   &  -- 			 \\
                                     & 9           &   20 			        &       -- 		     	 &   --  	   &  --  		          \\
                                     & 10         &   0.5 			        &       -- 		          &   --           &  --  		           \\
\hline
\multirow{3}{*}{40~cm} & 5           &   3$\times$10$^3$            &     0.02        &      --         &    --      \\
                                     & 6           & 	700		       		      &      --           &      --         &    --	    \\
                                     & 7           & 	90        			      &      --           &      -- 	      &    -- 	    \\
                                     & 8           &   8			   	      &   -- 		   &      --         &    --         \\
                                     & 9           &   0.4 			   	      &   -- 		   &      --         &    --         \\
                                     & 10         &   0.01			  	      &   -- 		   &     --          &    --          \\
\hline
\Xhline{2\arrayrulewidth}\\[-1.25em]\Xhline{2\arrayrulewidth}
\end{tabular}
\end{table}

\begin{table}[!h]
\centering
\caption{Approximate rates of fake Type II triggers due to neutron-capture backgrounds, where multiple background electrons appear in a single 5.4-ms bin.  A 20\% energy resolution is assumed.  Different shielding thicknesses and detector energy thresholds are used.  The rates are in units of month$^{-1}$~(10~kton)$^{-1}$; those lower than 1 event~(10 yr)$^{-1}$~(10~kton)$^{-1}$ are not shown.}
\label{neutron_typeII}
\centering
\setlength{\extrarowheight}{2.pt}
\begin{tabular}{cccccc}
\Xhline{2\arrayrulewidth}\\[-1.25em]\Xhline{2\arrayrulewidth}
\multicolumn{1}{C{1.35cm}}{Shielding} &
\multicolumn{1}{C{1.35cm}}{$\rm E_{thr}$} &
\multicolumn{4}{C{4.6cm}}{Number of electrons in 5.4~ms} \\
\cline{3-6}
\multicolumn{1}{C{1.35cm}}{[cm]} &
\multicolumn{1}{C{1.35cm}}{[MeV]} &
\multicolumn{1}{C{1.15cm}}{1} & 
\multicolumn{1}{C{1.15cm}}{2} & 
\multicolumn{1}{C{1.15cm}}{3} & 
\multicolumn{1}{C{1.15cm}}{4} \\
\Xhline{1.1\arrayrulewidth}\\[-1.25em]\Xhline{1.1\arrayrulewidth}
\multirow{3}{*}{0~cm} & 5           &   1$\times$10$^7$   			&          2$\times$10$^5$  		     &      2$\times$10$^3$                &      10           \\
                                   & 6           &   3$\times$10$^6$        &          8$\times$10$^3$                &      20     				       &      0.02        \\
                                   & 7           &   4$\times$10$^5$        &   	   100 		     	              &      0.04   				       &      -- 	       \\
                                   & 8           &   3$\times$10$^4$        &   	    1	 		              &       --  				                &       -- 	 	\\
                                   & 9           &   2$\times$10$^3$ 	&   	      -- 		    	               &      --  				       & 	-- 		 \\
                                   & 10         &   60			         &            -- 		                         &     --                                        &       --                 \\
\hline
\multirow{3}{*}{20~cm} & 5           &    1$\times$10$^5$      &     20              &     --      &      --               \\
                                     & 6           &    2$\times$10$^4$      &      0.6            &   --        &      --             \\
                                     & 7           &    3$\times$10$^3$	 &      0.01          &   --        &     --	       \\
                                     & 8           &    300 			          &   -- 		 &     --       &  --  \\
                                     & 9           &    20 			          &   -- 		 &     --       &  --  \\
                                     & 10         &   0.5			          &   -- 		 &     --       &  --  \\
\hline
\multirow{3}{*}{40~cm} & 5           &   3$\times$10$^3$            &      0.01        &      --       &   --      \\
                                     & 6           & 	700		      		      &      --            &     --        &    --	 \\
                                     & 7           & 	90       		              &      --            &      -- 	    &     -- 	  \\
                                     & 8           &   8 			              &   -- 		   &     --        &     --  \\
                                     & 9           &   0.4 			              &   -- 		   &     --        &  --  \\
                                     & 10         &   0.01 			              &   -- 		   &     --        &  --  \\
\hline
\Xhline{2\arrayrulewidth}\\[-1.25em]\Xhline{2\arrayrulewidth}
\end{tabular}
\end{table}

%+++++++++++++++++++++++++++++++++++++++++++++++++++++++++++++++++++++++++++++++%

\subsection{Radioactivity neutrons}
\label{sec: radio_neutron}

The dominant source of neutron backgrounds is due to radioactivities in the rock, primarily $^{238}$U and $^{232}$Th.  The neutrons from muons in the rock or from radioactivities in the detector (orders of magnitude lower U/Th concentrations) are much fewer.  The most relevant energy range is at or below a few MeV, as neutrons at these energies can easily reach all parts of the detector and get captured inside the detector.  (Neutrons between a few MeV and about 10~MeV can travel long distances and may escape;  neutrons at higher energies have short attenuation lengths due to inelastic interactions, and cannot travel far into the detector.)  Here we give simple estimates to understand neutron propagation; our results are based on full {\tt FLUKA} simulations.  (A recent measurement of the thermal-neutron capture cross section~\cite{Fischer:2019qfr} is in good agreement with the values used by {\tt FLUKA}; new measurements on neutron interactions with argon at higher energies would be very helpful.)  For neutrons at or below a few MeV, their mean free path due to elastic scattering is $\lambda \approx 15$~cm.  Because the elastic-scattering energy loss on $^{40}$Ar is very inefficient, a 1-MeV neutron needs to scatter $n \approx 400$ times to be thermalized.  Thus, with random walk approximation, the diameter of the neutron trajectory would be $2\,\sqrt{n}\,\lambda \approx 6$~m, comparable to the fiducial volume height or width.  Once a neutron is captured on $^{40}$Ar, a total energy of 6.1~MeV will be released in gamma rays.  These gamma rays will mostly Compton scatter, producing electrons that can be backgrounds for supernova neutrinos.

The background rate due to neutron captures is determined by two factors: the neutron capture rate and the number of electrons per capture that are above the energy threshold.  Given the radioisotope concentration in the rock, we find that the neutron capture rate is $\simeq 81$~s$^{-1}$ in each module.  With a 5-MeV energy threshold and a 20\% energy resolution, the per-module background electron rate is $\simeq 5$~s$^{-1}$.  As shown below, such a high background rate can be a concern for both types of trigger.

Tables~\ref{neutron_typeI} and~\ref{neutron_typeII} show our calculated neutron background multiplicity distributions relevant to the Type I and II triggers, respectively.  Because these backgrounds follow a Poisson distribution, the rates shown in the tables are determined by $\mu$, the expected counts in a 5.4-ms readout window.  For example, the background electron rate of $\simeq 5$~s$^{-1}$ corresponds to a $\mu \simeq 0.03$.  For a Type I trigger, if the definition is to have $n$ tracks in a row of 5.4-ms bins, then the fake trigger rate would be $(\mu^n \times 0.0054^{-1})$~s$^{-1}$.  For a Type II trigger, if it requires $n$ tracks in a single 5.4-ms bin, then the fake rate would be $(\mu^n / n! \times 0.0054^{-1})$~s$^{-1}$.  As shown in the tables, the fake trigger rates in both cases could be as large as hundreds per month per module for some choices of $n > 2$.   

To lower the fake trigger rate, one approach is to enforce a higher energy threshold.  When smeared with a 20\% energy resolution, the number of electrons per capture that are above $E_{thr} = 5$, 6, 7, 8, 9, and 10~MeV are about 0.066, 0.013, 0.002, 10$^{-4}$, 8.4$\times 10^{-6}$, and 2.6$\times 10^{-7}$, respectively.  The per-module background electron rates are accordingly about 5, 1, 0.2, 0.01, 7$\times 10^{-4}$, and 2$\times 10^{-5}$~s$^{-1}$, respectively.  With an 8 or 9~MeV threshold, the fake rate could be reduced significantly, as shown in Tables~\ref{neutron_typeI} and~\ref{neutron_typeII}.  {\it However, this increases the probability of missing a supernova with a small number of events or with a low-energy spectrum.}

Alternatively, to avoid sacrifice on the energy threshold, we propose to add passive water (/oil/plastic) shielding that would greatly reduce the neutron capture rate.  With no shielding, 20-cm shielding, and 40-cm shielding, the neutron capture rates in each module are about 81, 0.7, and 0.02~s$^{-1}$, respectively.  Assuming a 5-MeV threshold, the background electron rates are accordingly about 5, 0.05, and 0.001~s$^{-1}$, respectively.  With shielding, the fake trigger rate would be negligible, even for a threshold at 5~MeV, as shown in Tables~\ref{neutron_typeI} and~\ref{neutron_typeII}.  This strategy will benefit not only the supernova detection, but also the solar neutrino program proposed in Ref.~\cite{Capozzi:2018dat}. 

%%%%%%%%%%%%%%%%%%%%%%%%%%%%%%%%%%%%%%%%%%%%%%%%%%%%%%

\subsection{Radioactivities in the detector}

For intrinsic radioactivities in the detector, relevant beta-decay sources could be $^{39}$Ar and $^{42}$Ar in the atmospheric argon used in DUNE, $^{42}$K from $^{42}$Ar decay, and $^{214}$Bi from the decay chain of $^{222}$Rn.  In each module, the $^{39}$Ar activity is $\approx10^7$~Hz~\cite{Benetti:2006az}, and the $^{42}$Ar activity is $\approx10^3$~Hz~\cite{Agostini:2013tek, Barabash:2016uzl}.  Note that both $^{39}$Ar and $^{42}$Ar have Q values $\simeq 0.6$~MeV, so $^{42}$Ar itself is unimportant.  Similarly, both $^{42}$K and $^{214}$Bi have Q values $\simeq 3.5$~MeV.  The $^{42}$K activity is $\approx10^3$~Hz, due to its short half-life compared to $^{42}$Ar.  The activity of $^{214}$Bi (or $^{222}$Rn) has not yet been measured at DUNE.  Conservatively, we assume a $^{222}$Rn activity of $\approx10$~mBq/m$^3$, mainly from the detector materials, corresponding to $\approx10^2$~Hz per module.  (In Super-Kamiokande, after the detector was sealed and the initial radon decayed away, the $^{222}$Rn activity was $\approx20$~mBq/m$^3$ until water-purification procedures lowered it to a few mBq/m$^3$~\cite{Takeuchi:1999zq}.)  Thus, $^{214}$Bi should be unimportant compared to $^{42}$K in DUNE.

With the above arguments, the most important intrinsic radioactive isotopes are $^{39}$Ar ($Q \simeq 0.6$~MeV, $R \approx 10^7$~Hz) and $^{42}$K ($Q \simeq 3.5$~MeV, $R \approx 10^3$~Hz).  They are abundant, but due to their low Q values, they cannot trigger the detector unless there is pileup.  

%%%%%%%%%%%%%%%%%%%%%%%%%%%%%%%%%%%%%%%%%%%%%%%%%%%%%%

\subsection{Pileup}
\label{sec: pileup}

We first present a framework to calculate pileup rates due to sub-threshold events.  With specific numbers and conservative assumptions, we show that the expected pileup rates would be negligible for a nominal threshold near 5~MeV in electron energy.   

In general, the rate of $n$-event pileup is $R \times P\,(\mu, n-1)$, where $R$ is the single-event rate, and $P\,(\mu, n - 1)$ is the Poisson probability for another $(n - 1)$ events that satisfy the pileup conditions of coincidence within a space volume $\Delta x \times \Delta y \times \Delta z$ and a time window $\Delta t$.

We first consider how pileup would work at the reconstruction level, due to greater simplicity, though that is not the whole story.  At the reconstruction level, the resolution in the drift direction, which we call $x$, is $\simeq 0.8$~mm, determined by the $t_0$ resolution and the drift speed $v$, while the resolution in $y$ and $z$ is $\simeq 5$ mm, limited by the wire spacing.  A 5-MeV electron's track length is $\simeq 2.5$~cm long, crossing $\simeq 5$ wires.  For simplicity, we conservatively choose $\Delta x = \Delta y = \Delta z = 5\, \rm cm$.  The time window $\Delta t$ should be decided by the charge pulse width on the wire planes, i.e., $\delta t \simeq 50\, \rm \mu s$~\cite{Josh}, conservatively.  For $n$ events to appear as a single event, the maximum time separation between the first and the last pulse peak is $\Delta t = (n-1)\,\delta t$, which is very conservative given the $\delta t$ value we use.   

At the trigger level, $t_0$ might not be extractable, and one would have a higher pileup rate.  We start with the pileup of two events, with single-event rates $R_i$ and $R_j$, respectively.  We use the $\Delta y$ and $\Delta z$ defined above, and derive $\Delta x$.  Suppose the first event happens at $(x_0,\, t_0=0)$, where $x_0\in[0,\,x_{max}]$, i.e., anywhere along the drift direction.  If a second event, originated at $(x',\ t' > 0)$, reaches the wire plane within $\delta t$ of the first event, these two events would appear as one.  Thus, the pileup condition is as follows: for any $x_0$, all $(x',\ t')$ that satisfies $\mid x' - (x_0 + vt') \mid\ \leq v\,\delta t$, which is a band of area $f(x_{0}) = 2\,(x_{max} - x_0)\,\delta t$ on a $x' - t'$ plot.  The Poisson expectation would accordingly be 
%%%%%%%%%%%%%%%
\begin{align}
\mu_j  & = R_j\, (x_{max}\,y_{max}\,z_{max})^{-1}\,(\Delta y\, \Delta z)\int_{0}^{x_{max}}\frac{dx_0}{x_{max}}\,f(x_{0}) \nonumber \\
& = R_j\, (x_{max}\,y_{max}\,z_{max})^{-1}\,(\Delta y\, \Delta z\, x_{max})\,\delta t.
\end{align}
%%%%%%%%%%%%%%
This is equivalent to setting $\Delta x = x_{max}$ and $\Delta t = \delta t$.  The length scales we use are $y_{max} = 12$~m, $z_{max} = 58$~m, and $x_{max} = 14.5$~m (maximum total drift distance, to be conservative).  Given $R_i$ and $\mu_j$, the two-event pileup rate is $R_i \times P\,(\mu_j, 1) = R_i\,\mu_j$.  For n-event pileup at the trigger level, the above derivation can be simply generalized by requiring $\Delta x = x_{max}$ and $\Delta t = (n - 1)\,\delta t$.  For example, the pileup rate due to four $^{39}$Ar decays is $\simeq 0.3$~Hz per module, consistent with the $\simeq 1$~Hz from Ref.~\cite{Josh}.  

In this framework, we now calculate the pileup rates at the trigger level with conservative assumptions.  We assume that $^{39}$Ar and $^{42}$K always decay to betas with the endpoint energies.  For the sub-threshold electrons due to neutron captures, we assume that there are two 3-MeV electrons produced per capture~\cite{Hardell1970, Nesaraja:2016ktw, CapGam}, with a rate of $\approx200$~Hz per module without shielding.  Therefore, neutron capture would have a rate $\approx5$ times lower than that of $^{42}$K decay.  Thus, there are only two sources that could dominate the pileup rate: 0.6-MeV betas from $^{39}$Ar, with a rate of $\approx10^7$~Hz, and 3.5-MeV betas from $^{42}$K, with a rate of $\approx10^3$~Hz.

For pileups that appear as a single $\simeq 5$-MeV event, there are two types.  The first type is among multiple $^{39}$Ar or $^{42}$K events themselves.  There could be $\simeq 10$ decays of $^{39}$Ar, with a per-module rate of $\approx10^{-15}$~s$^{-1}$, or $\simeq 2$ decays of $^{42}$K, with a per-module rate of $\approx10^{-4}$~s$^{-1}$.  The second type is between $^{39}$Ar and $^{42}$K events, where the only scenario would be $\simeq 3$ decays of $^{39}$Ar with $\simeq 1$ decay of $^{42}$K, with a per-module rate of $\approx10^{-5}$~s$^{-1}$.          
 
Even such pileup rates at the trigger level would be negligible compared to neutron-capture background rate above 5~MeV, which is $\simeq 5$~s$^{-1}$ without shielding and $\simeq 10^{-3}$~s$^{-1}$ with 40-cm shielding (details in Appendix.~\ref{sec: radio_neutron}).  At the reconstruction level, or with a higher energy threshold, the pileup rates would be even smaller than above.  We conclude that pileup is irrelevant for $E > 5$~MeV.  Even for slightly lower thresholds, the results would hold, as we have unrealistically assumed that every decay beta has its full Q value.

%%%%%%%%%%%%%%%%%%%%%%%%%%%%%%%%%%%%%%%%%%%%%%%%%%%%%%%%
%%%%%%%%%%%%%%%%%%%%%%%%%%%%%%%%%%%%%%%%%%%%%%%%%%%%%%%%

\bibliographystyle{apsrev4-1}
\bibliography{references}

\begin{thebibliography}{89}%
\makeatletter
\providecommand \@ifxundefined [1]{%
 \@ifx{#1\undefined}
}%
\providecommand \@ifnum [1]{%
 \ifnum #1\expandafter \@firstoftwo
 \else \expandafter \@secondoftwo
 \fi
}%
\providecommand \@ifx [1]{%
 \ifx #1\expandafter \@firstoftwo
 \else \expandafter \@secondoftwo
 \fi
}%
\providecommand \natexlab [1]{#1}%
\providecommand \enquote  [1]{``#1''}%
\providecommand \bibnamefont  [1]{#1}%
\providecommand \bibfnamefont [1]{#1}%
\providecommand \citenamefont [1]{#1}%
\providecommand \href@noop [0]{\@secondoftwo}%
\providecommand \href [0]{\begingroup \@sanitize@url \@href}%
\providecommand \@href[1]{\@@startlink{#1}\@@href}%
\providecommand \@@href[1]{\endgroup#1\@@endlink}%
\providecommand \@sanitize@url [0]{\catcode `\\12\catcode `\$12\catcode
  `\&12\catcode `\#12\catcode `\^12\catcode `\_12\catcode `\%12\relax}%
\providecommand \@@startlink[1]{}%
\providecommand \@@endlink[0]{}%
\providecommand \url  [0]{\begingroup\@sanitize@url \@url }%
\providecommand \@url [1]{\endgroup\@href {#1}{\urlprefix }}%
\providecommand \urlprefix  [0]{URL }%
\providecommand \Eprint [0]{\href }%
\providecommand \doibase [0]{http://dx.doi.org/}%
\providecommand \selectlanguage [0]{\@gobble}%
\providecommand \bibinfo  [0]{\@secondoftwo}%
\providecommand \bibfield  [0]{\@secondoftwo}%
\providecommand \translation [1]{[#1]}%
\providecommand \BibitemOpen [0]{}%
\providecommand \bibitemStop [0]{}%
\providecommand \bibitemNoStop [0]{.\EOS\space}%
\providecommand \EOS [0]{\spacefactor3000\relax}%
\providecommand \BibitemShut  [1]{\csname bibitem#1\endcsname}%
\let\auto@bib@innerbib\@empty
%</preamble>
\bibitem [{\citenamefont {Bahcall}\ and\ \citenamefont
  {Ulrich}(1988)}]{Bahcall:1987jc}%
  \BibitemOpen
  \bibfield  {author} {\bibinfo {author} {\bibfnamefont {J.~N.}\ \bibnamefont
  {Bahcall}}\ and\ \bibinfo {author} {\bibfnamefont {R.~K.}\ \bibnamefont
  {Ulrich}},\ }\href {\doibase 10.1103/RevModPhys.60.297} {\bibfield  {journal}
  {\bibinfo  {journal} {Rev. Mod. Phys.}\ }\textbf {\bibinfo {volume} {60}},\
  \bibinfo {pages} {297} (\bibinfo {year} {1988})}\BibitemShut {NoStop}%
%%CITATION = RMPHA,60,297;%%
\bibitem [{\citenamefont {Haxton}\ \emph {et~al.}(2013)\citenamefont {Haxton},
  \citenamefont {Hamish~Robertson},\ and\ \citenamefont
  {Serenelli}}]{Robertson:2012ib}%
  \BibitemOpen
  \bibfield  {author} {\bibinfo {author} {\bibfnamefont {W.~C.}\ \bibnamefont
  {Haxton}}, \bibinfo {author} {\bibfnamefont {R.~G.}\ \bibnamefont
  {Hamish~Robertson}}, \ and\ \bibinfo {author} {\bibfnamefont {A.~M.}\
  \bibnamefont {Serenelli}},\ }\href {\doibase
  10.1146/annurev-astro-081811-125539} {\bibfield  {journal} {\bibinfo
  {journal} {Ann. Rev. Astron. Astrophys.}\ }\textbf {\bibinfo {volume} {51}},\
  \bibinfo {pages} {21} (\bibinfo {year} {2013})},\ \Eprint
  {http://arxiv.org/abs/1208.5723} {arXiv:1208.5723 [astro-ph.SR]} \BibitemShut
  {NoStop}%
%%CITATION = ARXIV:1208.5723;%%
\bibitem [{\citenamefont {Vissani}(2017)}]{Vissani:2017dto}%
  \BibitemOpen
  \bibfield  {author} {\bibinfo {author} {\bibfnamefont {F.}~\bibnamefont
  {Vissani}},\ }\href {\doibase 10.15407/jnpae2017.01.005} {\bibfield
  {journal} {\bibinfo  {journal} {Nucl. Phys. Atom. Energy}\ }\textbf {\bibinfo
  {volume} {18}},\ \bibinfo {pages} {5} (\bibinfo {year} {2017})},\ \Eprint
  {http://arxiv.org/abs/1706.05435} {arXiv:1706.05435 [nucl-th]} \BibitemShut
  {NoStop}%
%%CITATION = ARXIV:1706.05435;%%
\bibitem [{\citenamefont {Capozzi}\ \emph {et~al.}(2018)\citenamefont
  {Capozzi}, \citenamefont {Li}, \citenamefont {Zhu},\ and\ \citenamefont
  {Beacom}}]{Capozzi:2018dat}%
  \BibitemOpen
  \bibfield  {author} {\bibinfo {author} {\bibfnamefont {F.}~\bibnamefont
  {Capozzi}}, \bibinfo {author} {\bibfnamefont {S.~W.}\ \bibnamefont {Li}},
  \bibinfo {author} {\bibfnamefont {G.}~\bibnamefont {Zhu}}, \ and\ \bibinfo
  {author} {\bibfnamefont {J.~F.}\ \bibnamefont {Beacom}},\ }\href@noop {} {\
  (\bibinfo {year} {2018})},\ \Eprint {http://arxiv.org/abs/1808.08232}
  {arXiv:1808.08232 [hep-ph]} \BibitemShut {NoStop}%
%%CITATION = ARXIV:1808.08232;%%
\bibitem [{\citenamefont {Burrows}(1990)}]{Burrows:1990ts}%
  \BibitemOpen
  \bibfield  {author} {\bibinfo {author} {\bibfnamefont {A.}~\bibnamefont
  {Burrows}},\ }\href {\doibase 10.1146/annurev.ns.40.120190.001145} {\bibfield
   {journal} {\bibinfo  {journal} {Ann. Rev. Nucl. Part. Sci.}\ }\textbf
  {\bibinfo {volume} {40}},\ \bibinfo {pages} {181} (\bibinfo {year}
  {1990})}\BibitemShut {NoStop}%
%%CITATION = ARNUA,40,181;%%
\bibitem [{\citenamefont {Cei}(2002)}]{Cei:2002mq}%
  \BibitemOpen
  \bibfield  {author} {\bibinfo {author} {\bibfnamefont {F.}~\bibnamefont
  {Cei}},\ }\bibfield  {booktitle} {\emph {\bibinfo {booktitle} {{Matter,
  antimatter and dark matter. Proceedings, 2nd International Workshop, Trento,
  Italy, October 29-30, 2001}}},\ }\href {\doibase 10.1142/S0217751X02011266}
  {\bibfield  {journal} {\bibinfo  {journal} {Int. J. Mod. Phys.}\ }\textbf
  {\bibinfo {volume} {A 17}},\ \bibinfo {pages} {1765} (\bibinfo {year}
  {2002})},\ \Eprint {http://arxiv.org/abs/hep-ex/0202043}
  {arXiv:hep-ex/0202043 [hep-ex]} \BibitemShut {NoStop}%
%%CITATION = HEP-EX/0202043;%%
\bibitem [{\citenamefont {Mezzacappa}(2005)}]{Mezzacappa:2005ju}%
  \BibitemOpen
  \bibfield  {author} {\bibinfo {author} {\bibfnamefont {A.}~\bibnamefont
  {Mezzacappa}},\ }\href {\doibase 10.1146/annurev.nucl.55.090704.151608}
  {\bibfield  {journal} {\bibinfo  {journal} {Ann. Rev. Nucl. Part. Sci.}\
  }\textbf {\bibinfo {volume} {55}},\ \bibinfo {pages} {467} (\bibinfo {year}
  {2005})}\BibitemShut {NoStop}%
%%CITATION = ARNUA,55,467;%%
\bibitem [{\citenamefont {Janka}(2012)}]{Janka:2012wk}%
  \BibitemOpen
  \bibfield  {author} {\bibinfo {author} {\bibfnamefont {H.-T.}\ \bibnamefont
  {Janka}},\ }\href {\doibase 10.1146/annurev-nucl-102711-094901} {\bibfield
  {journal} {\bibinfo  {journal} {Ann. Rev. Nucl. Part. Sci.}\ }\textbf
  {\bibinfo {volume} {62}},\ \bibinfo {pages} {407} (\bibinfo {year} {2012})},\
  \Eprint {http://arxiv.org/abs/1206.2503} {arXiv:1206.2503 [astro-ph.SR]}
  \BibitemShut {NoStop}%
%%CITATION = ARXIV:1206.2503;%%
\bibitem [{\citenamefont {Scholberg}(2018)}]{Scholberg:2017czd}%
  \BibitemOpen
  \bibfield  {author} {\bibinfo {author} {\bibfnamefont {K.}~\bibnamefont
  {Scholberg}},\ }\href {\doibase 10.1088/1361-6471/aa97be} {\bibfield
  {journal} {\bibinfo  {journal} {J. Phys.}\ }\textbf {\bibinfo {volume} {G
  45}},\ \bibinfo {pages} {014002} (\bibinfo {year} {2018})},\ \Eprint
  {http://arxiv.org/abs/1707.06384} {arXiv:1707.06384 [hep-ex]} \BibitemShut
  {NoStop}%
%%CITATION = ARXIV:1707.06384;%%
\bibitem [{\citenamefont {Ando}\ and\ \citenamefont
  {Sato}(2004)}]{Ando:2004hc}%
  \BibitemOpen
  \bibfield  {author} {\bibinfo {author} {\bibfnamefont {S.}~\bibnamefont
  {Ando}}\ and\ \bibinfo {author} {\bibfnamefont {K.}~\bibnamefont {Sato}},\
  }\href {\doibase 10.1088/1367-2630/6/1/170} {\bibfield  {journal} {\bibinfo
  {journal} {New J. Phys.}\ }\textbf {\bibinfo {volume} {6}},\ \bibinfo {pages}
  {170} (\bibinfo {year} {2004})},\ \Eprint
  {http://arxiv.org/abs/astro-ph/0410061} {arXiv:astro-ph/0410061 [astro-ph]}
  \BibitemShut {NoStop}%
%%CITATION = ASTRO-PH/0410061;%%
\bibitem [{\citenamefont {Beacom}(2010)}]{Beacom:2010kk}%
  \BibitemOpen
  \bibfield  {author} {\bibinfo {author} {\bibfnamefont {J.~F.}\ \bibnamefont
  {Beacom}},\ }\href {\doibase 10.1146/annurev.nucl.010909.083331} {\bibfield
  {journal} {\bibinfo  {journal} {Ann. Rev. Nucl. Part. Sci.}\ }\textbf
  {\bibinfo {volume} {60}},\ \bibinfo {pages} {439} (\bibinfo {year} {2010})},\
  \Eprint {http://arxiv.org/abs/1004.3311} {arXiv:1004.3311 [astro-ph.HE]}
  \BibitemShut {NoStop}%
%%CITATION = ARXIV:1004.3311;%%
\bibitem [{\citenamefont {Lunardini}(2016)}]{Lunardini:2010ab}%
  \BibitemOpen
  \bibfield  {author} {\bibinfo {author} {\bibfnamefont {C.}~\bibnamefont
  {Lunardini}},\ }\href {\doibase 10.1016/j.astropartphys.2016.02.005}
  {\bibfield  {journal} {\bibinfo  {journal} {Astropart. Phys.}\ }\textbf
  {\bibinfo {volume} {79}},\ \bibinfo {pages} {49} (\bibinfo {year} {2016})},\
  \Eprint {http://arxiv.org/abs/1007.3252} {arXiv:1007.3252 [astro-ph.CO]}
  \BibitemShut {NoStop}%
%%CITATION = ARXIV:1007.3252;%%
\bibitem [{\citenamefont {Acciarri}\ \emph
  {et~al.}(2016{\natexlab{a}})\citenamefont {Acciarri} \emph
  {et~al.}}]{Acciarri:2016crz}%
  \BibitemOpen
  \bibfield  {author} {\bibinfo {author} {\bibfnamefont {R.}~\bibnamefont
  {Acciarri}} \emph {et~al.} (\bibinfo {collaboration} {DUNE}),\ }\href@noop {}
  {\  (\bibinfo {year} {2016}{\natexlab{a}})},\ \Eprint
  {http://arxiv.org/abs/1601.05471} {arXiv:1601.05471 [physics.ins-det]}
  \BibitemShut {NoStop}%
%%CITATION = ARXIV:1601.05471;%%
\bibitem [{\citenamefont {Acciarri}\ \emph {et~al.}(2015)\citenamefont
  {Acciarri} \emph {et~al.}}]{Acciarri:2015uup}%
  \BibitemOpen
  \bibfield  {author} {\bibinfo {author} {\bibfnamefont {R.}~\bibnamefont
  {Acciarri}} \emph {et~al.} (\bibinfo {collaboration} {DUNE}),\ }\href@noop {}
  {\  (\bibinfo {year} {2015})},\ \Eprint {http://arxiv.org/abs/1512.06148}
  {arXiv:1512.06148 [physics.ins-det]} \BibitemShut {NoStop}%
%%CITATION = ARXIV:1512.06148;%%
\bibitem [{\citenamefont {Strait}\ \emph {et~al.}(2016)\citenamefont {Strait}
  \emph {et~al.}}]{Strait:2016mof}%
  \BibitemOpen
  \bibfield  {author} {\bibinfo {author} {\bibfnamefont {J.}~\bibnamefont
  {Strait}} \emph {et~al.} (\bibinfo {collaboration} {DUNE}),\ }\href@noop {}
  {\  (\bibinfo {year} {2016})},\ \Eprint {http://arxiv.org/abs/1601.05823}
  {arXiv:1601.05823 [physics.ins-det]} \BibitemShut {NoStop}%
%%CITATION = ARXIV:1601.05823;%%
\bibitem [{\citenamefont {Acciarri}\ \emph
  {et~al.}(2016{\natexlab{b}})\citenamefont {Acciarri} \emph
  {et~al.}}]{Acciarri:2016ooe}%
  \BibitemOpen
  \bibfield  {author} {\bibinfo {author} {\bibfnamefont {R.}~\bibnamefont
  {Acciarri}} \emph {et~al.} (\bibinfo {collaboration} {DUNE}),\ }\href@noop {}
  {\  (\bibinfo {year} {2016}{\natexlab{b}})},\ \Eprint
  {http://arxiv.org/abs/1601.02984} {arXiv:1601.02984 [physics.ins-det]}
  \BibitemShut {NoStop}%
%%CITATION = ARXIV:1601.02984;%%
\bibitem [{\citenamefont {Abi}\ \emph {et~al.}(2018{\natexlab{a}})\citenamefont
  {Abi} \emph {et~al.}}]{Abi:2018dnh}%
  \BibitemOpen
  \bibfield  {author} {\bibinfo {author} {\bibfnamefont {B.}~\bibnamefont
  {Abi}} \emph {et~al.} (\bibinfo {collaboration} {DUNE}),\ }\href@noop {} {\
  (\bibinfo {year} {2018}{\natexlab{a}})},\ \Eprint
  {http://arxiv.org/abs/1807.10334} {arXiv:1807.10334 [physics.ins-det]}
  \BibitemShut {NoStop}%
%%CITATION = ARXIV:1807.10334;%%
\bibitem [{\citenamefont {Abi}\ \emph {et~al.}(2018{\natexlab{b}})\citenamefont
  {Abi} \emph {et~al.}}]{Abi:2018alz}%
  \BibitemOpen
  \bibfield  {author} {\bibinfo {author} {\bibfnamefont {B.}~\bibnamefont
  {Abi}} \emph {et~al.} (\bibinfo {collaboration} {DUNE}),\ }\href@noop {} {\
  (\bibinfo {year} {2018}{\natexlab{b}})},\ \Eprint
  {http://arxiv.org/abs/1807.10327} {arXiv:1807.10327 [physics.ins-det]}
  \BibitemShut {NoStop}%
%%CITATION = ARXIV:1807.10327;%%
\bibitem [{\citenamefont {Abi}\ \emph {et~al.}(2018{\natexlab{c}})\citenamefont
  {Abi} \emph {et~al.}}]{Abi:2018rgm}%
  \BibitemOpen
  \bibfield  {author} {\bibinfo {author} {\bibfnamefont {B.}~\bibnamefont
  {Abi}} \emph {et~al.} (\bibinfo {collaboration} {DUNE}),\ }\href@noop {} {\
  (\bibinfo {year} {2018}{\natexlab{c}})},\ \Eprint
  {http://arxiv.org/abs/1807.10340} {arXiv:1807.10340 [physics.ins-det]}
  \BibitemShut {NoStop}%
%%CITATION = ARXIV:1807.10340;%%
\bibitem [{\citenamefont {Barker}\ \emph {et~al.}(2012)\citenamefont {Barker},
  \citenamefont {Mei},\ and\ \citenamefont {Zhang}}]{Barker:2012nb}%
  \BibitemOpen
  \bibfield  {author} {\bibinfo {author} {\bibfnamefont {D.}~\bibnamefont
  {Barker}}, \bibinfo {author} {\bibfnamefont {D.~M.}\ \bibnamefont {Mei}}, \
  and\ \bibinfo {author} {\bibfnamefont {C.}~\bibnamefont {Zhang}},\ }\href
  {\doibase 10.1103/PhysRevD.86.054001} {\bibfield  {journal} {\bibinfo
  {journal} {Phys. Rev.}\ }\textbf {\bibinfo {volume} {D 86}},\ \bibinfo
  {pages} {054001} (\bibinfo {year} {2012})},\ \Eprint
  {http://arxiv.org/abs/1202.5000} {arXiv:1202.5000 [physics.ins-det]}
  \BibitemShut {NoStop}%
%%CITATION = ARXIV:1202.5000;%%
\bibitem [{\citenamefont {Convery}()}]{Schoberg}%
  \BibitemOpen
  \bibfield  {author} {\bibinfo {author} {\bibfnamefont {M.}~\bibnamefont
  {Convery}},\ }\href {http://lbne2-docdb.fnal.gov/cgi-bin/DocumentDatabase}
  {\enquote {\bibinfo {title} {{Quick study of fast neutron background to
  supernova neutrinos}},}\ }\bibinfo {note} {LBNE-doc-6119-v1
  (2012)}\BibitemShut {NoStop}%
\bibitem [{\citenamefont {Franco}\ \emph {et~al.}(2016)\citenamefont {Franco}
  \emph {et~al.}}]{Franco:2015pha}%
  \BibitemOpen
  \bibfield  {author} {\bibinfo {author} {\bibfnamefont {D.}~\bibnamefont
  {Franco}} \emph {et~al.},\ }\href {\doibase 10.1088/1475-7516/2016/08/017}
  {\bibfield  {journal} {\bibinfo  {journal} {JCAP}\ }\textbf {\bibinfo
  {volume} {1608}},\ \bibinfo {pages} {017} (\bibinfo {year} {2016})},\ \Eprint
  {http://arxiv.org/abs/1510.04196} {arXiv:1510.04196 [physics.ins-det]}
  \BibitemShut {NoStop}%
%%CITATION = ARXIV:1510.04196;%%
\bibitem [{\citenamefont {Li}\ and\ \citenamefont {Beacom}(2014)}]{Li:2014sea}%
  \BibitemOpen
  \bibfield  {author} {\bibinfo {author} {\bibfnamefont {S.~W.}\ \bibnamefont
  {Li}}\ and\ \bibinfo {author} {\bibfnamefont {J.~F.}\ \bibnamefont
  {Beacom}},\ }\href {\doibase 10.1103/PhysRevC.89.045801} {\bibfield
  {journal} {\bibinfo  {journal} {Phys. Rev.}\ }\textbf {\bibinfo {volume} {C
  89}},\ \bibinfo {pages} {045801} (\bibinfo {year} {2014})},\ \Eprint
  {http://arxiv.org/abs/1402.4687} {arXiv:1402.4687 [hep-ph]} \BibitemShut
  {NoStop}%
%%CITATION = ARXIV:1402.4687;%%
\bibitem [{\citenamefont {Li}\ and\ \citenamefont
  {Beacom}(2015{\natexlab{a}})}]{Li:2015kpa}%
  \BibitemOpen
  \bibfield  {author} {\bibinfo {author} {\bibfnamefont {S.~W.}\ \bibnamefont
  {Li}}\ and\ \bibinfo {author} {\bibfnamefont {J.~F.}\ \bibnamefont
  {Beacom}},\ }\href {\doibase 10.1103/PhysRevD.91.105005} {\bibfield
  {journal} {\bibinfo  {journal} {Phys. Rev.}\ }\textbf {\bibinfo {volume} {D
  91}},\ \bibinfo {pages} {105005} (\bibinfo {year} {2015}{\natexlab{a}})},\
  \Eprint {http://arxiv.org/abs/1503.04823} {arXiv:1503.04823 [hep-ph]}
  \BibitemShut {NoStop}%
%%CITATION = ARXIV:1503.04823;%%
\bibitem [{\citenamefont {Li}\ and\ \citenamefont
  {Beacom}(2015{\natexlab{b}})}]{Li:2015lxa}%
  \BibitemOpen
  \bibfield  {author} {\bibinfo {author} {\bibfnamefont {S.~W.}\ \bibnamefont
  {Li}}\ and\ \bibinfo {author} {\bibfnamefont {J.~F.}\ \bibnamefont
  {Beacom}},\ }\href {\doibase 10.1103/PhysRevD.92.105033} {\bibfield
  {journal} {\bibinfo  {journal} {Phys. Rev.}\ }\textbf {\bibinfo {volume} {D
  92}},\ \bibinfo {pages} {105033} (\bibinfo {year} {2015}{\natexlab{b}})},\
  \Eprint {http://arxiv.org/abs/1508.05389} {arXiv:1508.05389
  [physics.ins-det]} \BibitemShut {NoStop}%
%%CITATION = ARXIV:1508.05389;%%
\bibitem [{\citenamefont {Galbiati}\ \emph {et~al.}(2005)\citenamefont
  {Galbiati}, \citenamefont {Pocar}, \citenamefont {Franco}, \citenamefont
  {Ianni}, \citenamefont {Cadonati},\ and\ \citenamefont
  {Schonert}}]{Galbiati:2004wx}%
  \BibitemOpen
  \bibfield  {author} {\bibinfo {author} {\bibfnamefont {C.}~\bibnamefont
  {Galbiati}}, \bibinfo {author} {\bibfnamefont {A.}~\bibnamefont {Pocar}},
  \bibinfo {author} {\bibfnamefont {D.}~\bibnamefont {Franco}}, \bibinfo
  {author} {\bibfnamefont {A.}~\bibnamefont {Ianni}}, \bibinfo {author}
  {\bibfnamefont {L.}~\bibnamefont {Cadonati}}, \ and\ \bibinfo {author}
  {\bibfnamefont {S.}~\bibnamefont {Schonert}},\ }\href {\doibase
  10.1103/PhysRevC.71.055805} {\bibfield  {journal} {\bibinfo  {journal} {Phys.
  Rev.}\ }\textbf {\bibinfo {volume} {C 71}},\ \bibinfo {pages} {055805}
  (\bibinfo {year} {2005})},\ \Eprint {http://arxiv.org/abs/hep-ph/0411002}
  {arXiv:hep-ph/0411002 [hep-ph]} \BibitemShut {NoStop}%
%%CITATION = HEP-PH/0411002;%%
\bibitem [{\citenamefont {Galbiati}\ and\ \citenamefont
  {Beacom}(2005)}]{Galbiati:2005ft}%
  \BibitemOpen
  \bibfield  {author} {\bibinfo {author} {\bibfnamefont {C.}~\bibnamefont
  {Galbiati}}\ and\ \bibinfo {author} {\bibfnamefont {J.~F.}\ \bibnamefont
  {Beacom}},\ }\href {\doibase 10.1103/PhysRevC.73.049906,
  10.1103/PhysRevC.72.025807} {\bibfield  {journal} {\bibinfo  {journal} {Phys.
  Rev.}\ }\textbf {\bibinfo {volume} {C 72}},\ \bibinfo {pages} {025807}
  (\bibinfo {year} {2005})},\ \bibinfo {note} {[Erratum: Phys. Rev. C 73,049906
  (2006)]},\ \Eprint {http://arxiv.org/abs/hep-ph/0504227}
  {arXiv:hep-ph/0504227 [hep-ph]} \BibitemShut {NoStop}%
%%CITATION = HEP-PH/0504227;%%
\bibitem [{\citenamefont {Bays}\ \emph {et~al.}(2012)\citenamefont {Bays} \emph
  {et~al.}}]{Bays:2011si}%
  \BibitemOpen
  \bibfield  {author} {\bibinfo {author} {\bibfnamefont {K.}~\bibnamefont
  {Bays}} \emph {et~al.} (\bibinfo {collaboration} {Super-Kamiokande}),\ }\href
  {\doibase 10.1103/PhysRevD.85.052007} {\bibfield  {journal} {\bibinfo
  {journal} {Phys. Rev.}\ }\textbf {\bibinfo {volume} {D 85}},\ \bibinfo
  {pages} {052007} (\bibinfo {year} {2012})},\ \Eprint
  {http://arxiv.org/abs/1111.5031} {arXiv:1111.5031 [hep-ex]} \BibitemShut
  {NoStop}%
%%CITATION = ARXIV:1111.5031;%%
\bibitem [{\citenamefont {Patrignani}\ \emph {et~al.}(2016)\citenamefont
  {Patrignani} \emph {et~al.}}]{Patrignani:2016xqp}%
  \BibitemOpen
  \bibfield  {author} {\bibinfo {author} {\bibfnamefont {C.}~\bibnamefont
  {Patrignani}} \emph {et~al.} (\bibinfo {collaboration} {Particle Data
  Group}),\ }\href {\doibase 10.1088/1674-1137/40/10/100001} {\bibfield
  {journal} {\bibinfo  {journal} {Chin. Phys. C}\ }\textbf {\bibinfo {volume}
  {40}},\ \bibinfo {pages} {100001} (\bibinfo {year} {2016})}\BibitemShut
  {NoStop}%
%%CITATION = CHPHD,C40,100001;%%
\bibitem [{\citenamefont {Konopinski}(1950)}]{Konopinski1950}%
  \BibitemOpen
  \bibfield  {author} {\bibinfo {author} {\bibfnamefont {E.~J.}\ \bibnamefont
  {Konopinski}},\ }\href
  {https://archive.org/details/TheTheoryOfBetaRadioactivity} {\emph {\bibinfo
  {title} {The Theory of Beta Radioactivity}}}\ (\bibinfo  {publisher} {Oxford
  University Press},\ \bibinfo {year} {1950})\BibitemShut {NoStop}%
\bibitem [{\citenamefont {Ponomarev}(1973)}]{Ponomarev:1973ya}%
  \BibitemOpen
  \bibfield  {author} {\bibinfo {author} {\bibfnamefont {L.~I.}\ \bibnamefont
  {Ponomarev}},\ }\href {\doibase 10.1146/annurev.ns.23.120173.002143}
  {\bibfield  {journal} {\bibinfo  {journal} {Ann. Rev. Nucl. Part. Sci.}\
  }\textbf {\bibinfo {volume} {23}},\ \bibinfo {pages} {395} (\bibinfo {year}
  {1973})}\BibitemShut {NoStop}%
%%CITATION = ARNUA,23,395;%%
\bibitem [{\citenamefont {Measday}(2001)}]{Measday:2001yr}%
  \BibitemOpen
  \bibfield  {author} {\bibinfo {author} {\bibfnamefont {D.~F.}\ \bibnamefont
  {Measday}},\ }\href {\doibase 10.1016/S0370-1573(01)00012-6} {\bibfield
  {journal} {\bibinfo  {journal} {Phys. Rept.}\ }\textbf {\bibinfo {volume}
  {354}},\ \bibinfo {pages} {243} (\bibinfo {year} {2001})}\BibitemShut
  {NoStop}%
%%CITATION = PRPLC,354,243;%%
\bibitem [{\citenamefont {Zhang}\ \emph {et~al.}(2016)\citenamefont {Zhang}
  \emph {et~al.}}]{Super-Kamiokande:2015xra}%
  \BibitemOpen
  \bibfield  {author} {\bibinfo {author} {\bibfnamefont {Y.}~\bibnamefont
  {Zhang}} \emph {et~al.} (\bibinfo {collaboration} {Super-Kamiokande}),\
  }\href {\doibase 10.1103/PhysRevD.93.012004} {\bibfield  {journal} {\bibinfo
  {journal} {Phys. Rev.}\ }\textbf {\bibinfo {volume} {D 93}},\ \bibinfo
  {pages} {012004} (\bibinfo {year} {2016})},\ \Eprint
  {http://arxiv.org/abs/1509.08168} {arXiv:1509.08168 [hep-ex]} \BibitemShut
  {NoStop}%
%%CITATION = ARXIV:1509.08168;%%
\bibitem [{\citenamefont {Bellini}\ \emph {et~al.}(2013)\citenamefont {Bellini}
  \emph {et~al.}}]{Bellini:2013pxa}%
  \BibitemOpen
  \bibfield  {author} {\bibinfo {author} {\bibfnamefont {G.}~\bibnamefont
  {Bellini}} \emph {et~al.} (\bibinfo {collaboration} {Borexino}),\ }\href
  {\doibase 10.1088/1475-7516/2013/08/049} {\bibfield  {journal} {\bibinfo
  {journal} {JCAP}\ }\textbf {\bibinfo {volume} {1308}},\ \bibinfo {pages}
  {049} (\bibinfo {year} {2013})},\ \Eprint {http://arxiv.org/abs/1304.7381}
  {arXiv:1304.7381 [physics.ins-det]} \BibitemShut {NoStop}%
%%CITATION = ARXIV:1304.7381;%%
\bibitem [{\citenamefont {Marchionni}(2013)}]{Marchionni:2013tfa}%
  \BibitemOpen
  \bibfield  {author} {\bibinfo {author} {\bibfnamefont {A.}~\bibnamefont
  {Marchionni}},\ }\href {\doibase 10.1146/annurev.nucl.012809.104445}
  {\bibfield  {journal} {\bibinfo  {journal} {Ann. Rev. Nucl. Part. Sci.}\
  }\textbf {\bibinfo {volume} {63}},\ \bibinfo {pages} {269} (\bibinfo {year}
  {2013})},\ \Eprint {http://arxiv.org/abs/1307.6918} {arXiv:1307.6918
  [physics.ins-det]} \BibitemShut {NoStop}%
%%CITATION = ARXIV:1307.6918;%%
\bibitem [{\citenamefont {Acciarri}\ \emph {et~al.}(2019)\citenamefont
  {Acciarri} \emph {et~al.}}]{Acciarri:2018myr}%
  \BibitemOpen
  \bibfield  {author} {\bibinfo {author} {\bibfnamefont {R.}~\bibnamefont
  {Acciarri}} \emph {et~al.} (\bibinfo {collaboration} {ArgoNeuT}),\ }\href
  {\doibase 10.1103/PhysRevD.99.012002} {\bibfield  {journal} {\bibinfo
  {journal} {Phys. Rev.}\ }\textbf {\bibinfo {volume} {D 99}},\ \bibinfo
  {pages} {012002} (\bibinfo {year} {2019})},\ \Eprint
  {http://arxiv.org/abs/1810.06502} {arXiv:1810.06502 [hep-ex]} \BibitemShut
  {NoStop}%
%%CITATION = ARXIV:1810.06502;%%
\bibitem [{\citenamefont {Berger}\ \emph {et~al.}(2005)\citenamefont {Berger},
  \citenamefont {Coursey}, \citenamefont {Zucker},\ and\ \citenamefont
  {Chang}}]{NIST}%
  \BibitemOpen
  \bibfield  {author} {\bibinfo {author} {\bibfnamefont {M.}~\bibnamefont
  {Berger}}, \bibinfo {author} {\bibfnamefont {J.}~\bibnamefont {Coursey}},
  \bibinfo {author} {\bibfnamefont {M.}~\bibnamefont {Zucker}}, \ and\ \bibinfo
  {author} {\bibfnamefont {J.}~\bibnamefont {Chang}},\ }\href@noop {} {\
  (\bibinfo {year} {2005})},\ \bibinfo {note}
  {[\url{http://physics.nist.gov/Star}, retrieved Oct. 22, 2018]}\BibitemShut
  {NoStop}%
\bibitem [{\citenamefont {Acciarri}\ \emph {et~al.}(2017)\citenamefont
  {Acciarri} \emph {et~al.}}]{Acciarri:2017sjy}%
  \BibitemOpen
  \bibfield  {author} {\bibinfo {author} {\bibfnamefont {R.}~\bibnamefont
  {Acciarri}} \emph {et~al.} (\bibinfo {collaboration} {MicroBooNE}),\ }\href
  {\doibase 10.1088/1748-0221/12/09/P09014} {\bibfield  {journal} {\bibinfo
  {journal} {JINST}\ }\textbf {\bibinfo {volume} {12}},\ \bibinfo {pages}
  {P09014} (\bibinfo {year} {2017})},\ \Eprint
  {http://arxiv.org/abs/1704.02927} {arXiv:1704.02927 [physics.ins-det]}
  \BibitemShut {NoStop}%
%%CITATION = ARXIV:1704.02927;%%
\bibitem [{\citenamefont {Böhlen}\ \emph {et~al.}(2014)\citenamefont
  {Böhlen}, \citenamefont {Cerutti}, \citenamefont {Chin}, \citenamefont
  {Fassò}, \citenamefont {Ferrari}, \citenamefont {Ortega}, \citenamefont
  {Mairani}, \citenamefont {Sala}, \citenamefont {Smirnov},\ and\ \citenamefont
  {Vlachoudis}}]{Bohlen:2014buj}%
  \BibitemOpen
  \bibfield  {author} {\bibinfo {author} {\bibfnamefont {T.~T.}\ \bibnamefont
  {Böhlen}}, \bibinfo {author} {\bibfnamefont {F.}~\bibnamefont {Cerutti}},
  \bibinfo {author} {\bibfnamefont {M.~P.~W.}\ \bibnamefont {Chin}}, \bibinfo
  {author} {\bibfnamefont {A.}~\bibnamefont {Fassò}}, \bibinfo {author}
  {\bibfnamefont {A.}~\bibnamefont {Ferrari}}, \bibinfo {author} {\bibfnamefont
  {P.~G.}\ \bibnamefont {Ortega}}, \bibinfo {author} {\bibfnamefont
  {A.}~\bibnamefont {Mairani}}, \bibinfo {author} {\bibfnamefont {P.~R.}\
  \bibnamefont {Sala}}, \bibinfo {author} {\bibfnamefont {G.}~\bibnamefont
  {Smirnov}}, \ and\ \bibinfo {author} {\bibfnamefont {V.}~\bibnamefont
  {Vlachoudis}},\ }\href {\doibase 10.1016/j.nds.2014.07.049} {\bibfield
  {journal} {\bibinfo  {journal} {Nucl. Data Sheets}\ }\textbf {\bibinfo
  {volume} {120}},\ \bibinfo {pages} {211} (\bibinfo {year}
  {2014})}\BibitemShut {NoStop}%
%%CITATION = NDTSB,120,211;%%
\bibitem [{\citenamefont {Ferrari}\ \emph {et~al.}(2005)\citenamefont
  {Ferrari}, \citenamefont {Sala}, \citenamefont {Fasso},\ and\ \citenamefont
  {Ranft}}]{Ferrari:2005zk}%
  \BibitemOpen
  \bibfield  {author} {\bibinfo {author} {\bibfnamefont {A.}~\bibnamefont
  {Ferrari}}, \bibinfo {author} {\bibfnamefont {P.~R.}\ \bibnamefont {Sala}},
  \bibinfo {author} {\bibfnamefont {A.}~\bibnamefont {Fasso}}, \ and\ \bibinfo
  {author} {\bibfnamefont {J.}~\bibnamefont {Ranft}},\ }\href@noop {} {\
  (\bibinfo {year} {2005})}\BibitemShut {NoStop}%
%%CITATION = CERN-2005-010;%%
\bibitem [{\citenamefont {Fasso}\ \emph {et~al.}(1994)\citenamefont {Fasso},
  \citenamefont {Ferrari}, \citenamefont {Ranft},\ and\ \citenamefont
  {Sala}}]{fasso1994fluka}%
  \BibitemOpen
  \bibfield  {author} {\bibinfo {author} {\bibfnamefont {A.}~\bibnamefont
  {Fasso}}, \bibinfo {author} {\bibfnamefont {A.}~\bibnamefont {Ferrari}},
  \bibinfo {author} {\bibfnamefont {J.}~\bibnamefont {Ranft}}, \ and\ \bibinfo
  {author} {\bibfnamefont {P.}~\bibnamefont {Sala}},\ }in\ \href@noop {} {\emph
  {\bibinfo {booktitle} {Proceedings of the Specialists Meeting on Shielding
  Aspects of Accelerators, Targets \& Irradiation Facilities, Arlington,
  USA}}}\ (\bibinfo {year} {1994})\ p.\ \bibinfo {pages} {287}\BibitemShut
  {NoStop}%
\bibitem [{\citenamefont {Ferrari}\ and\ \citenamefont
  {Sala}(1993)}]{Ferrari:1993xr}%
  \BibitemOpen
  \bibfield  {author} {\bibinfo {author} {\bibfnamefont {A.}~\bibnamefont
  {Ferrari}}\ and\ \bibinfo {author} {\bibfnamefont {P.~R.}\ \bibnamefont
  {Sala}},\ }in\ \href@noop {} {\emph {\bibinfo {booktitle} {{International
  Conference on Monte Carlo Simulation in High-Energy and Nuclear Physics - MC
  93 Tallahassee, Florida, February 22-26, 1993}}}}\ (\bibinfo {year} {1993})\
  pp.\ \bibinfo {pages} {277--288}\BibitemShut {NoStop}%
%%CITATION = INSPIRE-367781;%%
\bibitem [{\citenamefont {{Vitaly Kudryavtsev}}()}]{Vitaly}%
  \BibitemOpen
  \bibfield  {author} {\bibinfo {author} {\bibnamefont {{Vitaly
  Kudryavtsev}}},\ }\href@noop {} {}\bibinfo {note} {Private
  communication}\BibitemShut {NoStop}%
\bibitem [{\citenamefont {Kudryavtsev}(2009)}]{Kudryavtsev:2008qh}%
  \BibitemOpen
  \bibfield  {author} {\bibinfo {author} {\bibfnamefont {V.~A.}\ \bibnamefont
  {Kudryavtsev}},\ }\href {\doibase 10.1016/j.cpc.2008.10.013} {\bibfield
  {journal} {\bibinfo  {journal} {Comput. Phys. Commun.}\ }\textbf {\bibinfo
  {volume} {180}},\ \bibinfo {pages} {339} (\bibinfo {year} {2009})},\ \Eprint
  {http://arxiv.org/abs/0810.4635} {arXiv:0810.4635 [physics.comp-ph]}
  \BibitemShut {NoStop}%
%%CITATION = ARXIV:0810.4635;%%
\bibitem [{\citenamefont {Klinger}\ \emph {et~al.}(2015)\citenamefont
  {Klinger}, \citenamefont {Kudryavtsev}, \citenamefont {Richardson},\ and\
  \citenamefont {Spooner}}]{Klinger:2015kva}%
  \BibitemOpen
  \bibfield  {author} {\bibinfo {author} {\bibfnamefont {J.}~\bibnamefont
  {Klinger}}, \bibinfo {author} {\bibfnamefont {V.~A.}\ \bibnamefont
  {Kudryavtsev}}, \bibinfo {author} {\bibfnamefont {M.}~\bibnamefont
  {Richardson}}, \ and\ \bibinfo {author} {\bibfnamefont {N.~J.~C.}\
  \bibnamefont {Spooner}},\ }\href {\doibase 10.1016/j.physletb.2015.04.054}
  {\bibfield  {journal} {\bibinfo  {journal} {Phys. Lett.}\ }\textbf {\bibinfo
  {volume} {B 746}},\ \bibinfo {pages} {44} (\bibinfo {year} {2015})},\ \Eprint
  {http://arxiv.org/abs/1504.06520} {arXiv:1504.06520 [physics.ins-det]}
  \BibitemShut {NoStop}%
%%CITATION = ARXIV:1504.06520;%%
\bibitem [{\citenamefont {Rogers}(1990)}]{rogers:1990geology}%
  \BibitemOpen
  \bibfield  {author} {\bibinfo {author} {\bibfnamefont {H.}~\bibnamefont
  {Rogers}},\ }in\ \href
  {https://pubs.geoscienceworld.org/books/book/1824/chapter/107705612/geology-of-precambrian-rocks-in-the-poorman}
  {\emph {\bibinfo {booktitle} {Metallogeny of Gold in the Black Hills, South
  Dakota}}}\ (\bibinfo  {publisher} {Society of Economic Geologists},\ \bibinfo
  {year} {1990})\ p.\ \bibinfo {pages} {204}\BibitemShut {NoStop}%
\bibitem [{\citenamefont {Gaisser}(1990)}]{Gaisser:1990vg}%
  \BibitemOpen
  \bibfield  {author} {\bibinfo {author} {\bibfnamefont {T.~K.}\ \bibnamefont
  {Gaisser}},\ }\href
  {http://www.cambridge.org/uk/catalogue/catalogue.asp?isbn=0521326672} {\emph
  {\bibinfo {title} {{Cosmic rays and particle physics}}}}\ (\bibinfo {year}
  {1990})\BibitemShut {NoStop}%
%%CITATION = INSPIRE-306864;%%
\bibitem [{\citenamefont {Aglietta}\ \emph {et~al.}(1998)\citenamefont
  {Aglietta} \emph {et~al.}}]{Aglietta:1998nx}%
  \BibitemOpen
  \bibfield  {author} {\bibinfo {author} {\bibfnamefont {M.}~\bibnamefont
  {Aglietta}} \emph {et~al.} (\bibinfo {collaboration} {LVD}),\ }\href
  {\doibase 10.1103/PhysRevD.58.092005} {\bibfield  {journal} {\bibinfo
  {journal} {Phys. Rev.}\ }\textbf {\bibinfo {volume} {D 58}},\ \bibinfo
  {pages} {092005} (\bibinfo {year} {1998})},\ \Eprint
  {http://arxiv.org/abs/hep-ex/9806001} {arXiv:hep-ex/9806001 [hep-ex]}
  \BibitemShut {NoStop}%
%%CITATION = HEP-EX/9806001;%%
\bibitem [{\citenamefont {Antonioli}\ \emph {et~al.}(1997)\citenamefont
  {Antonioli}, \citenamefont {Ghetti}, \citenamefont {Korolkova}, \citenamefont
  {Kudryavtsev},\ and\ \citenamefont {Sartorelli}}]{Antonioli:1997qw}%
  \BibitemOpen
  \bibfield  {author} {\bibinfo {author} {\bibfnamefont {P.}~\bibnamefont
  {Antonioli}}, \bibinfo {author} {\bibfnamefont {C.}~\bibnamefont {Ghetti}},
  \bibinfo {author} {\bibfnamefont {E.~V.}\ \bibnamefont {Korolkova}}, \bibinfo
  {author} {\bibfnamefont {V.~A.}\ \bibnamefont {Kudryavtsev}}, \ and\ \bibinfo
  {author} {\bibfnamefont {G.}~\bibnamefont {Sartorelli}},\ }\href {\doibase
  10.1016/S0927-6505(97)00035-2} {\bibfield  {journal} {\bibinfo  {journal}
  {Astropart. Phys.}\ }\textbf {\bibinfo {volume} {7}},\ \bibinfo {pages} {357}
  (\bibinfo {year} {1997})},\ \Eprint {http://arxiv.org/abs/hep-ph/9705408}
  {arXiv:hep-ph/9705408 [hep-ph]} \BibitemShut {NoStop}%
%%CITATION = HEP-PH/9705408;%%
\bibitem [{\citenamefont {Cherry}\ \emph {et~al.}(1983)\citenamefont {Cherry},
  \citenamefont {Deakyne}, \citenamefont {Lande}, \citenamefont {Lee},
  \citenamefont {Steinberg}, \citenamefont {Cleveland},\ and\ \citenamefont
  {Fenyves}}]{Cherry:1983dp}%
  \BibitemOpen
  \bibfield  {author} {\bibinfo {author} {\bibfnamefont {M.~L.}\ \bibnamefont
  {Cherry}}, \bibinfo {author} {\bibfnamefont {M.}~\bibnamefont {Deakyne}},
  \bibinfo {author} {\bibfnamefont {K.}~\bibnamefont {Lande}}, \bibinfo
  {author} {\bibfnamefont {C.~K.}\ \bibnamefont {Lee}}, \bibinfo {author}
  {\bibfnamefont {R.~I.}\ \bibnamefont {Steinberg}}, \bibinfo {author}
  {\bibfnamefont {B.~T.}\ \bibnamefont {Cleveland}}, \ and\ \bibinfo {author}
  {\bibfnamefont {E.~J.}\ \bibnamefont {Fenyves}},\ }\href {\doibase
  10.1103/PhysRevD.27.1444} {\bibfield  {journal} {\bibinfo  {journal} {Phys.
  Rev.}\ }\textbf {\bibinfo {volume} {D 27}},\ \bibinfo {pages} {1444}
  (\bibinfo {year} {1983})}\BibitemShut {NoStop}%
%%CITATION = PHRVA,D27,1444;%%
\bibitem [{\citenamefont {Abgrall}\ \emph {et~al.}(2017)\citenamefont {Abgrall}
  \emph {et~al.}}]{Abgrall:2016cfi}%
  \BibitemOpen
  \bibfield  {author} {\bibinfo {author} {\bibfnamefont {N.}~\bibnamefont
  {Abgrall}} \emph {et~al.} (\bibinfo {collaboration} {MAJORANA}),\ }\href
  {\doibase 10.1016/j.astropartphys.2017.01.013} {\bibfield  {journal}
  {\bibinfo  {journal} {Astropart. Phys.}\ }\textbf {\bibinfo {volume} {93}},\
  \bibinfo {pages} {70} (\bibinfo {year} {2017})},\ \Eprint
  {http://arxiv.org/abs/1602.07742} {arXiv:1602.07742 [nucl-ex]} \BibitemShut
  {NoStop}%
%%CITATION = ARXIV:1602.07742;%%
\bibitem [{NND()}]{NNDC}%
  \BibitemOpen
  \href@noop {} {}\bibinfo {note} {National Nuclear Data Center, NuDat 2
  database, \url{http://www.nndc.bnl.gov/nudat2/}, online version
  2.7}\BibitemShut {NoStop}%
\bibitem [{\citenamefont {Chu}\ \emph {et~al.}()\citenamefont {Chu},
  \citenamefont {Ekström},\ and\ \citenamefont {Firestone}}]{TOI}%
  \BibitemOpen
  \bibfield  {author} {\bibinfo {author} {\bibfnamefont {S.}~\bibnamefont
  {Chu}}, \bibinfo {author} {\bibfnamefont {L.}~\bibnamefont {Ekström}}, \
  and\ \bibinfo {author} {\bibfnamefont {R.}~\bibnamefont {Firestone}},\
  }\href@noop {} {\ }\bibinfo {note} {WWW Table of Radioactive Isotopes,
  database version 1999-02-28,
  \url{http://nucleardata.nuclear.lu.se/toi/}}\BibitemShut {NoStop}%
\bibitem [{Bac()}]{Bacon}%
  \BibitemOpen
  \href@noop {} {}\bibinfo {note} {Bottle of Argon Counting Neutrons,
  \url{http://svoboda.ucdavis.edu/experiments/bacon}}\BibitemShut {NoStop}%
\bibitem [{\citenamefont {Bahcall}\ \emph {et~al.}(1996)\citenamefont
  {Bahcall}, \citenamefont {Lisi}, \citenamefont {Alburger}, \citenamefont
  {De~Braeckeleer}, \citenamefont {Freedman},\ and\ \citenamefont
  {Napolitano}}]{Bahcall:1996qv}%
  \BibitemOpen
  \bibfield  {author} {\bibinfo {author} {\bibfnamefont {J.~N.}\ \bibnamefont
  {Bahcall}}, \bibinfo {author} {\bibfnamefont {E.}~\bibnamefont {Lisi}},
  \bibinfo {author} {\bibfnamefont {D.~E.}\ \bibnamefont {Alburger}}, \bibinfo
  {author} {\bibfnamefont {L.}~\bibnamefont {De~Braeckeleer}}, \bibinfo
  {author} {\bibfnamefont {S.~J.}\ \bibnamefont {Freedman}}, \ and\ \bibinfo
  {author} {\bibfnamefont {J.}~\bibnamefont {Napolitano}},\ }\href {\doibase
  10.1103/PhysRevC.54.411} {\bibfield  {journal} {\bibinfo  {journal} {Phys.
  Rev.}\ }\textbf {\bibinfo {volume} {C 54}},\ \bibinfo {pages} {411} (\bibinfo
  {year} {1996})},\ \Eprint {http://arxiv.org/abs/nucl-th/9601044}
  {arXiv:nucl-th/9601044 [nucl-th]} \BibitemShut {NoStop}%
%%CITATION = NUCL-TH/9601044;%%
\bibitem [{\citenamefont {Winter}\ \emph {et~al.}(2006)\citenamefont {Winter},
  \citenamefont {Freedman}, \citenamefont {Rehm},\ and\ \citenamefont
  {Schiffer}}]{Winter:2004kf}%
  \BibitemOpen
  \bibfield  {author} {\bibinfo {author} {\bibfnamefont {W.~T.}\ \bibnamefont
  {Winter}}, \bibinfo {author} {\bibfnamefont {S.~J.}\ \bibnamefont
  {Freedman}}, \bibinfo {author} {\bibfnamefont {K.~E.}\ \bibnamefont {Rehm}},
  \ and\ \bibinfo {author} {\bibfnamefont {J.~P.}\ \bibnamefont {Schiffer}},\
  }\href {\doibase 10.1103/PhysRevC.73.025503} {\bibfield  {journal} {\bibinfo
  {journal} {Phys. Rev.}\ }\textbf {\bibinfo {volume} {C 73}},\ \bibinfo
  {pages} {025503} (\bibinfo {year} {2006})},\ \Eprint
  {http://arxiv.org/abs/nucl-ex/0406019} {arXiv:nucl-ex/0406019 [nucl-ex]}
  \BibitemShut {NoStop}%
%%CITATION = NUCL-EX/0406019;%%
\bibitem [{\citenamefont {Bhattacharya}\ \emph {et~al.}(2006)\citenamefont
  {Bhattacharya}, \citenamefont {Adelberger},\ and\ \citenamefont
  {Swanson}}]{Bhattacharya:2006ah}%
  \BibitemOpen
  \bibfield  {author} {\bibinfo {author} {\bibfnamefont {M.}~\bibnamefont
  {Bhattacharya}}, \bibinfo {author} {\bibfnamefont {E.~G.}\ \bibnamefont
  {Adelberger}}, \ and\ \bibinfo {author} {\bibfnamefont {H.~E.}\ \bibnamefont
  {Swanson}},\ }\href {\doibase 10.1103/PhysRevC.73.055802} {\bibfield
  {journal} {\bibinfo  {journal} {Phys. Rev.}\ }\textbf {\bibinfo {volume} {C
  73}},\ \bibinfo {pages} {055802} (\bibinfo {year} {2006})}\BibitemShut
  {NoStop}%
%%CITATION = PHRVA,C73,055802;%%
\bibitem [{\citenamefont {Santorelli}\ \emph {et~al.}(2018)\citenamefont
  {Santorelli}, \citenamefont {Di~Luise}, \citenamefont {Sanchez~Garcia},
  \citenamefont {Abia}, \citenamefont {Lux}, \citenamefont {Pesudo},\ and\
  \citenamefont {Romero}}]{Santorelli:2017aut}%
  \BibitemOpen
  \bibfield  {author} {\bibinfo {author} {\bibfnamefont {R.}~\bibnamefont
  {Santorelli}}, \bibinfo {author} {\bibfnamefont {S.}~\bibnamefont
  {Di~Luise}}, \bibinfo {author} {\bibfnamefont {E.}~\bibnamefont
  {Sanchez~Garcia}}, \bibinfo {author} {\bibfnamefont {P.~G.}\ \bibnamefont
  {Abia}}, \bibinfo {author} {\bibfnamefont {T.}~\bibnamefont {Lux}}, \bibinfo
  {author} {\bibfnamefont {V.}~\bibnamefont {Pesudo}}, \ and\ \bibinfo {author}
  {\bibfnamefont {L.}~\bibnamefont {Romero}},\ }\bibfield  {booktitle} {\emph
  {\bibinfo {booktitle} {{Proceedings, Light Detection in Noble Elements
  (LIDINE 2017): Menlo Park, CA, USA, September 22-24, 2017}}},\ }\href
  {\doibase 10.1088/1748-0221/13/04/C04015} {\bibfield  {journal} {\bibinfo
  {journal} {JINST}\ }\textbf {\bibinfo {volume} {13}},\ \bibinfo {pages}
  {C04015} (\bibinfo {year} {2018})},\ \Eprint
  {http://arxiv.org/abs/1712.07971} {arXiv:1712.07971 [physics.ins-det]}
  \BibitemShut {NoStop}%
%%CITATION = ARXIV:1712.07971;%%
\bibitem [{\citenamefont {Audi}\ \emph {et~al.}(2017)\citenamefont {Audi},
  \citenamefont {Kondev}, \citenamefont {Wang}, \citenamefont {Huang},\ and\
  \citenamefont {Naimi}}]{Audi_2017}%
  \BibitemOpen
  \bibfield  {author} {\bibinfo {author} {\bibfnamefont {G.}~\bibnamefont
  {Audi}}, \bibinfo {author} {\bibfnamefont {F.~G.}\ \bibnamefont {Kondev}},
  \bibinfo {author} {\bibfnamefont {M.}~\bibnamefont {Wang}}, \bibinfo {author}
  {\bibfnamefont {W.}~\bibnamefont {Huang}}, \ and\ \bibinfo {author}
  {\bibfnamefont {S.}~\bibnamefont {Naimi}},\ }\href {\doibase
  10.1088/1674-1137/41/3/030001} {\bibfield  {journal} {\bibinfo  {journal}
  {Chinese Physics C}\ }\textbf {\bibinfo {volume} {41}},\ \bibinfo {pages}
  {030001} (\bibinfo {year} {2017})}\BibitemShut {NoStop}%
\bibitem [{\citenamefont {Wang}\ \emph {et~al.}(2017)\citenamefont {Wang},
  \citenamefont {Audi}, \citenamefont {Kondev}, \citenamefont {Huang},
  \citenamefont {Naimi},\ and\ \citenamefont {Xu}}]{Wang_2017}%
  \BibitemOpen
  \bibfield  {author} {\bibinfo {author} {\bibfnamefont {M.}~\bibnamefont
  {Wang}}, \bibinfo {author} {\bibfnamefont {G.}~\bibnamefont {Audi}}, \bibinfo
  {author} {\bibfnamefont {F.~G.}\ \bibnamefont {Kondev}}, \bibinfo {author}
  {\bibfnamefont {W.}~\bibnamefont {Huang}}, \bibinfo {author} {\bibfnamefont
  {S.}~\bibnamefont {Naimi}}, \ and\ \bibinfo {author} {\bibfnamefont
  {X.}~\bibnamefont {Xu}},\ }\href {\doibase 10.1088/1674-1137/41/3/030003}
  {\bibfield  {journal} {\bibinfo  {journal} {Chinese Physics C}\ }\textbf
  {\bibinfo {volume} {41}},\ \bibinfo {pages} {030003} (\bibinfo {year}
  {2017})}\BibitemShut {NoStop}%
\bibitem [{\citenamefont {{Alejandro Sonzogni}}()}]{Sonzogni}%
  \BibitemOpen
  \bibfield  {author} {\bibinfo {author} {\bibnamefont {{Alejandro
  Sonzogni}}},\ }\href@noop {} {}\bibinfo {note} {Private
  communication}\BibitemShut {NoStop}%
\bibitem [{\citenamefont {Li}\ \emph {et~al.}(2019)\citenamefont {Li},
  \citenamefont {Bustamante},\ and\ \citenamefont {Beacom}}]{Li:2016kra}%
  \BibitemOpen
  \bibfield  {author} {\bibinfo {author} {\bibfnamefont {S.~W.}\ \bibnamefont
  {Li}}, \bibinfo {author} {\bibfnamefont {M.}~\bibnamefont {Bustamante}}, \
  and\ \bibinfo {author} {\bibfnamefont {J.~F.}\ \bibnamefont {Beacom}},\
  }\href {\doibase 10.1103/PhysRevLett.122.151101} {\bibfield  {journal}
  {\bibinfo  {journal} {Phys. Rev. Lett.}\ }\textbf {\bibinfo {volume} {122}},\
  \bibinfo {pages} {151101} (\bibinfo {year} {2019})},\ \Eprint
  {http://arxiv.org/abs/1606.06290} {arXiv:1606.06290 [astro-ph.HE]}
  \BibitemShut {NoStop}%
%%CITATION = ARXIV:1606.06290;%%
\bibitem [{\citenamefont {Bahcall}\ \emph {et~al.}(1986)\citenamefont
  {Bahcall}, \citenamefont {Baldo-Ceolin}, \citenamefont {Cline},\ and\
  \citenamefont {Rubbia}}]{Bahcall:1986ry}%
  \BibitemOpen
  \bibfield  {author} {\bibinfo {author} {\bibfnamefont {J.~N.}\ \bibnamefont
  {Bahcall}}, \bibinfo {author} {\bibfnamefont {M.}~\bibnamefont
  {Baldo-Ceolin}}, \bibinfo {author} {\bibfnamefont {D.~B.}\ \bibnamefont
  {Cline}}, \ and\ \bibinfo {author} {\bibfnamefont {C.}~\bibnamefont
  {Rubbia}},\ }\href {\doibase 10.1016/0370-2693(86)91519-4} {\bibfield
  {journal} {\bibinfo  {journal} {Phys. Lett. B}\ }\textbf {\bibinfo {volume}
  {178}},\ \bibinfo {pages} {324} (\bibinfo {year} {1986})}\BibitemShut
  {NoStop}%
%%CITATION = PHLTA,B178,324;%%
\bibitem [{\citenamefont {Arneodo}\ \emph {et~al.}(2000)\citenamefont {Arneodo}
  \emph {et~al.}}]{Arneodo:2000fa}%
  \BibitemOpen
  \bibfield  {author} {\bibinfo {author} {\bibfnamefont {F.}~\bibnamefont
  {Arneodo}} \emph {et~al.},\ }\href {\doibase 10.1016/S0168-9002(00)00520-9}
  {\bibfield  {journal} {\bibinfo  {journal} {Nucl. Instrum. Meth. A}\ }\textbf
  {\bibinfo {volume} {455}},\ \bibinfo {pages} {376} (\bibinfo {year}
  {2000})}\BibitemShut {NoStop}%
%%CITATION = NUIMA,A455,376;%%
\bibitem [{\citenamefont {Ioannisian}\ \emph {et~al.}(2017)\citenamefont
  {Ioannisian}, \citenamefont {Smirnov},\ and\ \citenamefont
  {Wyler}}]{Ioannisian:2017dkx}%
  \BibitemOpen
  \bibfield  {author} {\bibinfo {author} {\bibfnamefont {A.}~\bibnamefont
  {Ioannisian}}, \bibinfo {author} {\bibfnamefont {A.}~\bibnamefont {Smirnov}},
  \ and\ \bibinfo {author} {\bibfnamefont {D.}~\bibnamefont {Wyler}},\ }\href
  {\doibase 10.1103/PhysRevD.96.036005} {\bibfield  {journal} {\bibinfo
  {journal} {Phys. Rev.}\ }\textbf {\bibinfo {volume} {D 96}},\ \bibinfo
  {pages} {036005} (\bibinfo {year} {2017})},\ \Eprint
  {http://arxiv.org/abs/1702.06097} {arXiv:1702.06097 [hep-ph]} \BibitemShut
  {NoStop}%
%%CITATION = ARXIV:1702.06097;%%
\bibitem [{\citenamefont {Mirizzi}\ \emph {et~al.}(2016)\citenamefont
  {Mirizzi}, \citenamefont {Tamborra}, \citenamefont {Janka}, \citenamefont
  {Saviano}, \citenamefont {Scholberg}, \citenamefont {Bollig}, \citenamefont
  {Hudepohl},\ and\ \citenamefont {Chakraborty}}]{Mirizzi:2015eza}%
  \BibitemOpen
  \bibfield  {author} {\bibinfo {author} {\bibfnamefont {A.}~\bibnamefont
  {Mirizzi}}, \bibinfo {author} {\bibfnamefont {I.}~\bibnamefont {Tamborra}},
  \bibinfo {author} {\bibfnamefont {H.-T.}\ \bibnamefont {Janka}}, \bibinfo
  {author} {\bibfnamefont {N.}~\bibnamefont {Saviano}}, \bibinfo {author}
  {\bibfnamefont {K.}~\bibnamefont {Scholberg}}, \bibinfo {author}
  {\bibfnamefont {R.}~\bibnamefont {Bollig}}, \bibinfo {author} {\bibfnamefont
  {L.}~\bibnamefont {Hudepohl}}, \ and\ \bibinfo {author} {\bibfnamefont
  {S.}~\bibnamefont {Chakraborty}},\ }\href {\doibase
  10.1393/ncr/i2016-10120-8} {\bibfield  {journal} {\bibinfo  {journal} {Riv.
  Nuovo Cim.}\ }\textbf {\bibinfo {volume} {39}},\ \bibinfo {pages} {1}
  (\bibinfo {year} {2016})},\ \Eprint {http://arxiv.org/abs/1508.00785}
  {arXiv:1508.00785 [astro-ph.HE]} \BibitemShut {NoStop}%
%%CITATION = ARXIV:1508.00785;%%
\bibitem [{\citenamefont {Ankowski}\ \emph {et~al.}(2016)\citenamefont
  {Ankowski} \emph {et~al.}}]{Ankowski:2016lab}%
  \BibitemOpen
  \bibfield  {author} {\bibinfo {author} {\bibfnamefont {A.}~\bibnamefont
  {Ankowski}} \emph {et~al.},\ }in\ \href
  {https://inspirehep.net/record/1484266/files/arXiv:1608.07853.pdf} {\emph
  {\bibinfo {booktitle} {{Supernova Physics at DUNE Blacksburg, Virginia, USA,
  March 11-12, 2016}}}}\ (\bibinfo {year} {2016})\ \Eprint
  {http://arxiv.org/abs/1608.07853} {arXiv:1608.07853 [hep-ex]} \BibitemShut
  {NoStop}%
%%CITATION = ARXIV:1608.07853;%%
\bibitem [{\citenamefont {Nikrant}\ \emph {et~al.}(2018)\citenamefont
  {Nikrant}, \citenamefont {Laha},\ and\ \citenamefont
  {Horiuchi}}]{Nikrant:2017nya}%
  \BibitemOpen
  \bibfield  {author} {\bibinfo {author} {\bibfnamefont {A.}~\bibnamefont
  {Nikrant}}, \bibinfo {author} {\bibfnamefont {R.}~\bibnamefont {Laha}}, \
  and\ \bibinfo {author} {\bibfnamefont {S.}~\bibnamefont {Horiuchi}},\ }\href
  {\doibase 10.1103/PhysRevD.97.023019} {\bibfield  {journal} {\bibinfo
  {journal} {Phys. Rev.}\ }\textbf {\bibinfo {volume} {D 97}},\ \bibinfo
  {pages} {023019} (\bibinfo {year} {2018})},\ \Eprint
  {http://arxiv.org/abs/1711.00008} {arXiv:1711.00008 [astro-ph.HE]}
  \BibitemShut {NoStop}%
%%CITATION = ARXIV:1711.00008;%%
\bibitem [{\citenamefont {Seadrow}\ \emph {et~al.}(2018)\citenamefont
  {Seadrow}, \citenamefont {Burrows}, \citenamefont {Vartanyan}, \citenamefont
  {Radice},\ and\ \citenamefont {Skinner}}]{Seadrow:2018ftp}%
  \BibitemOpen
  \bibfield  {author} {\bibinfo {author} {\bibfnamefont {S.}~\bibnamefont
  {Seadrow}}, \bibinfo {author} {\bibfnamefont {A.}~\bibnamefont {Burrows}},
  \bibinfo {author} {\bibfnamefont {D.}~\bibnamefont {Vartanyan}}, \bibinfo
  {author} {\bibfnamefont {D.}~\bibnamefont {Radice}}, \ and\ \bibinfo {author}
  {\bibfnamefont {M.~A.}\ \bibnamefont {Skinner}},\ }\href {\doibase
  10.1093/mnras/sty2164} {\bibfield  {journal} {\bibinfo  {journal} {Mon. Not.
  Roy. Astron. Soc.}\ }\textbf {\bibinfo {volume} {480}},\ \bibinfo {pages}
  {4710} (\bibinfo {year} {2018})},\ \Eprint {http://arxiv.org/abs/1804.00689}
  {arXiv:1804.00689 [astro-ph.HE]} \BibitemShut {NoStop}%
%%CITATION = ARXIV:1804.00689;%%
\bibitem [{\citenamefont {Cocco}\ \emph {et~al.}(2004)\citenamefont {Cocco},
  \citenamefont {Ereditato}, \citenamefont {Fiorillo}, \citenamefont
  {Mangano},\ and\ \citenamefont {Pettorino}}]{Cocco:2004ac}%
  \BibitemOpen
  \bibfield  {author} {\bibinfo {author} {\bibfnamefont {A.~G.}\ \bibnamefont
  {Cocco}}, \bibinfo {author} {\bibfnamefont {A.}~\bibnamefont {Ereditato}},
  \bibinfo {author} {\bibfnamefont {G.}~\bibnamefont {Fiorillo}}, \bibinfo
  {author} {\bibfnamefont {G.}~\bibnamefont {Mangano}}, \ and\ \bibinfo
  {author} {\bibfnamefont {V.}~\bibnamefont {Pettorino}},\ }\href {\doibase
  10.1088/1475-7516/2004/12/002} {\bibfield  {journal} {\bibinfo  {journal}
  {JCAP}\ }\textbf {\bibinfo {volume} {0412}},\ \bibinfo {pages} {002}
  (\bibinfo {year} {2004})},\ \Eprint {http://arxiv.org/abs/hep-ph/0408031}
  {arXiv:hep-ph/0408031 [hep-ph]} \BibitemShut {NoStop}%
%%CITATION = HEP-PH/0408031;%%
\bibitem [{\citenamefont {Jeong}\ \emph {et~al.}(2018)\citenamefont {Jeong},
  \citenamefont {Palomares-Ruiz}, \citenamefont {Reno},\ and\ \citenamefont
  {Sarcevic}}]{Jeong:2018yts}%
  \BibitemOpen
  \bibfield  {author} {\bibinfo {author} {\bibfnamefont {Y.~S.}\ \bibnamefont
  {Jeong}}, \bibinfo {author} {\bibfnamefont {S.}~\bibnamefont
  {Palomares-Ruiz}}, \bibinfo {author} {\bibfnamefont {M.~H.}\ \bibnamefont
  {Reno}}, \ and\ \bibinfo {author} {\bibfnamefont {I.}~\bibnamefont
  {Sarcevic}},\ }\href {\doibase 10.1088/1475-7516/2018/06/019} {\bibfield
  {journal} {\bibinfo  {journal} {JCAP}\ }\textbf {\bibinfo {volume} {1806}},\
  \bibinfo {pages} {019} (\bibinfo {year} {2018})},\ \Eprint
  {http://arxiv.org/abs/1803.04541} {arXiv:1803.04541 [hep-ph]} \BibitemShut
  {NoStop}%
%%CITATION = ARXIV:1803.04541;%%
\bibitem [{\citenamefont {Møller}\ \emph {et~al.}(2018)\citenamefont {Møller},
  \citenamefont {Suliga}, \citenamefont {Tamborra},\ and\ \citenamefont
  {Denton}}]{Moller:2018kpn}%
  \BibitemOpen
  \bibfield  {author} {\bibinfo {author} {\bibfnamefont {K.}~\bibnamefont
  {Møller}}, \bibinfo {author} {\bibfnamefont {A.~M.}\ \bibnamefont {Suliga}},
  \bibinfo {author} {\bibfnamefont {I.}~\bibnamefont {Tamborra}}, \ and\
  \bibinfo {author} {\bibfnamefont {P.~B.}\ \bibnamefont {Denton}},\ }\href
  {\doibase 10.1088/1475-7516/2018/05/066} {\bibfield  {journal} {\bibinfo
  {journal} {JCAP}\ }\textbf {\bibinfo {volume} {1805}},\ \bibinfo {pages}
  {066} (\bibinfo {year} {2018})},\ \Eprint {http://arxiv.org/abs/1804.03157}
  {arXiv:1804.03157 [astro-ph.HE]} \BibitemShut {NoStop}%
%%CITATION = ARXIV:1804.03157;%%
\bibitem [{\citenamefont {Cline}\ \emph {et~al.}(1994)\citenamefont {Cline},
  \citenamefont {Fuller}, \citenamefont {Hong}, \citenamefont {Meyer},\ and\
  \citenamefont {Wilson}}]{Cline:1993rx}%
  \BibitemOpen
  \bibfield  {author} {\bibinfo {author} {\bibfnamefont {D.~B.}\ \bibnamefont
  {Cline}}, \bibinfo {author} {\bibfnamefont {G.~M.}\ \bibnamefont {Fuller}},
  \bibinfo {author} {\bibfnamefont {W.~P.}\ \bibnamefont {Hong}}, \bibinfo
  {author} {\bibfnamefont {B.}~\bibnamefont {Meyer}}, \ and\ \bibinfo {author}
  {\bibfnamefont {J.}~\bibnamefont {Wilson}},\ }\href {\doibase
  10.1103/PhysRevD.50.720} {\bibfield  {journal} {\bibinfo  {journal} {Phys.
  Rev.}\ }\textbf {\bibinfo {volume} {D 50}},\ \bibinfo {pages} {720} (\bibinfo
  {year} {1994})}\BibitemShut {NoStop}%
%%CITATION = PHRVA,D50,720;%%
\bibitem [{\citenamefont {Duba}\ \emph {et~al.}(2008)\citenamefont {Duba} \emph
  {et~al.}}]{Duba:2008zz}%
  \BibitemOpen
  \bibfield  {author} {\bibinfo {author} {\bibfnamefont {C.~A.}\ \bibnamefont
  {Duba}} \emph {et~al.},\ }\bibfield  {booktitle} {\emph {\bibinfo {booktitle}
  {{Proceedings, 23rd International Conference on Neutrino Physics and
  Astrophysics (Neutrino 2008): Christchurch, New Zealand, May 26-31, 2008}}},\
  }\href {\doibase 10.1088/1742-6596/136/4/042077} {\bibfield  {journal}
  {\bibinfo  {journal} {J. Phys. Conf. Ser.}\ }\textbf {\bibinfo {volume}
  {136}},\ \bibinfo {pages} {042077} (\bibinfo {year} {2008})}\BibitemShut
  {NoStop}%
%%CITATION = 00462,136,042077;%%
\bibitem [{\citenamefont {Malek}\ \emph {et~al.}(2003)\citenamefont {Malek}
  \emph {et~al.}}]{Malek:2002ns}%
  \BibitemOpen
  \bibfield  {author} {\bibinfo {author} {\bibfnamefont {M.}~\bibnamefont
  {Malek}} \emph {et~al.} (\bibinfo {collaboration} {Super-Kamiokande}),\
  }\href {\doibase 10.1103/PhysRevLett.90.061101} {\bibfield  {journal}
  {\bibinfo  {journal} {Phys. Rev. Lett.}\ }\textbf {\bibinfo {volume} {90}},\
  \bibinfo {pages} {061101} (\bibinfo {year} {2003})},\ \Eprint
  {http://arxiv.org/abs/hep-ex/0209028} {arXiv:hep-ex/0209028 [hep-ex]}
  \BibitemShut {NoStop}%
%%CITATION = HEP-EX/0209028;%%
\bibitem [{\citenamefont {Zhang}\ \emph {et~al.}(2015)\citenamefont {Zhang}
  \emph {et~al.}}]{Zhang:2013tua}%
  \BibitemOpen
  \bibfield  {author} {\bibinfo {author} {\bibfnamefont {H.}~\bibnamefont
  {Zhang}} \emph {et~al.} (\bibinfo {collaboration} {Super-Kamiokande}),\
  }\href {\doibase 10.1016/j.astropartphys.2014.05.004} {\bibfield  {journal}
  {\bibinfo  {journal} {Astropart. Phys.}\ }\textbf {\bibinfo {volume} {60}},\
  \bibinfo {pages} {41} (\bibinfo {year} {2015})},\ \Eprint
  {http://arxiv.org/abs/1311.3738} {arXiv:1311.3738 [hep-ex]} \BibitemShut
  {NoStop}%
%%CITATION = ARXIV:1311.3738;%%
\bibitem [{\citenamefont {Horiuchi}\ \emph {et~al.}(2009)\citenamefont
  {Horiuchi}, \citenamefont {Beacom},\ and\ \citenamefont
  {Dwek}}]{Horiuchi:2008jz}%
  \BibitemOpen
  \bibfield  {author} {\bibinfo {author} {\bibfnamefont {S.}~\bibnamefont
  {Horiuchi}}, \bibinfo {author} {\bibfnamefont {J.~F.}\ \bibnamefont
  {Beacom}}, \ and\ \bibinfo {author} {\bibfnamefont {E.}~\bibnamefont
  {Dwek}},\ }\href {\doibase 10.1103/PhysRevD.79.083013} {\bibfield  {journal}
  {\bibinfo  {journal} {Phys. Rev.}\ }\textbf {\bibinfo {volume} {D 79}},\
  \bibinfo {pages} {083013} (\bibinfo {year} {2009})},\ \Eprint
  {http://arxiv.org/abs/0812.3157} {arXiv:0812.3157 [astro-ph]} \BibitemShut
  {NoStop}%
%%CITATION = ARXIV:0812.3157;%%
\bibitem [{\citenamefont {Hopkins}\ and\ \citenamefont
  {Beacom}(2006)}]{Hopkins:2006bw}%
  \BibitemOpen
  \bibfield  {author} {\bibinfo {author} {\bibfnamefont {A.~M.}\ \bibnamefont
  {Hopkins}}\ and\ \bibinfo {author} {\bibfnamefont {J.~F.}\ \bibnamefont
  {Beacom}},\ }\href {\doibase 10.1086/506610} {\bibfield  {journal} {\bibinfo
  {journal} {Astrophys. J.}\ }\textbf {\bibinfo {volume} {651}},\ \bibinfo
  {pages} {142} (\bibinfo {year} {2006})},\ \Eprint
  {http://arxiv.org/abs/astro-ph/0601463} {arXiv:astro-ph/0601463 [astro-ph]}
  \BibitemShut {NoStop}%
%%CITATION = ASTRO-PH/0601463;%%
\bibitem [{\citenamefont {Gil~Botella}\ and\ \citenamefont
  {Rubbia}(2003)}]{GilBotella:2003sz}%
  \BibitemOpen
  \bibfield  {author} {\bibinfo {author} {\bibfnamefont {I.}~\bibnamefont
  {Gil~Botella}}\ and\ \bibinfo {author} {\bibfnamefont {A.}~\bibnamefont
  {Rubbia}},\ }\href {\doibase 10.1088/1475-7516/2003/10/009} {\bibfield
  {journal} {\bibinfo  {journal} {JCAP}\ }\textbf {\bibinfo {volume} {0310}},\
  \bibinfo {pages} {009} (\bibinfo {year} {2003})},\ \Eprint
  {http://arxiv.org/abs/hep-ph/0307244} {arXiv:hep-ph/0307244 [hep-ph]}
  \BibitemShut {NoStop}%
%%CITATION = HEP-PH/0307244;%%
\bibitem [{\citenamefont {Scholberg}(2012)}]{Scholberg:2012id}%
  \BibitemOpen
  \bibfield  {author} {\bibinfo {author} {\bibfnamefont {K.}~\bibnamefont
  {Scholberg}},\ }\href {\doibase 10.1146/annurev-nucl-102711-095006}
  {\bibfield  {journal} {\bibinfo  {journal} {Ann. Rev. Nucl. Part. Sci.}\
  }\textbf {\bibinfo {volume} {62}},\ \bibinfo {pages} {81} (\bibinfo {year}
  {2012})},\ \Eprint {http://arxiv.org/abs/1205.6003} {arXiv:1205.6003
  [astro-ph.IM]} \BibitemShut {NoStop}%
%%CITATION = ARXIV:1205.6003;%%
\bibitem [{\citenamefont {{Josh Klein}}()}]{Josh}%
  \BibitemOpen
  \bibfield  {author} {\bibinfo {author} {\bibnamefont {{Josh Klein}}},\
  }\href@noop {} {}\bibinfo {note} {Private communication}\BibitemShut
  {NoStop}%
\bibitem [{\citenamefont {Fischer}\ \emph {et~al.}(2019)\citenamefont {Fischer}
  \emph {et~al.}}]{Fischer:2019qfr}%
  \BibitemOpen
  \bibfield  {author} {\bibinfo {author} {\bibfnamefont {V.}~\bibnamefont
  {Fischer}} \emph {et~al.},\ }\href@noop {} {\  (\bibinfo {year} {2019})},\
  \Eprint {http://arxiv.org/abs/1902.00596} {arXiv:1902.00596 [nucl-ex]}
  \BibitemShut {NoStop}%
%%CITATION = ARXIV:1902.00596;%%
\bibitem [{\citenamefont {Benetti}\ \emph {et~al.}(2007)\citenamefont {Benetti}
  \emph {et~al.}}]{Benetti:2006az}%
  \BibitemOpen
  \bibfield  {author} {\bibinfo {author} {\bibfnamefont {P.}~\bibnamefont
  {Benetti}} \emph {et~al.} (\bibinfo {collaboration} {WARP}),\ }\href
  {\doibase 10.1016/j.nima.2007.01.106} {\bibfield  {journal} {\bibinfo
  {journal} {Nucl. Instrum. Meth.}\ }\textbf {\bibinfo {volume} {A 574}},\
  \bibinfo {pages} {83} (\bibinfo {year} {2007})},\ \Eprint
  {http://arxiv.org/abs/astro-ph/0603131} {arXiv:astro-ph/0603131 [astro-ph]}
  \BibitemShut {NoStop}%
%%CITATION = ASTRO-PH/0603131;%%
\bibitem [{\citenamefont {Agostini}\ \emph {et~al.}(2014)\citenamefont
  {Agostini} \emph {et~al.}}]{Agostini:2013tek}%
  \BibitemOpen
  \bibfield  {author} {\bibinfo {author} {\bibfnamefont {M.}~\bibnamefont
  {Agostini}} \emph {et~al.} (\bibinfo {collaboration} {GERDA}),\ }\href
  {\doibase 10.1140/epjc/s10052-014-2764-z} {\bibfield  {journal} {\bibinfo
  {journal} {Eur. Phys. J.}\ }\textbf {\bibinfo {volume} {C 74}},\ \bibinfo
  {pages} {2764} (\bibinfo {year} {2014})},\ \Eprint
  {http://arxiv.org/abs/1306.5084} {arXiv:1306.5084 [physics.ins-det]}
  \BibitemShut {NoStop}%
%%CITATION = ARXIV:1306.5084;%%
\bibitem [{\citenamefont {Barabash}\ \emph {et~al.}(2016)\citenamefont
  {Barabash}, \citenamefont {Saakyan},\ and\ \citenamefont
  {Umatov}}]{Barabash:2016uzl}%
  \BibitemOpen
  \bibfield  {author} {\bibinfo {author} {\bibfnamefont {A.~S.}\ \bibnamefont
  {Barabash}}, \bibinfo {author} {\bibfnamefont {R.~R.}\ \bibnamefont
  {Saakyan}}, \ and\ \bibinfo {author} {\bibfnamefont {V.~I.}\ \bibnamefont
  {Umatov}},\ }\href {\doibase 10.1016/j.nima.2016.09.042} {\bibfield
  {journal} {\bibinfo  {journal} {Nucl. Instrum. Meth.}\ }\textbf {\bibinfo
  {volume} {A 839}},\ \bibinfo {pages} {39} (\bibinfo {year} {2016})},\ \Eprint
  {http://arxiv.org/abs/1609.08890} {arXiv:1609.08890 [nucl-ex]} \BibitemShut
  {NoStop}%
%%CITATION = ARXIV:1609.08890;%%
\bibitem [{\citenamefont {Takeuchi}\ \emph {et~al.}(1999)\citenamefont
  {Takeuchi} \emph {et~al.}}]{Takeuchi:1999zq}%
  \BibitemOpen
  \bibfield  {author} {\bibinfo {author} {\bibfnamefont {Y.}~\bibnamefont
  {Takeuchi}} \emph {et~al.} (\bibinfo {collaboration} {SuperKamiokade}),\
  }\href {\doibase 10.1016/S0370-2693(99)00311-1} {\bibfield  {journal}
  {\bibinfo  {journal} {Phys. Lett.}\ }\textbf {\bibinfo {volume} {B 452}},\
  \bibinfo {pages} {418} (\bibinfo {year} {1999})},\ \Eprint
  {http://arxiv.org/abs/hep-ex/9903006} {arXiv:hep-ex/9903006 [hep-ex]}
  \BibitemShut {NoStop}%
%%CITATION = HEP-EX/9903006;%%
\bibitem [{\citenamefont {Hardell}\ and\ \citenamefont
  {Beer}(1970)}]{Hardell1970}%
  \BibitemOpen
  \bibfield  {author} {\bibinfo {author} {\bibfnamefont {R.}~\bibnamefont
  {Hardell}}\ and\ \bibinfo {author} {\bibfnamefont {C.}~\bibnamefont {Beer}},\
  }\href {http://iopscience.iop.org/1402-4896/1/2-3/003} {\bibfield  {journal}
  {\bibinfo  {journal} {Physica Scripta}\ }\textbf {\bibinfo {volume} {1}},\
  \bibinfo {pages} {85} (\bibinfo {year} {1970})}\BibitemShut {NoStop}%
\bibitem [{\citenamefont {Nesaraja}\ and\ \citenamefont
  {McCutchan}(2016)}]{Nesaraja:2016ktw}%
  \BibitemOpen
  \bibfield  {author} {\bibinfo {author} {\bibfnamefont {C.~D.}\ \bibnamefont
  {Nesaraja}}\ and\ \bibinfo {author} {\bibfnamefont {E.~A.}\ \bibnamefont
  {McCutchan}},\ }\href {\doibase 10.1016/j.nds.2016.02.001} {\bibfield
  {journal} {\bibinfo  {journal} {Nucl. Data Sheets}\ }\textbf {\bibinfo
  {volume} {133}},\ \bibinfo {pages} {1} (\bibinfo {year} {2016})}\BibitemShut
  {NoStop}%
%%CITATION = NDTSB,133,1;%%
\bibitem [{\citenamefont {{National Nuclear Data Center}}(2013)}]{CapGam}%
  \BibitemOpen
  \bibfield  {author} {\bibinfo {author} {\bibnamefont {{National Nuclear Data
  Center}}},\ }\href {https://www.nndc.bnl.gov/capgam/index.html} {\enquote
  {\bibinfo {title} {{CapGam}},}\ } (\bibinfo {year} {2013}),\ \bibinfo {note}
  {[Online; accessed 2017-9-30]}\BibitemShut {NoStop}%
\end{thebibliography}%

\end{document}